\documentclass[10pt,journal,compsoc]{IEEEtran}
\ifCLASSOPTIONcompsoc
  \usepackage[nocompress]{cite}
\else
  \usepackage{cite}
\fi
\ifCLASSINFOpdf
\else
\fi
\usepackage{multicol}
\usepackage{latexsym}
\usepackage[usenames,dvipsnames]{xcolor}
\definecolor{mygreen}{rgb}{0,0.6,0}
\definecolor{mygray}{rgb}{0.5,0.5,0.5}
\definecolor{mymauve}{rgb}{0.58,0,0.82}
\usepackage{listings}
\usepackage{amssymb}
\usepackage{wrapfig}
\usepackage{balance}
\usepackage{listings, xcolor}

\definecolor{verylightgray}{rgb}{.97,.97,.97}

\lstdefinelanguage{Solidity}{
	keywords=[1]{anonymous, assembly, assert, balance, break, call, callcode, case, catch, class, constant, continue, constructor, contract, debugger, default, delegatecall, delete, do, else, emit, event, experimental, export, external, false, finally, for, function, gas, if, implements, import, in, indexed, instanceof, interface, internal, is, length, library, log0, log1, log2, log3, log4, memory, modifier, new, payable, pragma, private, protected, public, pure, push, require, return, returns, revert, selfdestruct, send, solidity, storage, struct, suicide, super, switch, then, this, throw, transfer, true, try, typeof, using, view, while, with, addmod, ecrecover, keccak256, mulmod, ripemd160, sha256, sha3}, % generic keywords including crypto operations
	keywordstyle=[1]\color{blue}\bfseries,
	keywords=[2]{address, bool, byte, bytes, bytes1, bytes2, bytes3, bytes4, bytes5, bytes6, bytes7, bytes8, bytes9, bytes10, bytes11, bytes12, bytes13, bytes14, bytes15, bytes16, bytes17, bytes18, bytes19, bytes20, bytes21, bytes22, bytes23, bytes24, bytes25, bytes26, bytes27, bytes28, bytes29, bytes30, bytes31, bytes32, enum, int, int8, int16, int24, int32, int40, int48, int56, int64, int72, int80, int88, int96, int104, int112, int120, int128, int136, int144, int152, int160, int168, int176, int184, int192, int200, int208, int216, int224, int232, int240, int248, int256, mapping, string, uint, uint8, uint16, uint24, uint32, uint40, uint48, uint56, uint64, uint72, uint80, uint88, uint96, uint104, uint112, uint120, uint128, uint136, uint144, uint152, uint160, uint168, uint176, uint184, uint192, uint200, uint208, uint216, uint224, uint232, uint240, uint248, uint256, var, void, ether, finney, szabo, wei, days, hours, minutes, seconds, weeks, years},	% types; money and time units
	keywordstyle=[2]\color{teal}\bfseries,
	keywords=[3]{block, blockhash, coinbase, difficulty, gaslimit, number, timestamp,msg, gas, sender, sig, now, tx, gasprice, origin},	% environment variables
	keywordstyle=[3]\color{violet}\bfseries,
	keywords=[4]{calldataload, mload, calldatacopy, div,push, eq, jump, jumpi, push4, push1, and, stop,push20, mstore, iszero, push2, push21, lt, dup2},
	keywordstyle=[4]\color{blue}\bfseries,
	identifierstyle=\color{black},
	sensitive=false,
	comment=[l]{//},
	morecomment=[s]{/*}{*/},
	commentstyle=\color[RGB]{166,41,41}\bfseries,
	stringstyle=\color{red}\ttfamily,
	morestring=[b]',
	morestring=[b]"
}

\usepackage{graphicx}
\usepackage{url}

\usepackage{comment}
\graphicspath{{./figs/}}
\newcommand{\ignore}[1]{}

\begin{document}
\title{\texttt{SigRec}: Automatic Recovery of Function Signatures in Smart Contracts}

\author{Ting Chen,
	Zihao Li,
	Xiapu Luo,
	Xiaofeng Wang,
	Ting Wang,
	Zheyuan He,
	Kezhao Fang,
	Yufei Zhang,
	Hang Zhu,
	Hongwei Li,
	Yan Cheng,
	and Xiaosong Zhang% <-this % stops a space
	\IEEEcompsocitemizethanks{\IEEEcompsocthanksitem Ting Chen (brokendragon@uestc.edu.cn), Zheyuan He, Kezhao Fang, Yufei Zhang, Hang Zhu, Hongwei Li and Xiaosong Zhang are with University of Electronic Science and Technology of China, China.\protect
		\IEEEcompsocthanksitem Zihao Li and Xiapu Luo (csxluo@comp.polyu.edu.hk) are with Hong Kong Polytechnic University, Hong Kong. (Corresponding Author: Xiapu Luo.)
		\IEEEcompsocthanksitem Xiaofeng Wang is with Indiana University, Bloomington, USA.
		\IEEEcompsocthanksitem Ting Wang is with Pennsylvania State University, USA.
		\IEEEcompsocthanksitem Yan Cheng is with Ant Group, China.\\ Postal Code: 310013
	}% 
}

\IEEEtitleabstractindextext{%
\begin{abstract}
	Millions of smart contracts have been deployed onto Ethereum for providing various services, whose functions can be invoked. For this purpose, the caller needs to know the \textit{function signature} of a callee, which includes its function id and parameter types.
	Such signatures are \textit{critical} to many applications focusing on smart contracts, e.g., reverse engineering, fuzzing, attack detection, and profiling. 
	Unfortunately, it is challenging to recover the function signatures from contract bytecode, since neither debug information nor type information is present in the bytecode.
	To address this issue, prior approaches rely on source code, or a collection of known signatures from incomplete databases or incomplete heuristic rules, which, however, are far from adequate and cannot cope with the rapid growth of new contracts. 
	In this paper, we propose a novel solution that leverages how functions are handled by Ethereum virtual machine (EVM) to automatically recover function signatures.
	In particular, we exploit how smart contracts determine the functions to be invoked to locate and extract function ids, and 
	propose a new approach named \emph{type-aware} symbolic execution (TASE) that utilizes the semantics of EVM operations on parameters to identify the number and the types of parameters. 
	Moreover, we develop \texttt{\footnotesize SigRec}, a new tool for recovering function signatures from contract bytecode without the need of source code and function signature databases. 
	The extensive experimental results show that \texttt{\footnotesize SigRec} outperforms all existing tools, achieving an unprecedented 98.7\% accuracy within 0.074 seconds. 
	We further demonstrate that the recovered function signatures are useful in attack detection, fuzzing and reverse engineering of EVM bytecode.
\end{abstract}
\begin{IEEEkeywords}
smart contract, function signature, Ethereum, automatic recovery, type-aware symbolic execution.
\end{IEEEkeywords}}

\maketitle

\IEEEdisplaynontitleabstractindextext

\IEEEpeerreviewmaketitle

\IEEEraisesectionheading{\section{Introduction}}
\label{sec_intro}
\IEEEPARstart{M}ore than 30 million smart contracts have been deployed on Ethereum, the largest contract-hosting blockchain. An Ethereum \emph{smart contract} is a program (i.e., a collection of code and data) running on Ethereum virtual machine (EVM), which can be accessed through its address in Ethereum~\cite{solidity-intro}. A smart contract is typically written in a high-level language (e.g., Solidity~\cite{solidity}) and then compiled into EVM bytecode. After the bytecode is deployed, its public and external functions can be invoked. To invoke a contract function, the caller needs to know its function signature which consists of a function id and the list of parameter types~\cite{solidity-sig}\footnote{A function signature defined by Ethereum includes a function name and the list of parameter types~\cite{solidity-sig}. Since function id instead of function name is required to invoke a contract function, in this paper we just consider function id, parameter number and parameter types.}.
A \textit{function id} refers to the first 4 bytes in the Keccak-256 hash of a function name and the list of parameter types~\cite{solidity-sig}. 

Function signatures play a critical role in many applications focusing on smart contracts because their functions (in terms of function signatures) need to be first identified before their behaviors can be evaluated. For example, some studies recognize wallet contracts~\cite{wallet}, token contracts~\cite{token1,token2,chen2019tokenscope}, and the contracts for Ethereum name service~\cite{token2} based on function signatures. \S \ref{sec_app} shows that function signatures can be used to enhance the results of reverse engineering results contract bytecode by adding types, meaningful variable names and simplifying the code for accessing parameters. 
Besides, knowing the parameter list of a target function, a fuzzer can strategically mutate the test cases~\cite{contractfuzzer,ILF} for better vulnerability detection. Our experimental results (\S \ref{sec_app}) show that function signatures enable ContractFuzzer~\cite{contractfuzzer} to find 23\% more bugs than it could without these signatures. Function signatures are also important to the detection of attacks against smart contracts. A prominent case is the \textit{short address} attack that exploits an EVM vulnerability in handling a malformed actual argument of the \textsf{\small address} type, which is shorter than a valid \textsf{\small address}~\cite{short}. We develop ParChecker, a new tool based on the recovered function signatures to detect malformed actual arguments including short address attacks (\S \ref{sec_app}).

\noindent\textbf{Challenges in signature discovery.} It is challenging to recover function signatures from bytecode since there is no debug information in contract bytecode and the bytecode is \textit{untyped}, i.e., parameters are kept in 256-bit words without type information~\cite{safeevm}. All existing approaches fail to effectively recover function signatures  (\S \ref{sec_compare}). 
More precisely, an intuitive method is to extract the function signatures from the source code of smart contracts, which, however, is only available for a small proportion of contracts ($<1\%$ until 2017~\cite{immutability}). Another method retrieves signatures from existing databases, like Ethereum Function Signature Database (EFSD)~\cite{database}, through function ids. For example, Gigahorse~\cite{gigahorse}, Eveem~\cite{eveem_web}, and Online Solidity Decompiler (OSD)~\cite{osd_web} rely on EFSD, while EVM Bytecode Decompiler (EBD)~\cite{ebd_web} and JEB~\cite{jeb_web} maintain their own databases. However, these databases are incomplete, only covering 31.7\% of the function signatures in the wild (Supplementary material A) and quickly become out of date without constantly updating. 

Alternatively, one could reverse engineering a function id by enumerating all possible parameter type combinations to compute a hash, together with a possible function name. Given the huge space of possible combinations, this attempt is fragile in general. Abi Decompiler~\cite{abi_decompiler} adopts this approach, which only covers 12.3\% of function signatures (Supplementary material A).  
Finally, one could derive function signatures using some heuristics: Gigahorse infers the number of parameters based upon the number of items that a function pops from stack~\cite{gigahorse}; Eveem regards a type as an \textsf{\small array} if its layout contains an \textit{offset} field and a \textit{num} field. Although this direction is promising, existing approaches turn out to be ineffective due to the incompleteness of their heuristics. For example, Eveem can only achieve an accuracy of 58.1\% and 18.3\% to infer the function signatures in closed-source smart contracts and synthesized contracts, respectively (\S \ref{sec_compare}), even it leverages an existing database, EFSD. 

\noindent\textbf{Our work.} In this paper, we propose a new solution called \texttt{\footnotesize SigRec} to automatically recover function signatures from EVM bytecode compiled from two mainstream compilers (i.e., Solidity and Vyper) for smart contracts \emph{without} the need of source code and other databases. Our key observation is that even in the absence of type information, the way EVM bytecode handles a function call and its inputs uniquely characterizes different parameter types. For example, a \textsf{\small uint32} argument will be extracted using a 32-bits mask applied to the input data (R11, \S \ref{rule_other}). 
Based upon this observation, we first generalize the semantics of such type-related operations into rules (\S \ref{sec_rule}). Then, we design \emph{type-aware} symbolic execution (TASE) to explore the EVM instructions that manipulate parameters, and use the rules for inferring the types of parameters (\S \ref{sec_sigrec}). 
Since the EVM instructions handling parameters are typically near a function's entry point, TASE can handle them very effectively and efficiently 
to keep up with the ever-growing Ethereum smart contracts.  

Studies on reverse engineering of variable types from binary executables by leveraging the semantics of instructions~\cite{caballero_ndss,lightweight,rewards,simple,tie,array,retypd,katz,xu,howard} %, which infer types from the semantics of instructions, 
are related to our work. However, TASE has several differences with existing approaches. First, to recover complicated types that cannot be handled by existing techniques, TASE infers how variables are affected by parameters through symbolic execution (SE). For example, to identify a multidimensional \textsf{\small array}, TASE checks whether the read location is calculated by adding the value of the \emph{offset} field to the base (\S \ref{rule_calldataload}). Second, semantic knowledge summarized from binary code cannot be applied to 
EVM bytecode, due to architecture differences (\S \ref{sec_related_binary}) and unique parameter types defined in smart contracts, e.g., 256-bit integers (no more than 64 bit in x64 binaries), the 20-byte \textsf{\small address} type, and reversed \textsf{\small array} notation (\S \ref{sec_types}). New knowledge, therefore, needs to be discovered to support type inference through TASE. Third, by exploiting the fact that the code responsible for handling parameters is usually around a function's entry point, TASE runs extremely fast (0.074s to recover one function signature on average, \S \ref{sec_efficiency}). 

We conduct extensive experiments to evaluate \texttt{\footnotesize SigRec} and compare it with all existing approaches. First, we collect all 119,404 unique (i.e., duplicates are eliminated) open-source smart contracts on Ethereum, which include 210,869 public/external functions with unique function signatures, and use them to evaluate the accuracy of \texttt{\footnotesize SigRec}.
The experimental result shows that \texttt{\footnotesize SigRec} achieves an average accuracy of 98.7\% (\S \ref{sec_accuracy}), and the accuracy never goes below 96\% across all compilers (from V 0.1.1 to V 0.8.0) with or without optimization (\S \ref{sec_compiler}). 
Second, we compare \texttt{\footnotesize SigRec} with all existing approaches, i.e., OSD~\cite{osd_web}, EBD~\cite{ebd_web}, JEB~\cite{jeb_web}, Gigahorse~\cite{gigahorse} and Eveem~\cite{eveem_web}. The experimental results show that \texttt{\footnotesize SigRec} correctly recovers much more signatures, outperforming them by at least 22.5                      \%, 40.1\% and 80.5\% in processing open-source, closed-source and synthesized smart contracts, respectively (\S \ref{sec_compare}). Manual investigation shows that for many function signatures, existing approaches report wrong types, produce \textit{nonexistent} types, output an \textit{unspecific} type, add \textit{nonexistent} parameters, miss parameters or fail to generate function signatures (\S \ref{sec_compare}). 
Third, we apply \texttt{\footnotesize SigRec} to the bytecode of all 37,009,570 smart contracts deployed on Ethereum with 47,329,149 public/external functions, and found that it only takes 0.074 seconds on average to recover a function signature (\S \ref{sec_efficiency}).

We further demonstrate how the recovered function signatures can enable the detection of stealthy short address attacks, empower the state-of-the-art fuzzer to discover more vulnerabilities, and enhance the result of reverse engineering contract bytecode (\S \ref{sec_app}). 

\noindent\textbf{Contributions.} The major contributions of the paper are summarized as follows:

\vspace{1pt}\noindent$\bullet$ We propose a novel solution that exploits the semantics of EVM instructions to correctly and efficiently recover function signatures from the bytecode of smart contracts.

\vspace{1pt}\noindent$\bullet$ We develop \texttt{\footnotesize SigRec} based on our new solution, and conduct extensive experiments to evaluate it. % with the functions of all deployed smart contracts. 
The experimental results show that \texttt{\footnotesize SigRec} achieves nearly 100\% accuracy in signature recovery, under different compiler versions, within 0.074 seconds on average, and it significantly outperforms existing methods.  We have deployed \texttt{\footnotesize SigRec} as an online web service  \url{http://bit.ly/SigRecWS} and will release its code after paper publication.

\vspace{1pt}\noindent$\bullet$ We use three important applications to demonstrate the usage of recovering function signatures, including attack detection, fuzzing and reverse engineering of EVM bytecode. 

\section{Accessing Patterns of Parameters}
\label{sec_calldata}
This section first introduces how function invocation is performed in EVM and then elaborate more on the EVM instructions and variables involved in a function invocation. 
The concepts of account, bytecode of smart contracts, and public/external function can be found in Supplementary material B.

\noindent\noindent\textbf{Function invocation}. To invoke a public/external function, a message will be sent by an EOA account or a smart contract account\cite{yellow}. The message contains the address of the smart contract whose function will be invoked and the \emph{call data} field which indicates the function to be invoked and carries actual arguments. The first 4 bytes of the call data is the function id of the function being called, which is followed by the arguments~\cite{solidity-sig}.
For example, to call a function whose signature is ``transfer(address, uint256)'', the call data begins with its function id 0xa9059cbb, followed by two arguments, including an \textsf{\small address} and a 256-bits unsigned integer.

\subsection{EVM Instructions to Read the Call Data}
\begin{figure}
	\centering
	\includegraphics[width=0.35\textwidth]{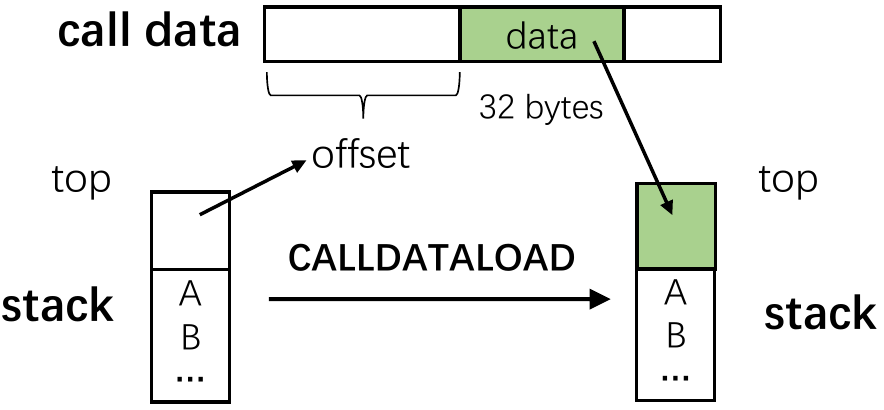}
	\vspace{-2pt}
	\caption{The process of executing a \textsf{\scriptsize CALLDATALOAD}}
	\label{fig_calldataload}
\end{figure}

\noindent\textbf{\textsf{\footnotesize CALLDATALOAD}}. \textsf{\footnotesize CALLDATALOAD} reads 32 bytes into the top of the stack~\cite{yellow}. As shown in Fig. \ref{fig_calldataload}, \textsf{\footnotesize CALLDATALOAD} first reads the top item from the stack, which is the offset used to locate the data. After that, the top item of the stack is removed. It then reads 32 bytes data from the call data starting from the offset. Finally, the data is pushed to the stack. 

\hspace{2pt}
\begin{figure}
	\centering
	\includegraphics[width=0.35\textwidth]{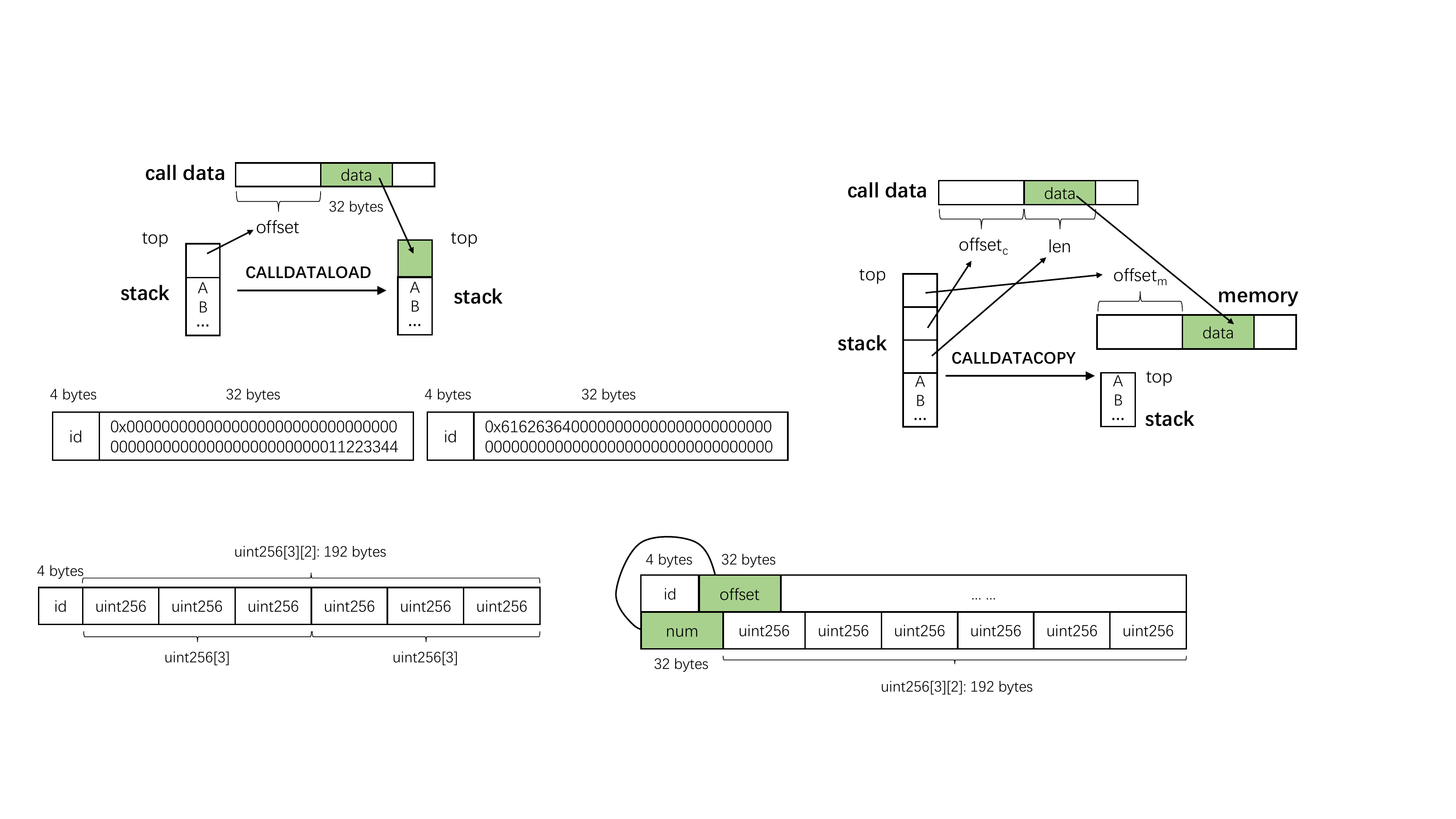}
	\vspace{-2pt}
	\caption{The process of running a \textsf{\scriptsize CALLDATACOPY}}
	\label{fig_calldatacopy}
\end{figure}

\noindent\textbf{\textsf{\footnotesize CALLDATACOPY}}. \textsf{\footnotesize CALLDATACOPY} reads variable-length data from the call data into memory~\cite{yellow}.
As shown in Fig. \ref{fig_calldatacopy}, \textsf{\footnotesize CALLDATACOPY} consumes top three stack items. The second item denotes the offset ($\textit{offset}_c$) in the call data from where the data will be copied. The first item is the offset ($\textit{offset}_m$) in the memory to where data will be copied. The third item indicates the data length.

\subsection{Accessing Function ID}
\label{sec_overview_calldata}
To invoke a public/external function in a smart contract, the caller should provide the \emph{function id}. 
The callee extracts it from the call data to determine the function to be invoked 
by executing a \textsf{\footnotesize CALLDATALOAD} with the offset being 0 to read 32 bytes from the beginning of the call data, whose highest 4 bytes is the function id. 
Then, the callee uses a \textsf{\footnotesize DIV} instruction (i.e., unsigned integer division~\cite{yellow}) or an \textsf{\footnotesize  SHR} instruction (i.e., bitwise right shift~\cite{eip145}) to move the function id to the lowest 4 bytes. 

\subsection{Accessing Different Types of Parameters}
\label{sec_types}
Our work support two mainstream compilers, Solidity and Vyper which have different parameter types.

\subsubsection{Solidity}
We classify all parameter types supported by Solidity into five categories: basic types, \textsf{\small array}
, \textsf{\small bytes} and \textsf{\small string}, and \textsf{\small struct}. %Currently, this work 
\noindent\textbf{1. Basic Types:}
There are five basic types.

\begin{figure}
	\centering
	\includegraphics[width=0.35\textwidth]{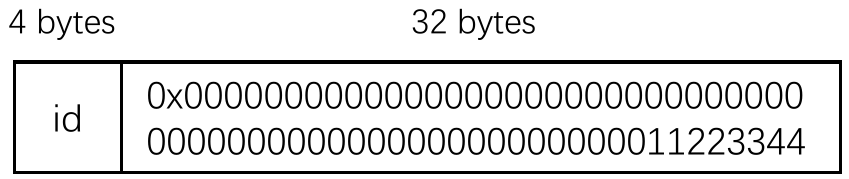}
	\vspace{-2pt}
	\caption{The call data layout of a \textsf{\footnotesize uint32}, 0x11223344}   
	\label{fig_uint} 
\end{figure}

\noindent\textbf{(1)}~\textbf{\textsf{\small uint$\langle M\rangle$}, $8\le M\le 256, M\%8==0$}. \textsf{\small uint$\langle M\rangle$} is an unsigned integer of $M$ bits~\cite{solidity-sig}. 
If a public/external function has a {\small uint$\langle M\rangle$} type parameter, it will be extended on the higher-order (left) side with zeros to make the length be 32 bytes~\cite{solidity-sig}.
Fig. \ref{fig_uint} shows the call data layout of a public/external function with one \textsf{\small uint32} argument whose value is 0x11223344. The call data begins with a 4-byte function id, followed by the extended 32-bytes value. To read a \textsf{\small uint$\langle M\rangle$}, the smart contract executes a \textsf{\footnotesize CALLDATALOAD} with the offset indicating the start of the extended value. After that, the extended value will be masked by an \textsf{\footnotesize AND} instruction, and the result is the \textsf{\small uint$\langle M\rangle$} argument before padding. The masking is not needed to access a \textsf{\small uint256} because it is not extended. 

\hspace{2pt}
\begin{figure}
	\centering
	\includegraphics[width=0.35\textwidth]{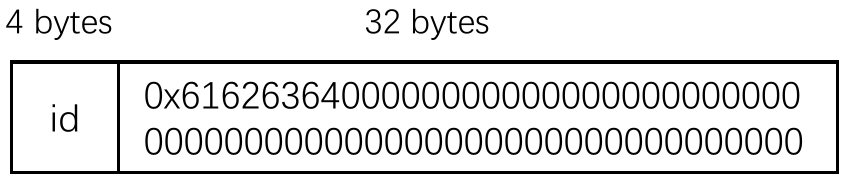}
	\vspace{-2ex}
	\caption{The call data layout of a \textsf{\footnotesize bytes4}, `abcd'}   
	\label{fig_bytes4}
\end{figure}

\noindent\textbf{(2)}~\textbf{\textsf{\small int$\langle M\rangle$}, $8\le M\le 256, M\%8==0$}. \textsf{\small int$\langle M\rangle$} is a signed integer of $M$ bits~\cite{solidity-sig}. The layout and accessing pattern of an \textsf{\small int$\langle M\rangle$} is similar to that of a \textsf{\small uint$\langle M\rangle$}, except that a \textsf{\footnotesize SIGNEXTEND} rather than an \textsf{\footnotesize AND} is used to mask an extended value because \textsf{\footnotesize SIGNEXTEND} is responsible for sign extension~\cite{yellow}. Similarly, the \textsf{\footnotesize SIGNEXTEND} is not needed to access an \textsf{\small int256} because the highest bit of an \textsf{\small int256} indicates the sign.
To distinguish an \textsf{\small int256} from a \textsf{\small uint256}, both of which are not extended, we leverage some instructions (e.g., \textsf{\footnotesize SDIV}) that can only access signed integers.

\noindent\textbf{(3)}~\textbf{\textsf{\small address}}.
Every account has a unique 20-bytes \textsf{\small address}~\cite{yellow}. We find that the call data layout of an \textsf{\small address} and reading an \textsf{\small address} from the call data are the same as a \textsf{\small uint160}. To differentiate them, we exploit the fact that mathematic operations can involve a \textsf{\small uint160} rather than an \textsf{\small address}. 

\noindent\textbf{(4)}~\textbf{\textsf{\small bool}}. It can be true or false. The layout and reading a \textsf{\small bool} from the call data are similar to that of a \textsf{\small uint$\langle M\rangle$}, except that two consecutive \textsf{\footnotesize ISZERO}s are used for masking. An \textsf{\footnotesize ISZERO} pushes a one to the stack if the top stack item is zero and a zero otherwise~\cite{yellow}. Hence, two consecutive \textsf{\footnotesize ISZERO}s push a one to the stack if the top stack item is not zero, otherwise push a zero. 

\noindent\textbf{(5)}~\textbf{\textsf{\small bytes$\langle M\rangle$}, $0< M\le 32$}. A \textsf{\small bytes$\langle M\rangle$} is a byte sequence of $M$ bytes~\cite{solidity-sig}. 
If  a public/external function has a \textsf{\small bytes$\langle M\rangle$} type parameter, it will be extended on the lower-order (right) side with zero to make the length be 32 bytes~\cite{solidity-sig}. 
Fig. \ref{fig_bytes4} shows the call data layout of a public/external function with one \textsf{\small bytes4} argument whose value is `abcd'. A \textsf{\footnotesize CALLDATALOAD} is needed to read a \textsf{\small bytes$\langle M\rangle$} from the call data, and an \textsf{\footnotesize AND} is used for masking. After masking, the higher-order bytes of a \textsf{\small bytes$\langle M\rangle$} is retained. 
By contrast, for a \textsf{\small uint$\langle M\rangle$}, the masking retains the lower-order bytes since a \textsf{\small uint} is extended on the higher-order side. 
Since \textsf{\small bytes32} and \textsf{\small uint256} are not extended, we differentiate them by exploiting the fact that a \textsf{\footnotesize BYTE} instruction is used for accessing a single byte of a \textsf{\small bytes32}, while an \textsf{\footnotesize AND} masks a \textsf{\small uint256} for the same purpose.

\noindent\textbf{2. \textsf{\small Array}:}~%static container types
An \textsf{\small array} hosts array items of the same basic type. An \textsf{\small array} can be static, dynamic or nested. The size of a static \textsf{\small array} is known during compilation whereas the size of a dynamic \textsf{\small array} and a nested \textsf{\small array} is determined by the actual argument at runtime.

\begin{figure}[ht]
	\centering
	\vspace{-2ex}
	\includegraphics[width=0.45\textwidth]{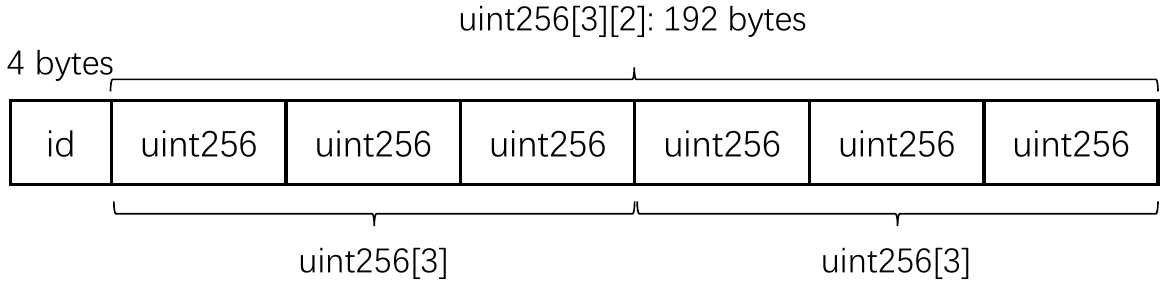}
	\vspace{-1ex}
	\caption{The call data layout of a static array, \textsf{\footnotesize uint256[3][2]}}
	\vspace{-2ex}
	\label{fig_static_array}
\end{figure}

\noindent\textbf{(1)}~\textbf{\textsf{\small Static array}}. We denote it as \textsf{\small T}$[X_1]...[X_n]$, where \textsf{\small T} is one of the basic types, $X_1$, ..., $X_n$ are constant numbers known in compilation, and $n$ is the dimension. If $n>1$, \textsf{\small T$[X_1]...[X_n]$} is a multidimensional \textsf{\small array}. The size of a static \textsf{\small array} is known in compilation since the number of items in each dimension is fixed. As a basic type, each \textsf{\small array} item is extended through the process described above.
Different from other languages (e.g., C), the notation of an \textsf{\small array} in EVM is reversed and the access is in the opposite direction of the declaration~\cite{type}.
For example, \textsf{\small uint256[3][2] \textit{x}} means an \textsf{\small array} \textsf{\small \textit{x}} of two arrays of three \textsf{\small uint256}, and \textsf{\small \textit{x}[1][0]} accesses the first item of the second \textsf{\small uint256[3]}. All items of a static \textsf{\small array} are stored consecutively in the call data. Fig. \ref{fig_static_array} shows the call data layout of a public/external function with one \textsf{\small uint256[3][2]} argument.

The accessing patterns of a static \textsf{\small array} are different for public and external functions. 

\noindent(a)~\textit{Public function.} Its static \textsf{\small array} will be read into the memory by one or more \textsf{\footnotesize CALLDATACOPY}s. 
Then, each array item is accessed by an \textsf{\footnotesize MLOAD}, which reads 32 bytes from the memory to the stack~\cite{yellow}. 
We find that a nested loop is used to read parameters from the call data, and the level of the nested loop is equal to the dimension of the static \textsf{\small array} minus one. In the most inner layer, a \textsf{\footnotesize CALLDATACOPY} is executed. 
A \textsf{\footnotesize CALLDATACOPY} can read a one-dimensional static \textsf{\small array}. 

Listing \ref{code_array_copy} shows the pseudocode (we do not show the EVM instructions for the ease of presentation) to read a \textsf{\small uint256[3][2]} from the call data to the memory. The execution of each \textsf{\footnotesize CALLDATACOPY} copies the \textsf{\small array} of the lowest dimension (Line 3). The loop guard (Line 2) determines how many times the \textsf{\footnotesize CALLDATACOPY} should be executed. The loop guard is compiled into an \textsf{\footnotesize LT} instruction, which checks whether the current loop count is smaller than the item number of the highest dimension.
\lstset{
    xleftmargin=10pt
}
\vspace{-1ex}
\begin{lstlisting}[language=C++, caption={Pseudocode for reading a \textsf{\footnotesize uint256{[3][2]}} parameter to memory}, label={code_array_copy}]
i = 0;
while (i < 2){ //how many uint256[3] needs to copy
  calldatacopy(uint256[3]);
  i++;
}
\end{lstlisting}
\vspace{-1ex}

\noindent(b)~\textit{External function.} Its static \textsf{\small array} will not be entirely copied. 
Instead, \textsf{\small array} items will be read from the call data to the stack using \textsf{\footnotesize CALLDATALOAD} on demand. 
We find that if the smart contract is compiled without optimization or the index of the \textsf{\small array} is a variable, the \textsf{\footnotesize CALLDATALOAD} is preceded by bound checks to prevent array overrun. For instance, before accessing the item \textsf{\small \textit{x}[i][j]} of the \textsf{\small array} \textsf{\small \textit{x}[3][2]}, two bound checks $i<2$ and $j<3$ are executed. Only after 
they are passed, \textsf{\small \textit{x}[i][j]} can be accessed. Such runtime bound check is not needed if the array index is a constant and the smart contract is compiled with optimization due to the compile-time bound checks. 

\begin{figure}[ht]
	\centering
	\vspace{-1ex}
	\includegraphics[width=0.48\textwidth]{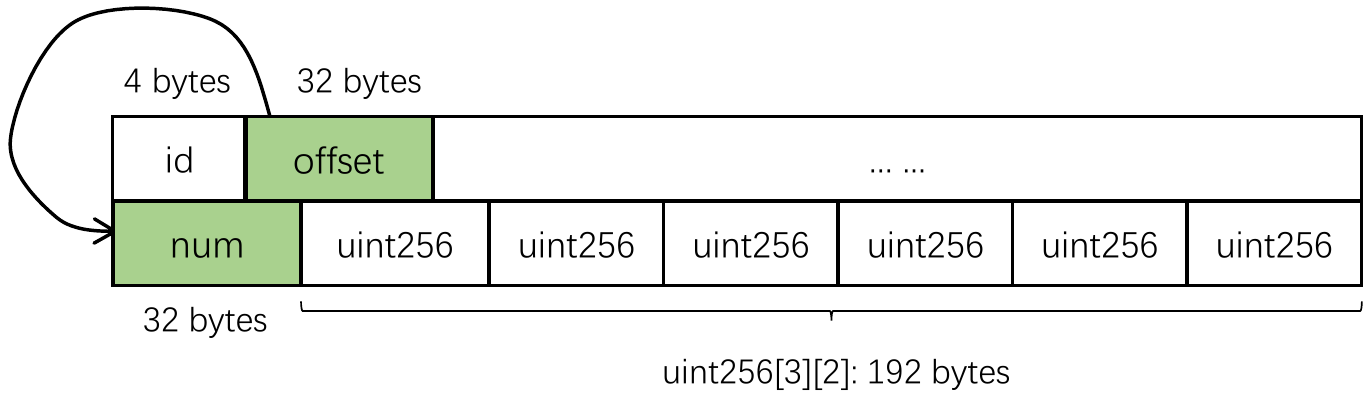}
	\vspace{-2ex}
	\caption{The call data layout of a dynamic array \textsf{\footnotesize uint256[3][]}, and the actual argument is \textsf{\footnotesize uint256[3][2]}}
	\vspace{-2ex}
	\label{fig_dynamic_array}
\end{figure}

\noindent\textbf{(2)}~\textbf{\textsf{\small Dynamic array}}. We denote it as \textsf{\small T}$[X_1]...[X_{n-1}][]$, where \textsf{\small T} is one of basic types, $X_1$, ..., $X_{n-1}$ are constant numbers known in compilation, $n$ is the dimension. The size of a dynamic \textsf{\small array} is unknown in compilation because the item number of the highest dimension is unknown. Instead, the item number should be provided in the call data so that the invoked function can know it at runtime. 
Fig. \ref{fig_dynamic_array} shows the call data layout of a dynamic \textsf{\small array} \textsf{\small uint256[3][]}, which is the first parameter, and the actual argument is \textsf{\small uint256[3][2]}. A 32-bytes value, i.e., the \emph{offset} field, is located right after the function id, and this value is an offset relative to the first byte after the function id. The 32 bytes, i.e., the \emph{num} field, pointed by the offset stores the number of items in the highest dimension. All \textsf{\small array} items are located after the \emph{num} field.

The accessing patterns of a dynamic \textsf{\small array} are different for public and external functions.

\noindent(a)~\textit{Public function.} Its dynamic \textsf{\small array} will be read into the memory by one or more \textsf{\footnotesize CALLDATACOPY}s. 
More precisely, the smart contract uses a \textsf{\footnotesize CALLDATALOAD} to read the offset (i.e., the \textit{offset} field), and another \textsf{\footnotesize CALLDATALOAD} to read the item number of the highest dimension (i.e., the \textit{num} field). Then, the item number will be read from the stack to the memory by an \textsf{\footnotesize MSTORE}, which saves 32 bytes from the stack into the memory~\cite{yellow}. After that, the smart contract uses a nested loop to copy all \textsf{\small array} items into the memory right after the item number. Only one \textsf{\footnotesize CALLDATACOPY} is needed to read a one-dimensional dynamic \textsf{\small array}. 

\noindent(b)~\textit{External function.} Two \textsf{\footnotesize CALLDATALOAD}s are  used to read the offset and the item number of the highest dimension. Different from the process for public functions, an \textsf{\small array} item will be read from the call data to the stack by a \textsf{\footnotesize CALLDATALOAD} on demand. The operand of the \textsf{\footnotesize CALLDATALOAD}, i.e., the location from where to copy, is computed from the value of the \textit{offset} field at runtime, and the result contains the multiplication of 32, because each \textsf{\small array} item is extended to 32 bytes. For example, the location of the third \textsf{\small array} item is computed by $\textrm{\textit{offset}}+4+32+2\times 32$ (the function id is of 4 bytes, an item number is of 32 bytes, the first two \textsf{\small array} items are of 64 bytes). The bound check of the highest dimension exists since its item number is unknown in compilation. %We define the rules R1 and R2 (\S \ref{rule_calldataload}) to recognize a dynamic \textsf{\small array} for external functions.

\begin{figure}[ht]
	\centering
	\vspace{-1ex}
	\includegraphics[width=0.48\textwidth]{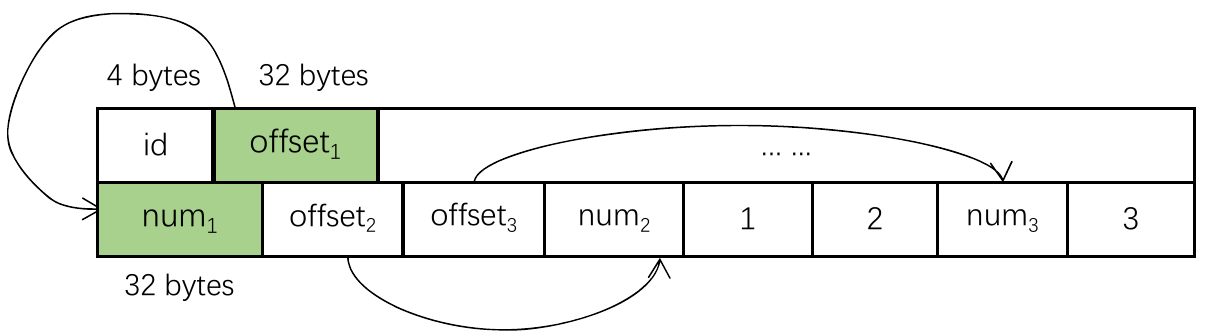}
	\vspace{-2ex}
	\caption{The call data layout of a nested array \textsf{\footnotesize uint[][]}, and the actual argument is \textsf{\footnotesize [[1, 2], [3]]}}
	\vspace{-2ex}
	\label{fig_nested_array_layout}
\end{figure}

\noindent\textbf{(3)}~\textbf{Nested \textsf{\small array}}.
%Nested array hosts array items of the same basic type, the dimensions of nested array is determined by the actual argument at compilation time. 
We denote a \textit{n}-dimensional nested \textsf{\small array} as \textsf{\small T}$[X_1]...[X_n]$, where \textsf{\small T} is one of basic types. At least one of $X_1$ to $X_{n-1}$ should be empty indicating the corresponding dimension is dynamic. The key difference between a \textit{n}-dimensional nested \textsf{\small array} and a \textit{n}-dimensional dynamic \textsf{\small array} is that at least one dimension of the lower $n-1$ dimensions of the former can be dynamic 
whereas the lower $n-1$ dimensions of the latter must be static~\cite{solidity-sig}. Hence, each dimension of a \textit{n}-dimensional nested \textsf{\small array} associates an \textit{offset} field and a \textit{num} field to enable dynamic dimension size.
Fig. \ref{fig_nested_array_layout} shows the call data layout of a nested \textsf{\small array} uint[][], which is the first parameter of a function. Let's assume that the actual argument is [[0x1, 0x2], [0x3]] for the ease of explanation. 
\textit{offset}$_1$ is the \textit{offset} field of this nested \textsf{\small array}, and \textit{num}$_1$ is the \textit{num} field of its highest dimension. Two items (i.e., [0x1, 0x2] and [0x3]) in the highest dimension are placed after \textit{num}$_1$. Because both the two items are dynamic \textsf{\small array}s, their \textit{offset} fields will be placed immediately after \textit{num}$_1$. In this example, \textit{offset}$_2$ and \textit{num}$_2$ are the \textit{offset} and \textit{num} fields of the item [0x1, 0x2], and \textit{offset}$_3$ and \textit{num}$_3$ are the \textit{offset} and \textit{num} fields of the item [0x3]. 0x1 and 0x2 which are two items of the \textsf{\small array} [0x1, 0x2] are placed immediately after \textit{num}$_2$. Similarly 0x3, the only item of the \textsf{\small array} [0x3] is placed immediately after \textit{num}$_3$. 

The accessing pattern of a nested \textsf{\small array} in the public mode is the same as that in the external mode, and an array item in nested \textsf{\small array} will be read from the call data to the stack by a \textsf{\footnotesize CALLDATALOAD} on demand. 
More specifically, two \textsf{\footnotesize CALLDATALOAD}s are used to read \textit{offset}$_1$ and \textit{num}$_1$, the item number of the highest dimension. 
For example, to read the \textsf{\small array} item 0x3 as an example, three additional \textsf{\footnotesize CALLDATALOAD}s are required to read \textit{offset}$_3$, \textit{num}$_3$ and the item 0x3.
To access an \textsf{\small array} item, there is a bound check for each dimension.

\noindent\textbf{3. \textsf{\small bytes}:}~
It is a dynamic byte sequence whose size is determined at runtime~\cite{solidity-sig}.
If a public/external function has a \textsf{\small bytes} type parameter,
it will be extended on the lower-order (right) side with the minimum number of zero to make the length of the extended result be multiple of 32 bytes~\cite{solidity-sig}. 
For example, if an un-extended \textsf{\small bytes} is `abcd', the extended result is shown in Fig. \ref{fig_bytes4}. The layout of a \textsf{\small bytes} is similar to that of a one-dimensional dynamic \textsf{\small array}, as shown in Fig. \ref{fig_dynamic_array}, except that the \textit{num} field stores the size (in byte) of the \textsf{\small bytes} before padding.
Note that the \textit{num} field of a one-dimensional dynamic \textsf{\small array} records the number of \textsf{\small array} items. The extended value is located right after the \textit{num} field. 

The accessing patterns of a \textsf{\small bytes} value are different for public and external functions.

\noindent(a)~\textit{Public function.} The accessing pattern of a \textsf{\small bytes} in a public function is the same as the accessing pattern of a one-dimensional dynamic \textsf{\small array}  in a public function except the computation of read size, because every single byte of a \textsf{\small bytes} is not extended. 

\noindent(b)~\textit{External function.} The accessing pattern of a \textsf{\small bytes} in the external mode is the same as the accessing pattern of a one-dimensional dynamic \textsf{\small array}  in the external mode, except that accessing a single byte of a \textsf{\small bytes} does not need the multiplication of 32 because every single byte of a \textsf{\small bytes} is not extended. 

\noindent\textbf{4. \textsf{\small string}:}~
It is a dynamic Unicode string, whose size is determined at runtime~\cite{solidity-sig}. The call data layout of a \textsf{\small string} and reading a \textsf{\small string} from the call data are the same as that of a \textsf{\small bytes}. We distinguish them based on the observation that a \textsf{\small bytes} supports reading/writing its individual byte whereas a \textsf{\small string} does not support such operation.

\begin{figure}[ht]
	\centering
	\vspace{-1ex}
	\includegraphics[width=0.20\textwidth]{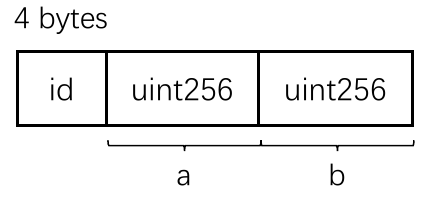}
	\vspace{-2ex}
	\caption{The layout of a \textsf{\footnotesize struct} containing two \textsf{\footnotesize uint256} items is the same to that of two individual \textsf{\footnotesize uint256} parameters}
	\vspace{-2ex}
	\label{fig_same_layout}
\end{figure}

\noindent\textbf{5. \textsf{\small struct}:} 
We denote a \textsf{\small struct} as $(T_1,...,T_n)$, where \textit{n} is a constant number, indicating \textit{n} \textsf{\small struct} items whose types can be any types in Solidity. The size of a \textsf{\small struct} can be dynamic or static. For a static \textsf{\small struct} whose size is known before compilation, the type of each \textsf{\small struct} item should be one of basic types, static \textsf{\small array}, or static \textsf{\small struct}. A dynamic \textsf{\small struct} whose size is unknown in compilation can host all possible Solidity types. The accessing pattern of a \textsf{\small struct} in the public mode is the same as that in the external mode. 
We find that the layout of a static \textsf{\small struct} is the same as that of all its items as if they were not inside the static \textsf{\small struct}.
For example, Listing \ref{struct_def} is the definition of a \textsf{\small struct} (\textsf{\small uint256}, \textsf{\small uint256}) in Solidity, Listing \ref{function_def} is the definition of a function with two \textsf{\small uint256} parameters in Solidity. Fig. \ref{fig_same_layout} shows the same call data layout of them.
Moreover, the bytecode to read an item of a static \textsf{\small struct} is the same as the bytecode to read the item as if it was not placed inside the \textsf{\small struct}. Therefore, there is no sufficient hint from the bytecode to distinguish these two different 
situations.
\vspace{-1ex}
\begin{lstlisting}[language=C++, caption={A function taking in a \textsf{\footnotesize struct} with two \textsf{\footnotesize uint256} items}, label={struct_def}]
struct StructSample {
  uint256 a;
  uint256 b;
}
function Func(StructSample var1) {}
\end{lstlisting}
\vspace{-1ex}
\begin{lstlisting}[language=C++, caption={A function taking in two \textsf{\footnotesize uint256} parameters}, label={function_def}]
function FuncSample(uint256 a, uint256 b) {}
\end{lstlisting}
\vspace{-2ex}

\begin{figure}[ht]
	\centering
	\vspace{-1ex}
	\includegraphics[width=0.48\textwidth]{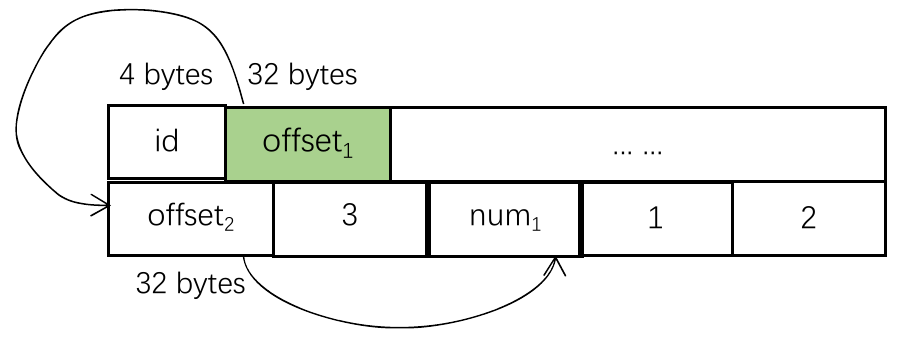}
	\vspace{-2ex}
	\caption{The call data layout of a dynamic \textsf{\small struct} \textsf{\footnotesize (uint[],uint)}, and the actual argument is \textsf{\footnotesize ([1, 2], 3)}}
	\vspace{-2ex}
	\label{fig_dynamic_struct_layout}
\end{figure}

The layout of a dynamic \textsf{\small struct} has an \textit{offset} field, and all \textsf{\small struct} items are placed after the \textit{offset} field.
Fig. \ref{fig_dynamic_struct_layout} shows the call data layout of a dynamic \textsf{\small struct} (uint[], uint), which is the first parameter of a function, and we assume that the actual argument is ([0x1, 0x2], 0x3). 
In this example, \textit{offset}$_1$ is the \textit{offset} field of this \textsf{\small struct}. Two items (i.e., [0x1, 0x2] and 0x3) of this \textsf{\small struct} are placed after the position which \textit{offset}$_1$ indicated. Because the item [0x1, 0x2] is \textsf{\small array}, its \textit{offset} field will be placed immediately after the position which \textit{offset}$_1$ indicated, and the value of item 0x3 will be placed immediately after it. In this example, \textit{offset}$_2$ and \textit{num}$_1$ are the \textit{offset} and \textit{num} fields of item [0x1, 0x2], and items of [0x1, 0x2] are placed immediately after \textit{num}$_1$.
For example, to access the \textsf{\small struct} item 0x3, one \textsf{\footnotesize CALLDATALOAD} is used for reading \textit{offset}$_1$ and one additional \textsf{\footnotesize CALLDATALOAD} is required to read the item 0x3 directly.

\subsubsection{Vyper}
Vyper supports ten parameter types, including \textsf{\small bool}, \textsf{\small int128}, \textsf{\small uint256}, \textsf{\small address}, \textsf{\small bytes32}, decimal, fixed-size \textsf{\small list}, fixed-size \textsf{\small byte array}, fixed-size \textsf{\small string}, and \textsf{\small struct}~\cite{vyper}. 
The first five types are also supported by Solidity, and the layouts of these five types are the same with the layouts of Solidity types. Different with Solidity, the instructions to read these five types use comparison instructions (e.g., \textsf{\footnotesize LT}) rather than mask instructions (e.g., \textsf{\footnotesize AND}) to ensure that the values of these five types are in the valid ranges.
For example, Listing \ref{Solidity_address} shows the accessing pattern for \textsf{\small address} parameter in Solidity and Listing \ref{Vyper_address} shows the accessing pattern for \textsf{\small address} parameter in Vyper.
We also learn that Vyper generates the same bytecode for public and external functions. 
We describe the remaining five types below. 
\vspace{-1ex}
\begin{lstlisting}[caption={Accessing pattern for an \textsf{\footnotesize address} parameter in Solidity}, label={Solidity_address}]
CALLDATALOAD //read the parameter
...
PUSH20 0xffffffffffffffffffffffffffffffffffffffff
AND //mask by 0xff...ff
\end{lstlisting}
\vspace{-1ex}
\begin{lstlisting}[caption={Accessing pattern for an \textsf{\footnotesize address} parameter in Vyper}, label={Vyper_address}]
PUSH21 0x010000000000000000000000000000000000000000
PUSH1 0x20
MSTORE
...
CALLDATALOAD //read the parameter
PUSH1 0x20
MLOAD //load 0x01...00
DUP2
LT //compare to 0x01...00  
\end{lstlisting}
\vspace{-1ex}

\noindent\textbf{1.}~\textbf{\textsf{\small decimal}}. 
A decimal is a fixed-point value with a precision of 10 decimal, whose value ranges from $-2^{127}$ and  $2^{127}-1$~\cite{vyper}. We find that the layout and the accessing pattern of a \textsf{\small decimal} is similar to that of signed integer in Solidity, except that two comparisons rather than a \textsf{\footnotesize SIGNEXTEND} mask are used for ensuring that the value of the \textsf{\small decimal} is between $-2^{127}$ and $2^{127}-1$. If the decimal value is out of the range, the execution of smart contract will be 
aborted.

\noindent\textbf{2.}~\textbf{Fixed-size \textsf{\small list}}. 
Fixed-size \textsf{\small list} records a constant number of items.
~\cite{vyper}. We denote a fixed-size \textsf{\small list} as $T[X_1]...[X_n]$, where T is one of \textsf{\small bool}, \textsf{\small int128}, \textsf{\small uint256}, \textsf{\small address}, \textsf{\small bytes32} and \textsf{\small decimal}. $X_1,...,X_n$ are constant numbers known in compilation, and \textit{n} is the dimension and $n\geq 1$. The layout and accessing pattern of a fixed-size \textsf{\small list} is the same as that of a static \textsf{\small array} in the external mode of Solidity. To access a list item, bound checks are used to prevent the overrun of the \textsf{\small list} if the index is a variable. 

\noindent\textbf{3.}~\textbf{Fixed-size \textsf{\small byte array}}. A fixed-size \textsf{\small byte array} is a byte sequence with a maximum length, which is given in the source code. Its real length which is provided at runtime, should not be longer than the maximum size. We denote a fixed-size \textsf{\small byte array} as \textsf{\small bytes}[\textit{maxLen}], \textit{maxLen} is a constant value denoting the maximum length. The layout and accessing pattern for a fixed-size byte \textsf{\small array} are similar to a \textsf{\small bytes} in the public mode of Solidity, except that 32 (the size of the \textit{num} field) $+$ \textit{maxLen} are read from the call data. That is, the extended bytes of a fixed-size \textsf{\small byte array} are not read. 

\noindent\textbf{4.}~\textbf{Fixed-size \textsf{\small string}}. We denote a fixed-size \textsf{\small string} as \textsf{\small string}[\textit{maxLen}], where \textit{maxLen} is a constant value representing the maximum length of the \textsf{\small string}. The layout and the way to read a fixed-size \textsf{\small string} from the call data are the same as that of a fixed-size \textsf{\small byte array}. The difference between the two types lies in that a fixed-size \textsf{\small byte array} allows accessing its individual byte whereas a fixed-size \textsf{\small byte array} does not allow it.

\begin{figure}[ht]
	\centering
	\vspace{-1ex}
	\includegraphics[width=0.20\textwidth]{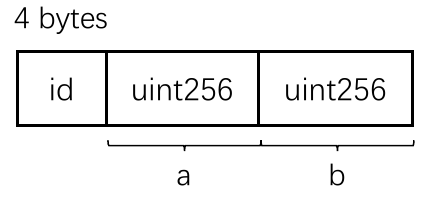}
	\vspace{-2ex}
	\caption{The layout of a \textsf{\footnotesize struct} containing two \textsf{\footnotesize uint256} items is the same to that of two individual \textsf{\footnotesize uint256} parameters}
	\vspace{-2ex}
	\label{fig_vyper_same_layout}
\end{figure}

\noindent\textbf{5.}~\textbf{\textsf{\small struct}}. 
\textsf{\small struct} is a container type that can host several variables.
We denote a \textsf{\small struct} as $(T_1,...,T_n)$, where \textit{n} is a constant value indicating \textit{n} \textsf{\small struct} items and $T_i$ can be one of the following types, \textsf{\small bool}, \textsf{\small int128}, \textsf{\small uint256}, \textsf{\small address}, \textsf{\small bytes32} and \textsf{\small decimal}~\cite{vyper}. The layout of a \textsf{\small struct} is the same as that of all its items as if they were not placed inside the \textsf{\small struct}. The bytecode to read a \textsf{\small struct} item is also the same as the bytecode to read the item as if it was not inside the \textsf{\small struct}. Therefore, there is no sufficient hint from the bytecode to distinguish these two different types.
% hao add
Listing \ref{vyper_struct_def} shows a function signature which takes in a \textsf{\small struct} containing two \textsf{\small uint256} items, and Listing \ref{vyper_function_def} presents a function taking in two \textsf{\small uint256} parameters. The layouts of the \textsf{\small struct} and two individual \textsf{\small uint256} parameters are the same, as shown in Fig. \ref{fig_vyper_same_layout}.
\vspace{-1ex}
\begin{lstlisting}[language=C++, caption={A function taking in a \textsf{\footnotesize struct} with two \textsf{\footnotesize uint256} items}, label={vyper_struct_def}]
struct StructSample:
  a: uint256
  b: uint256
def Func(var1: StructSample)
\end{lstlisting}
\vspace{-1ex}
\begin{lstlisting}[language=C++, caption={A function taking in two \textsf{\footnotesize uint256} parameters}, label={vyper_function_def}]
def FuncSample(a: uint256, b: uint256)
\end{lstlisting}
\vspace{-2ex}

\section{Rules for Type Inference}
\label{sec_rule}
This section first introduces the method to generate rules, and then describes the first four rules. We briefly introduce the remaining 27 rules, and elaborate more on them in Supplementary material C.

\subsection{Rules Generation}
\label{rule_extract}
We propose a systematic approach to generate rules. The basic idea is to first learn the accessing patterns automatically from the bytecode of self-generated smart contracts and then summarize the rules manually, which are used by \texttt{\small SigRec} to recover function signatures from the bytecode of other smart contracts. Our approach has five steps. Since the first four steps are automated, it can be easily extended to handle new accessing patterns due to new compilers/programming language/obfuscation techniques and new parameter types (discussed in \S \ref{sec_discuss}). 

\noindent\textbf{Step 1. Preparing smart contracts and their bytecode.} 
We develop a tool to automatically construct smart contracts and compile them into bytecode, from which the rules will be learned. We develop smart contracts in Solidity and Vyper since they are the most popular languages for developing Ethereum smart contracts. 
To ensure the completeness of the derived rules, we consider
all parameter types supported in Solidity and Vyper, all major versions of compilers as well as various optimization levels. More specifically, each smart contract contains one public or external function. Since we find that smart contracts handle the different parameters of a function independently, each generated function just has one parameter to ease rule learning. To expose the accessing patterns of parameters, the body of each generated function contains statements to access the parameter, e.g., assignments, arithmetic operations, comparisons, reading one byte from \textsf{\small bytes} and \textsf{\small bytes32}, and reading one \textsf{\small array} item.

We consider all basic types with all possible widths, e.g., \textsf{\small uint8}, \textsf{\small uint16}, ..., \textsf{\small uint256}. 
Since the upper bound of the array dimension and the upper bound of the size of a static dimension are undocumented, we first identify the patterns with concrete number of dimensions and then generalize the pattern for all possible dimensions. 
Specifically, we set \textsf{\small array} dimensions from 1 to 5 and each dimension can be static or dynamic. For each static dimension, we set its size from 1 to 10. The items of an \textsf{\small array} are set to one of supported basic types with all possible widths (e.g., \textsf{\small uint8}, \textsf{\small uint16}, ..., \textsf{\small uint256}). 

\noindent\textbf{Step 2. Collecting the accessing pattern of each parameter type.} We conduct data dependence analysis and control dependence analysis to locate the EVM instruction sequence used to access the parameter of each generated function, which are required in subsequent steps. We consider such instruction sequence as an accessing pattern. Specifically, the instructions which are data dependent with parameters are needed in step 4 for collecting the symbolic expressions of variables.  Step 5 needs 
the control dependency relations to infer the structure of some complicated parameter types. For example, TASE infers the loop structure from the EVM instructions that are control dependent on the instruction for reading the call data to discover \textit{n}-dimensional dynamic \textsf{\small array}.   

\noindent\textbf{Step 3. Extracting common accessing patterns.} 
We observe that the access patterns for some parameters types (e.g., \textsf{\small uint8} and \textsf{\small uint16}) are similar, and thus this step extracts common access patterns from the accessing patterns obtained in step 2. We regard an instruction sequence as the \textit{common} accessing pattern of some accessing patterns if it appears in all these accessing patterns. For example, from step 2 we obtain the accessing patterns of \textsf{\small uint8}, \textsf{\small uint16}, ..., \textsf{\small uint256} in Solidity, and then in this step we try to extract the common accessing pattern from these patterns to facilitate the derivation of the rule for \textsf{\small uint} with all possible widths. On the contrary, \textsf{\small bool}, \textsf{\small int128}, \textsf{\small uint256}, \textsf{\small address}, \textsf{\small bytes32}, \textsf{\small decimal} of Vyper are basic types, we skip the step for them because these types have fixed widths. Besides, we omit the description about the fixed-size \textsf{\small list} of Vyper because it is equivalent to the static \textsf{\small array} of Solidity.  

\noindent\textit{Basic types of Solidity.} We use unsigned integers as an example, and the other basic types are handled similarly. We extract the common accessing pattern from the accessing patterns of \textsf{\small uint8}, \textsf{\small uint16}, ..., \textsf{\small uint256}, which is used in step 5 to generate a general rule for inferring the unsigned integers with all possible widths, \textsf{\small uint$\langle M\rangle$}, $8\le M\le 256, M\%8==0$. 

\noindent\textit{One-dimensional static \textsf{\small array} of Solidity.} We extract the common accessing pattern from the accessing patterns of \textsf{\small uint8[1]}, \textsf{\small uint8[2]}, ..., \textsf{\small uint8[10]}. 
By retaining the instructions in the common accessing pattern but not in the accessing pattern of \textsf{\small uint8}, we further obtain the accessing pattern of \textsf{\small T}[$N$], $1\le N \le 10$, \textsf{\small T} is one of the basic types.

\noindent\textit{Multidimensional static \textsf{\small array} of Solidity.} We first extract the common accessing pattern  from the accessing patterns of \textsf{\small uint8}[$N_1$][$N_2$], $1\le N_1, N_2 \le 10$. 
Then we obtain the accessing pattern of \textsf{\small T}[$N_1$][$N_2$], $1\le N_1, N_2 \le 10$ by retaining the instructions in the common accessing pattern but not in the accessing pattern of \textsf{\small uint8}. Similarly, we obtain the accessing patterns of \textsf{\small T}[$N_1$][$N_2$][$N_3$], \textsf{\small T}[$N_1$][$N_2$][$N_3$][$N_4$], \textsf{\small T}[$N_1$][$N_2$][$N_3$][$N_4$][$N_5$], $1\le N_1, N_2, N_3, N_4, N_5 \le 10$.

\noindent\textit{One-dimensional dynamic \textsf{\small array} of Solidity.} We regard the EVM instructions that are in the accessing pattern of \textsf{\small uint8[]} but not in the accessing pattern of \textsf{\small uint8} as the common accessing pattern of \textsf{\small T}[].

\noindent\textit{Multidimensional dynamic \textsf{\small array} of Solidity.} We extract the common accessing pattern from accessing patterns of \textsf{\small uint8}[$N_1$][], $1\le N_1 \le 10$. By retaining the EVM instructions that appear in such common accessing pattern but not in the accessing pattern of \textsf{\small uint8}, we obtain the accessing pattern of \textsf{\small T}[$N_1$][], $1\le N_1 \le 10$. We also obtain the accessing patterns of \textsf{\small T}[$N_1$][$N_2$][], \textsf{\small T}[$N_1$][$N_2$][$N_3$][], \textsf{\small T}[$N_1$][$N_2$][$N_3$][$N_4$][], $1\le N_1, N_2, N_3, N_4 \le 10$ through the same process.

\noindent\textit{Nested \textsf{\small array} of Solidity.} We extract the common accessing pattern from the accessing patterns of \textsf{\small uint8}[][$N_1$], 1 $\leq N_1 \leq$ 10 or $N_1$ is empty. By retaining the EVM instructions that appear in such common accessing pattern but not in the accessing pattern of \textsf{\small uint8}, we obtain the accessing pattern of T[][$N_1$], 1 $\leq N_1 \leq$ 10 or $N_1$ is empty. We also obtain the accessing pattern of T[$N_1$][$N_2$][$N_3$]][$N_4$]][$N_5$], 1$\leq N_1, N_2, N_3, N_4, N_5 \leq$ 10 and at least one of $N_1$ to $N_4$ is empty through the same process.

\noindent\textit{Fixed-size \textsf{\small byte array} of Vyper.} We obtain the accessing pattern of \textsf{\small bytes}[\textit{maxLen}] by extracting the common accessing pattern from the accessing patterns of \textsf{\small bytes}[1], \textsf{\small bytes}[2],..., \textsf{\small bytes}[50]. 

\noindent\textit{Fixed-size \textsf{\small string} of Vyper.} We obtain the accessing pattern of \textsf{\small string}[\textit{maxLen}] by extracting the common accessing pattern from the accessing patterns of \textsf{\small string}[1], \textsf{\small string}[2],..., \textsf{\small string}[50]. 

\noindent\textit{\textsf{\small struct} of Solidity.} we extract the accessing pattern of (\textsf{\small uint8}[]). By retaining the EVM instructions that appear in such accessing pattern but not in the accessing pattern of \textsf{\small uint8}[], which is extracted above, we obtain the accessing pattern of the \textsf{\small struct} type of Solidity. 

\noindent\textit{\textsf{\small struct} of Vyper.}  we extract the common accessing pattern from accessing patterns of (\textsf{\small uint256}). By retaining the EVM instructions that appear in such common accessing pattern but not in the accessing pattern of \textsf{\small uint256}, we obtain the accessing pattern of the \textsf{\small struct} type of Vyper. 

\noindent\textbf{Step 4. Generating symbolic expressions from parameters.} To characterize how parameters are handled by EVM instructions, we conduct symbolic execution on the common accessing patterns obtained in step 3 by treating the call data as symbols, and collect the symbolic expressions for each variable. Step 5 needs symbolic expressions to summarize rules.

\noindent\textbf{Step 5. Summarizing rules.} We summarize the rules according to the common accessing patterns obtained in step 3 and the symbolic expressions collected in step 4, and then organize them into a decision tree (Fig. \ref{fig_rule_tree}) for fast rule checking. To determine the rules for \textsf{\small array}s, we investigate how the accessing patterns change when its dimension increases from 1 to 5 and the number of items per-dimension increases from 1 to 10.
Eventually, we generate 18 rules and divide them into three categories for \textsf{\footnotesize CALLDATALOAD} (\S \ref{rule_calldataload}), \textsf{\footnotesize CALLDATACOPY} (\S \ref{rule_calldatacopy}), and other instructions (\S \ref{rule_other}), respectively.

\subsection{Rules for \textsf{\normalsize CALLDATALOAD}}

\begin{figure*}[ht]
	\centering
    \frame{\includegraphics[width=0.75\textwidth]{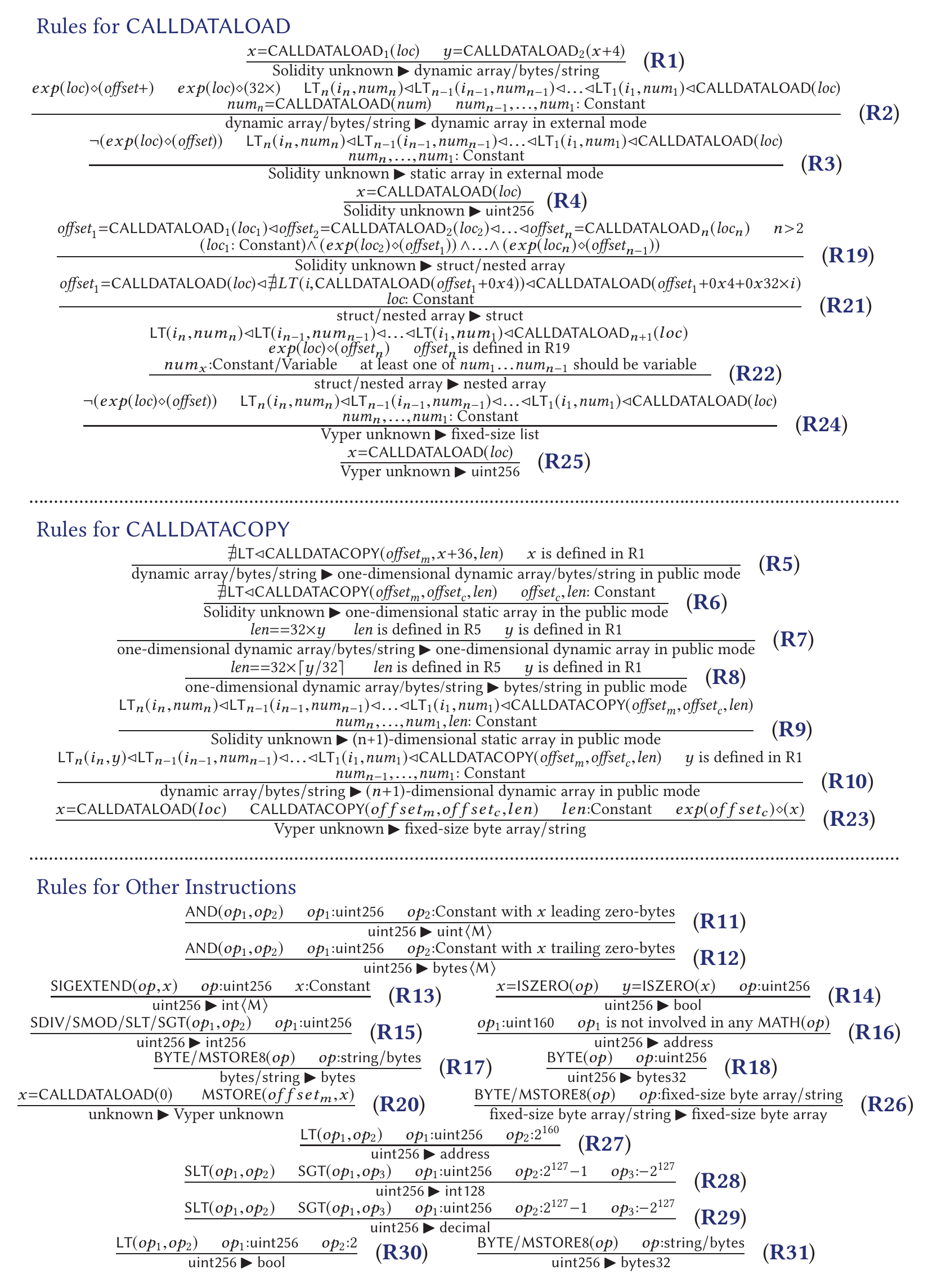}}
	\caption{Rules}
	\label{fig_formal_rule}
	%\vspace{-1ex}
\end{figure*}

\label{rule_calldataload}
A \textsf{\footnotesize CALLDATALOAD} can read 3 kinds of data from the call data (\S \ref{sec_calldata}): (1) the value of a parameter; (2) the \emph{offset} field of a parameter; (3) the \emph{num} field of a parameter.

\noindent\textbf{R1:} R1 is used to infer a dynamic \textsf{\small array}/\textsf{\small bytes}/\textsf{\small string}. It holds if two \textsf{\footnotesize CALLDATALOAD}s (termed by \textsf{\footnotesize CALLDATALOAD}$_1$ and \textsf{\footnotesize CALLDATALOAD}$_2$) satisfy that $x=\textsf{\footnotesize CALLDATALOAD}_1(\rm{\textit{loc}})$$\wedge y=\textsf{\footnotesize CALLDATALOAD}_2(x+0\rm{x}4)$, where $x=\textsf{\footnotesize CALLDATALOAD}_1(\rm{\textit{loc}})$ means that a 32-bytes value is read from the $\rm{\textit{loc}}$ byte of the call data into $x$ by a \textsf{\footnotesize CALLDATALOAD}. Actually, $x$ is the \emph{offset} field of a dynamic \textsf{\small array}/\textsf{\small bytes}/\textsf{\small string} parameter. \textsf{\footnotesize CALLDATALOAD}$_2$ reads the \emph{num} field of the parameter, because the location of the \emph{num} field is the value of the \emph{offset} field plus the length of the function id (i.e., 4 bytes).

\noindent\textbf{R2:} R2 is used to infer an $n$-dimensional ($n>1$) dynamic \textsf{\small array} in an external function. 
It holds if three requirements, $v$1, $v$2, $v$3 are satisfied.
Let $exp(\textit{loc})$ represent the symbolic expression of the location \textit{loc}, which describes how \textit{loc} is computed from symbols. Let $exp(p)\diamond q$ denote that the symbolic expression of \textit{p} (e.g., $x+y\times 5$) contains the symbolic expression \textit{q} (e.g., $y\times 5$).
$v$1 is defined as $exp(\textit{loc})\diamond (\textit{offset}+)$, meaning that the read location is computed by adding the value of the \emph{offset} field, because array items are placed after the \emph{num} field, which is pointed by the \emph{offset} field.
Hence, if a  \textsf{\footnotesize CALLDATALOAD} reads an item from a dynamic \textsf{\small array} in the external model, $v$1 is satisfied.

$v$2 is defined as $exp(\textit{loc})\diamond (32\times)$, meaning that the symbolic expression of \textit{loc} contains the multiplication of 32.  If a \textsf{\footnotesize CALLDATALOAD} reads an item from a dynamic \textsf{\small array} parameter in an external function, $v$2 is satisfied, because each array item is extended to 32 bytes so that the multiplication of 32 is needed to access an array item.

Before introducing $v$3, we let
$\textit{num}_\textit{n}=\textsf{\footnotesize CALLDATALOAD}(\textit{num})$ indicate that a \textsf{\footnotesize CALLDATALOAD} reads the value of the \textit{num} field and assigns it to $\textit{num}_\textit{n}$.
$\textit{num}_{n-1}, ..., \textit{num}_1$ are constant numbers, and $\textsf{\footnotesize LT}_j(i_j,\textit{num}_j)$ means that an \textsf{\footnotesize LT} instruction compares a number $i_j$ with a number $\textit{num}_j$.
Let $\textit{ins}_i\lhd \textit{ins}_j$ indicate that the instruction $\textit{ins}_j$ is control-flow dependent on the instruction $\textit{ins}_i$.
$v$3 is defined as $\textsf{\footnotesize LT}_n(i_n,\textit{num}_n)\lhd \textsf{\footnotesize LT}_{n-1}(i_{n-1},\textit{num}_{n-1})\lhd ... \lhd \textsf{\footnotesize LT}_1(i_1,\textit{num}_1)\lhd \textsf{\footnotesize CALLDATALOAD}(\textit{loc})$. $v$3 is satisfied if the \textsf{\footnotesize CALLDATALOAD} for reading the array item is within a nested loop.
Hence, if a  \textsf{\footnotesize CALLDATALOAD}$(\textit{loc})$ reads an item from a dynamic \textsf{\small array} in an external function, $v$3 is satisfied, because \textit{n} bound checks (i.e., \textit{n} \textsf{\footnotesize LT}s) for preventing array overrun are located between the two \textsf{\footnotesize CALLDATALOAD}s. 
Hence, if $v$1, $v$2, and $v$3 are all fulfilled, the dynamic \textsf{\small array} is $n$-dimensional, and $\textit{num}_{n-1}, ..., \textit{num}_1$ are the sizes of the lower $n-1$ dimensions, respectively. We give an example in Supplementary material D to explain this rule. 

\noindent\textbf{R3:} R3 is used to infer an $n$-dimensional ($n>1$) static \textsf{\small array} in an external function. It depends on two requirements $v$1 and $v$2.
$v$1 is defined as $\neg (exp(\textit{loc})\diamond (\textit{offset}))$, where ``$\neg$'' means negation. $v$1 means that the read location is not computed from the \emph{offset} field. If a  \textsf{\footnotesize CALLDATALOAD} reads an item from a static \textsf{\small array} in an external function, $v$1 holds because the layout of a static \textsf{\small array} does not have an \textit{offset} field (Fig. \ref{fig_static_array}). $v$2 is defined as $\textsf{\footnotesize LT}_n(i_n,\textit{num}_n)\lhd \textsf{\footnotesize LT}_{n-1}(i_{n-1},\textit{num}_{n-1})\lhd ... \lhd \textsf{\footnotesize LT}_1(i_1,\textit{num}_1)\lhd \textsf{\footnotesize CALLDATALOAD}(\textit{loc})$. $v$2 is the same as $v$3 of R2. If a \textsf{\footnotesize CALLDATALOAD}$(\textit{loc})$ reads an item from an $n$-dimensional static \textsf{\small array} parameter in an external function, $v$2 holds because there are \textit{n} bound checks for each dimension in a nested loop to prevent array overrun before reading the array item. Hence, if $v$1 and $v$2 are satisfied, the static \textsf{\small array} is $n$-dimensional, and the item numbers from the highest dimension to the lowest dimension are $\textit{num}_n, ..., \textit{num}_1$, which are all constants since the item number of each dimension in a static \textsf{\small array} is known in compilation. Supplementary material D uses an example to explain this rule  due to page limit. 

\noindent\textbf{R4:}
$x$ is regarded as a \textsf{\small uint256}, if R1, R2 and R3 are not fulfilled. R4 means that without sufficient hints we just know that the length of $x$ is 32 bytes and thus regard a 32-bytes parameter as a \textsf{\small uint256}. We will refine it to a specific type after using other rules to get more hints. 

R19 is used for inferring a \textsf{\small struct} nested \textsf{\small array} parameter. 
R21 and R22 are used for inferring a \textsf{\small struct} parameter (R21) and a nested \textsf{\small array} parameter (R22). 
R24 and R25 are used for inferring a fixed-size \textsf{\small list} parameter (R24) and a \textsf{\small uint256} parameter (R25) after R20 is satisfied.

\subsection{Rules for \textsf{\small CALLDATACOPY}}
\label{rule_calldatacopy}
A \textsf{\footnotesize CALLDATACOPY} reads the value of a parameter(\S \ref{sec_calldata}).
R5 -- R10 and R23 are used for inferring a one-dimensional dynamic \textsf{\small array}/\textsf{\small bytes}/\textsf{\small string} parameter in a public function (R5), a one-dimensional static \textsf{\small array} in a public function (R6), a one-dimensional dynamic \textsf{\small array} in a public function (R7), a \textsf{\small bytes}/\textsf{\small string} in a public function (R8), an ($n+1$)-dimensional ($n>0$) static \textsf{\small array} in a public function (R9), an ($n+1$)-dimensional ($n>0$) dynamic \textsf{\small array} in a public function (R10) and a fixed-size \textsf{\small byte array} and \textsf{\small string} in Vyper (R23).

\subsection{Rules for Other Instructions}
\label{rule_other}
R11 -- R18 will be applied after R4 is satisfied (Fig. \ref{fig_rule_tree}). By leveraging them, we can refine a \textsf{\small uint256} parameter into a \textsf{\small uint}$\langle 256-8\times x\rangle, 0<x<32$ (R11), a \textsf{\small bytes}$\langle32-x\rangle, 0<x<32$ (R12), an \textsf{\small int}$\langle (x+1)\times 8 \rangle, 0\le x<31$ (R13 ), a \textsf{\small bool} (R14), an \textsf{\small int256} (R15), an \textsf{\small address} (R16), a \textsf{\small bytes} (R17), and a \textsf{\small bytes32} (R18).
R20 is used to distinguish Vyper bytecode from Solidity bytecode.
R26 is used for inferring a fixed-size \textsf{\small byte array} parameter in Vyper after R23 is satisfied.
R27 -- R31 will be applied after R25 is satisfied (Fig. \ref{fig_rule_tree}). By leveraging them, we can refine a \textsf{\small uint256} parameter into an \textsf{\small address} (R27), an \textsf{\small int128} (R28), a \textsf{\small decimal} (R29), a \textsf{\small bool} (R30), a \textsf{\small bytes32} (R31) in Vyper.

\section{Design and Implementation of \texttt{SigRec}}
\label{sec_sigrec}
\subsection{Overview}
\label{sec_overview}
\begin{figure*}[ht]
	\centering
	\vspace{-2ex}
	\includegraphics[width=0.9\textwidth]{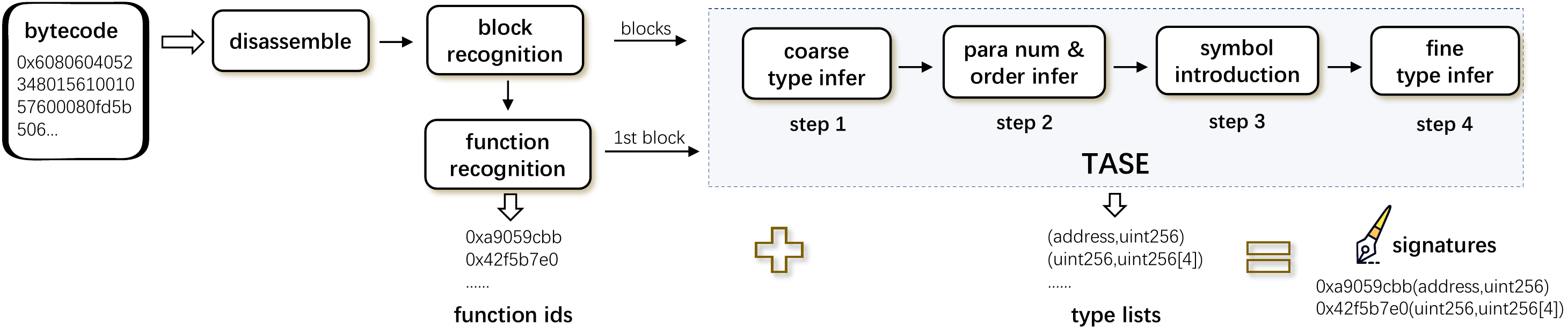}
	\vspace{-2ex}
	\caption{Architecture of \texttt{SigRec}}
	\vspace{-1ex}
	\label{fig_archi}
	%\vspace{-1ex}
\end{figure*}

Fig. \ref{fig_archi} shows the architecture of \texttt{\footnotesize SigRec} which takes in the \textit{runtime} bytecode of a smart contract and outputs the function signatures of all public/external functions in it. \texttt{\footnotesize SigRec} first disassembles the bytecode using Geth disassembler~\cite{geth} and recognizes basic blocks from them, then extracts function ids from the bytecode. Technical details are presented in Supplementary material E. After that,
it uses TASE to infer the types of all parameters (\S \ref{sec_tase}). Finally, it outputs the function ids as well as the list of parameter types. The reasons for using TASE rather than conventional SE and other methods are explained in Supplementary material F.

\subsection{TASE: Type-aware Symbolic Execution}
\label{sec_tase}

Being the core of \texttt{\footnotesize SigRec}, TASE has four steps. First, it conducts coarse-grained type inference to recognize Solidity and Vyper bytecode,
and then it recognizes \textsf{\small struct}, \textsf{\small array}s, \textsf{\small bytes}, \textsf{\small string}s and basic types of Solidity,
and fixed-size \textsf{\small list}s, fixed-size \textsf{\small byte array}s, fixed-size \textsf{\small string}s and basic types of Vyper.
This step only determines whether a parameter is a basic type instead of deciding the specific basic type, and this step does not infer the type of \textsf{\small array} items, the type of \textsf{\small struct} items, and this step does not distinguish a \textsf{\small bytes} from a \textsf{\small string}.
Second, it infers the number and the order of parameters. Third, by introducing parameter-related symbols when reading arguments from the call data, TASE determines whether an instruction operates on an argument and which argument is operated if that is the case. Finally, it conducts fine-grained type inference to distinguish different basic types for Solidity and Vyper, recognizes the item types of \textsf{\small array}s,
\textsf{\small nested array}s, and each item type of \textsf{\small struct} for Solidity and the item types of fixed-size \textsf{\small list} for Vyper,
and refines the types (e.g., from \textsf{\small uint256} to \textsf{\small bytes32}) for Solidity and Vyper.

TASE treats the call data as symbols and maintains the symbolic expressions of all variables depending on the call data in order to apply the rules. %, which are needed when applying the rules. 
TASE enhances conventional symbolic execution (CSE) with the ability to infer parameter types. 
It conducts symbolic execution (SE) to statically explores the paths of the smart contract
and stops if the jump target is determined by inputs (e.g., function parameters), because inputs are unknown in static analysis. This restriction is not a big problem in practice since we only find 5 smart contracts deployed in Ethereum containing such kind of jump instructions, and TASE usually does not need to analyze the code deep inside smart contracts because parameters are usually handled near the entry point of each function. 
TASE treats each value read from the environment as a free symbol, because \texttt{\footnotesize SigRec} focuses on how a smart contract processes the parameters rather than its program logic. 
Experimental results show that these two restrictions do not affect \texttt{\footnotesize SigRec}'s accuracy (\S \ref{sec_accuracy}). 
After describing the details of TASE, we 
present the differences between TASE and CSE, as well as the rationale of using TASE instead of other approaches.

\begin{figure*}[ht]
	\centering
	\vspace{-1ex}
	\includegraphics[width=1\textwidth]{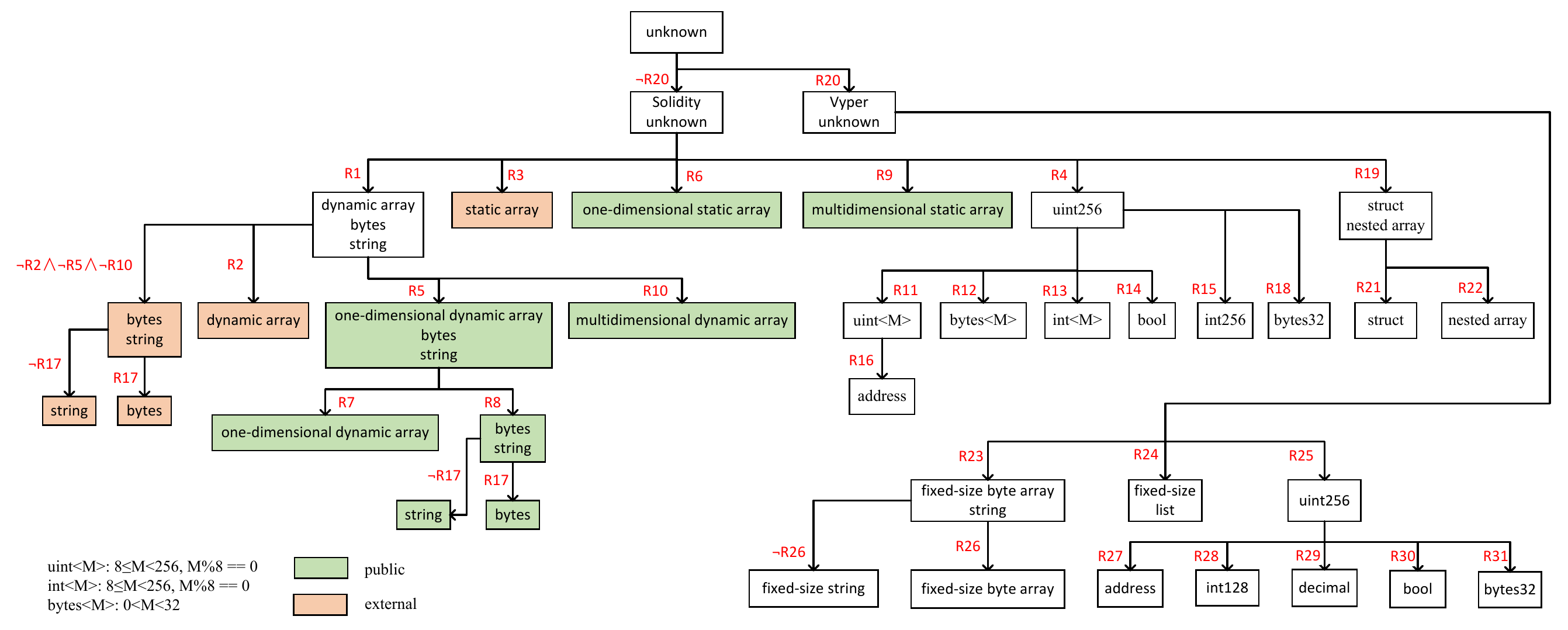}
	\vspace{-3ex}
	\caption{Hierarchy of rules applied by \texttt{SigRec}}
	\vspace{-2ex}
	\label{fig_rule_tree}
\end{figure*}

\noindent\textbf{Step 1. Coarse-grained type inference.}

At this step, TASE recognizes 
\textsf{\small struct}, \textsf{\small array}s, \textsf{\small bytes}, \textsf{\small string}s and basic types in Solidity,
and fixed-size \textsf{\small list}s, fixed-size \textsf{\small byte array}s, fixed-size \textsf{\small string}s and basic types in Vyper
according to the rules R1 -- R10 and R19 -- R25.
This step only determines whether a parameter is a basic type instead of deciding the specific basic type. %Moreover, this step 
This step does not infer the type of \textsf{\small array} items, the type of \textsf{\small struct} items, and this step does not distinguish a \textsf{\small bytes} from a \textsf{\small string}. Instead, these tasks are accomplished by step 4.

Fig. \ref{fig_rule_tree} shows the decision tree used by \texttt{\footnotesize SigRec} to determine the type of a parameter. A rectangle represents one or more types and an edge indicates applying one or more rules.
The root rectangle represents the unknown type. Two colors are used to differentiate between a public function and an external function. The $\neg$ indicates that the requirements of the corresponding rule are not satisfied.
\texttt{\footnotesize SigRec} infers the type of a parameter if all the rules on a path from the root rectangle to a leaf node are satisfied. 

We use the following example to explain the derivation process.
\texttt{\footnotesize SigRec} regards a parameter as a \textsf{\small bytes} in a public function if R1, R5, R8, and R17 are fulfilled in order.
More precisely, if R1 holds, the parameter should be a dynamic \textsf{\small array}/\textsf{\small bytes}/\textsf{\small string}, because the smart contract uses two consecutive \textsf{\footnotesize CALLDATALOAD}s to read the \textit{offset} field and the \textit{num} field. Since R5 is also satisfied, the type should be a one-dimensional dynamic \textsf{\small array} or a \textsf{\small bytes} or a \textsf{\small string} in a public function, because exactly one \textsf{\footnotesize CALLDATACOPY} is used to copy the parameter.
Then, R8 refines the type to a \textsf{\small bytes} or a \textsf{\small string} in a public function, because a \textsf{\small bytes}/\textsf{\small string} is extended to the length of a multiple of 32 bytes. Finally, R17 further refines the type to a \textsf{\small bytes} in the public mode, because the smart contract accesses a single byte of the parameter.

\noindent\textbf{Step 2. Determining the number and order of parameters.}

TASE infers the number of parameters by counting the number of rules R1, R3, R4, R6, R9, R21, R22, R23, R24, R25 which are used in coarse-grained type inference. Specifically, by counting the applied number of R1, TASE knows the number of dynamic \textsf{\small array}s/\textsf{\small bytes}/\textsf{\small string}s for Solidity (termed by \textit{n}1) since two consecutive \textsf{\footnotesize CALLDATALOAD}s are used before reading a dynamic \textsf{\small array}/\textsf{\small byte}s/\textsf{\small string}. 
By counting the applied number of R3, R6 and R9, TASE knows the number of static \textsf{\small array}s for Solidity (termed by \textit{n}2). 
Since R4 regards all basic types as \textsf{\small uint256}, TASE knows the number of basic types for Solidity (termed by \textit{n}3) by counting the applied number of R4.
By counting the applied number of R21 and R22, TASE knows the number of \textsf{\small nested array} and \textsf{\small struct} for Solidity (termed by \textit{n}4). 
By counting the applied number of R23 and R24, TASE knows the number of fixed-size \textsf{\small list}, \textsf{\small byte array} and \textsf{\small string} for Vyper (termed by \textit{n}5).
By counting the applied number of R25, TASE knows the number of basic types for Vyper (termed by \textit{n}6).
Therefore, TASE knows the number of parameters, which is $n1+n2+n3+n4+n5+n6$.

Then, TASE determines the order of parameters according to the locations of the corresponding arguments in the call data in ascending order. For example, if an argument $x$ locates before an argument $y$ in the call data, $x$ is on the left-hand side of $y$ in the parameter list.
The location of a basic type is given by the operand of the \textsf{\footnotesize CALLDATALOAD}.
The location of a dynamic \textsf{\small array}/\textsf{\small bytes}/\textsf{\small string} is indicated by the location of its \emph{offset} field, which is the operand of the first \textsf{\footnotesize CALLDATALOAD} instruction in the two consecutive \textsf{\footnotesize CALLDATALOAD} instructions (R1, \S \ref{rule_calldataload}). The location of a static \textsf{\small array} in a public function is given in the operand of the first \textsf{\footnotesize CALLDATACOPY} because it copies the inner-most dimension.
The start location of a static \textsf{\small array} in an external function cannot be directly obtained because it is not involved in reading \textsf{\small array} items.
Instead, a \textsf{\footnotesize CALLDATALOAD} is used to read an \textsf{\small array} item whose location is a constant number determined during compilation. Hence, we get the start location indirectly, which is after the end of the parameter immediately before the static \textsf{\small array}. If the static \textsf{\small array} is the first parameter, the start location is 0x4 since the function id precedes it. 

\begin{figure}[ht]
	\centering
	\vspace{-1ex}
	\includegraphics[width=0.48\textwidth]{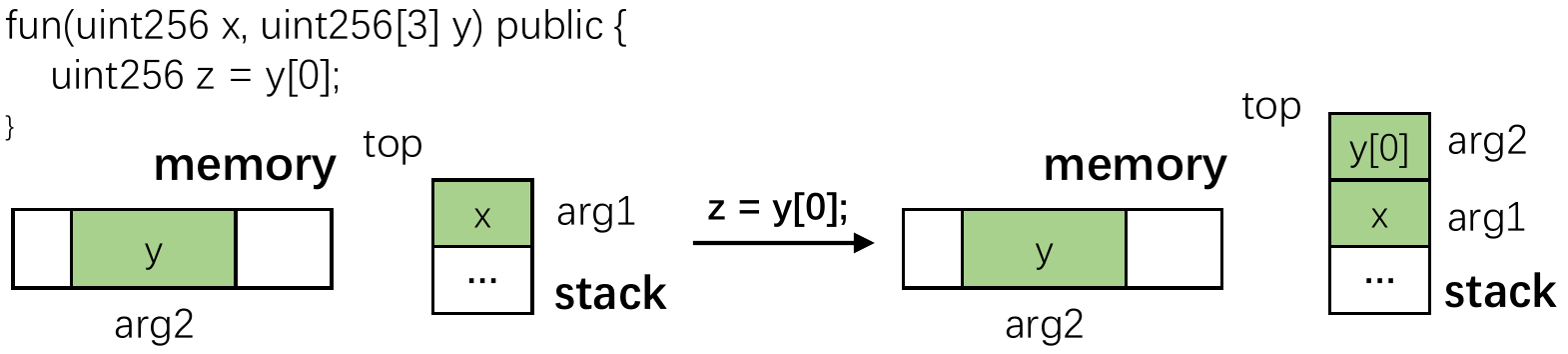}
	\vspace{-2ex}
	\caption{Marking parameter-related symbols}
	\vspace{-1ex}
	\label{fig_para_example}
\end{figure}

\noindent\textbf{Step 3. Introducing parameter-related symbols.}
When a smart contract reads arguments, TASE introduces parameter-related symbols by marking all bytes of an argument with the same symbol. After executing a \textsf{\footnotesize CALLDATALOAD} and a \textsf{\footnotesize CALLDATACOPY}, we mark the corresponding stack item and the memory region with symbols. Besides, we will copy the symbol of the memory region to the top item of the stack after executing an \textsf{\footnotesize MLOAD} if it reads from the memory region that stores arguments. Fig. \ref{fig_para_example} shows an example of marking parameter-related symbols. The public function, \textit{fun}(), takes in a \textsf{\small uint256} parameter \textit{x} and a \textsf{\small uint256[3]} parameter \textit{y}, and assigns a local variable \textit{z} as \textit{y}[0]. Before the assignment, \textit{x} and \textit{y} are read from the call data to the stack top and the memory, respectively. Since \textit{x} and \textit{y} are arguments, we mark the stack top and the corresponding memory region as symbols \textit{arg}1 and \textit{arg}2, respectively. The assignment copies the first item of \textit{y} from the memory to the stack top, and therefore we mark stack top with the same symbol of \textit{y} (i.e., \textit{arg}2).

\noindent\textbf{Step 4. Fine-grained type inference.}

Using R11--R18 and R26--R31 according to Fig. \ref{fig_rule_tree}, this step distinguishes basic types for Solidity and Vyper, infers the type of an \textsf{\small array} item and \textsf{\small nested array} item for Solidity, 
infers each item type of \textsf{\small struct} for Solidity,
and infers the type of a fixed-size \textsf{\small list} item for Vyper,
and refines the type of a parameter to a specific one for Solidity and Vyper.
Since TASE knows the structure of an \textsf{\small array} in step 1, it further knows the type of an \textsf{\small array} item in this step and then determines the type of the \textsf{\small array}.
For example, given a public function with a \textsf{\small uint8[]} parameter, TASE knows that the parameter is a one-dimensional dynamic \textsf{\small array} in a public function by applying R1, R5 and R7 in order in step 1.
Then, using R11 in this step, TASE learns that the type of an \textsf{\small array} item is \textsf{\small uint8}, and thus \texttt{\footnotesize SigRec} can recover the correct parameter type. 

\noindent\textbf{An example to illustrate the four steps.} Listing \ref{code_source_tase} shows a public function with two parameters: \textit{values} and \textit{to}, whose types are \textsf{\small uint8[]} and \textsf{\small address}, respectively. 
Line 2 reads the first item of \textit{values} and \textit{to}. Listing \ref{code_bytecode_tase} shows the corresponding EVM bytecode. For the ease of presentation, we just keep the EVM instructions that are needed by TASE to infer parameter types.

\begin{lstlisting}[caption={An example to explain the process of TASE}, label={code_source_tase}]
function test(uint8[] values, address to) public {
  to.send(values[0]);
}
\end{lstlisting}
\vspace{-1ex}

\begin{lstlisting}[caption={The bytecode code of this example}, label={code_bytecode_tase}]
CALLDATALOAD //read offset
......
CALLDATALOAD //read num
......
CALLDATACOPY //read values
......
CALLDATALOAD //read to
PUSH20 0xff...ff //20byte 0xFF
AND //mask
......
MLOAD //read values[0]
PUSH1 0xff
AND //mask
\end{lstlisting}
\vspace{-1ex}

\noindent\emph{Step 1: Coarse-grained type inference}. In Listing \ref{code_bytecode_tase}, Line 1 reads 32 bytes, termed by \textit{x}, from the 4th byte of the call data, and Line 3 reads 32 bytes, termed by \textit{y}, from the $(x+4)$-th byte of the call data.
Since these two instructions satisfy R1, 
this parameter is a dynamic \textsf{\small array}/\textsf{\small bytes}/\textsf{\small string}.
Since Line 5 reads $32\times y$ bytes from the $(36+x)$-th byte of the call data to memory from the 160th byte without using a loop, R5 is fulfilled and thus this parameter is refined to one-dimensional dynamic \textsf{\small array}/\textsf{\small bytes}/\textsf{\small string}.
\texttt{\footnotesize SigRec} further confirms that this parameter is a one-dimensional dynamic \textsf{\small array} because R7 is satisfied. Specially, the read length is the multiplication of 32 bytes (i.e., the length of each \textsf{\small array} item) and the item number (i.e., \textit{y}).
As Line 7 reads 32 bytes, termed by \textit{z}, from the 36th byte of the call data to the stack, R4 is fulfilled and thus \textit{z} is a basic-type parameter.

\noindent\emph{Step 2: Determining the number and order of parameters}. Line 3 reads a one-dimensional dynamic \textsf{\small array}, and its \textit{offset} field is located at the 4th byte of the call data. Line 7 reads a basic-type parameter located at the 36th byte of the call data. Since no other parameters are read in this example, this function takes in two parameters. Moreover, the one-dimensional dynamic \textsf{\small array} is the first parameter and the basic-type parameter is the second parameter, because the \textit{offset} field of the \textsf{\small array} precedes the basic-type parameter.

\noindent\emph{Step 3: Introducing parameter-related symbols}. Since Line 5 reads the first parameter to the memory, 
\texttt{\footnotesize SigRec} marks the corresponding memory region with the symbol \textit{arg}1. 
Similarly, since Line 7 reads the second parameter to the stack, \texttt{\footnotesize SigRec} marks the stack's top item with the symbol \textit{arg}2.

\noindent\emph{Step 4: Fine-grained type inference}. Since Line 9 masks the stack top, which is marked with \textit{arg}2, with 20-bytes 0xFF, R11 is satisfied and thus the second parameter should be a \textsf{\small uint160}. Since the second parameter is not involved in any mathematics operations, R16 is held, and thus \texttt{\footnotesize SigRec} refines the type of the second parameter to \textsf{\small address}. 
Line 11 reads 32 bytes from the 160th byte of the memory to stack top. Since such memory region is marked with \textit{arg}1 in step 3, \texttt{\footnotesize SigRec} marks stack top with \textit{arg}1. Since the first parameter is an \textsf{\small array}, the stack top stores an \textsf{\small array} item. As Line 13 masks the stack top with 0xFF, R11 is satisfied and thus we learn that the type of the \textsf{\small array} item is \textsf{\small uint8}. Eventually, TASE infers that the type list is ``\textsf{\small uint8}[], \textsf{\small address}'', the same as the source code.

\section{Evaluation}
\label{sec_experiments}
We implement \texttt{SigRec} in 8,327 lines of Python code and conduct extensive experiments to evaluate it by answering five research questions.
\noindent\textbf{RQ1}: How is the accuracy of \texttt{\footnotesize SigRec} in recovering function signatures (\S \ref{sec_accuracy})?
\noindent\textbf{RQ2:} Will \texttt{\footnotesize SigRec} be affected by different compiler versions and optimizations (\S \ref{sec_compiler})?
\noindent\textbf{RQ3}: How much time is required by \texttt{\footnotesize SigRec} to recover function signatures (\S \ref{sec_efficiency})?
\noindent\textbf{RQ4}: How frequently is each rule used for recovering function signatures (\S \ref{sec_use})?
\noindent\textbf{RQ5}: Is \texttt{\footnotesize SigRec} superior to existing tools (\S \ref{sec_compare})?

\subsection{Data Collection}
\label{sec_data}

To evaluate the accuracy of \texttt{\footnotesize SigRec}, we download the source code of all open-source smart contracts which were deployed before Jan. 6, 2021 from Etherscan because the ground-truth (i.e., function signatures) can be obtained from the source code. 
We collect 119,404 unique open-source smart contracts including 119,126 Solidity contracts and 278 Vyper contracts. 210,869 unique public/external function signatures are found in Solidity open-source smart contracts and 1,076 unique function signatures are found in Vyper open-source smart contracts.
To evaluate the efficiency of \texttt{\footnotesize SigRec} and the usefulness of heuristic rules, we collect the bytecode of all deployed smart contracts by instrumenting an Ethereum full node as suggested by~\cite{ChenINFOCOM18, chen2019dataether}. 
Eventually, we download 11,600,000 blocks (the last block was mined on Jan. 6, 2021) and obtain the bytecode of 37,009,570 smart contracts. 368,679 out of them are unique. There are 47,329,149 public/external functions in all smart contracts with 383,522 unique function signatures. Note that these 37,009,570 smart contracts include all open-source smart contracts.

\subsection{RQ1: How is the accuracy of \texttt{SigRec}?}
\label{sec_accuracy}

We evaluate the accuracy of \texttt{\footnotesize SigRec} with all 210,869 and 1,076 unique function signatures in Solidity and Vyper open-source smart contracts with ground-truth, respectively. 
A recovered function signature is \textit{correct}, if and only if the recovered function id, the number and the order of parameters, and the types of all parameters are the same as the ground-truth. The \textit{accuracy} is the proportion of correctly recovered function signatures to the total number of function signatures.
\texttt{\footnotesize SigRec} correctly recovers 208,218 and 1,052 function signatures for Solidity and Vyper, and hence its accuracy is 98.738\% ($(208,218+1,052)/(210,869+1,076)$). More specifically, \texttt{\footnotesize SigRec} correctly recovers 208,218 from all function signatures in Solidity smart contracts, and thus its accuracy for Solidity contracts is 98.743\% ($208,218/210,869$).
Besides, \texttt{\footnotesize SigRec} correctly recovers 1,052 from all 1,076 function signatures in Vyper smart contracts, and thus its accuracy for Vyper contracts is 97.770\% ($1,052/1,076$).

By manually investigate 2,651 incorrect function signatures in Solidity and 24 incorrect function signatures in Vyper, we reveal five cases of inaccuracies as follows. The value in $<>$ is the number of inaccurately recovered function signatures in each case. Please note that one incorrect function signature may be affected by multiple cases. We discuss how to further increase the accuracy of \texttt{\footnotesize SigRec} in \S \ref{sec_discuss}. 

\noindent\textbf{Case 1 $<498>$:} These smart contracts read the parameters that are not declared in function signatures by inline assembly.  
Listing \ref{code_inline} shows a practical case. The function start() has no parameters in declaration (Line 1), but it reads two parameters \textit{foo} (Line 7) and \textit{bar} (Line 8) using inline assembly. \texttt{\footnotesize SigRec} discovers these two parameters read by inline assembly, because it infers parameters by investigating how parameters are used.

\begin{lstlisting}[caption={An inaccurately recovered function signature in case 1}, label={code_inline}]
function start() auth note{
  stopped=false;
}
modifier note{
......
  assembly{
    foo:=calldataload(4)
    bar:=calldataload(36)
  }
......}
\end{lstlisting}

\noindent\textbf{Case 2 $<387>$:} These smart contracts forcibly convert the type of a parameter before using it. 
The recovered types may be more useful than those declared in function signatures, and the smart contract uses the parameter as the converted type. Listing \ref{code_conversion} shows a practical case. The function setGenOStat() has one parameter whose type is declared as \textsf{\small uint256[6]} (Line 1). \texttt{\footnotesize SigRec} recovers the type of \textit{\_gen0Stat} as \textsf{\small uint8[6]}, because the high-order 31 bytes of \textsf{\small uint256} are discarded by type conversion (Lines 4 -- 9). 
\begin{lstlisting}[caption={An inaccurately recovered function signature in case 2}, label={code_conversion}]
function setGen0Stat(uint256[6] _gen0Stat) public onlyCOO
{
  gen0State=Gen0Stat({
  retiredAge:uint8(_gen0Stat[0]),
  maxSpeed:uint8(_gen0Stat[1]),
  maxStamina:uint8(_gen0Stat[2]),
  maxStart:uint8(_gen0Stat[3]),
  maxBurst:uint8(_gen0Stat[4]),
  maxTemperament:uint8(_gen0Stat[5])});
}
\end{lstlisting}
\vspace{-1ex}

\noindent\textbf{Case 3 $<314>$:} These smart contracts have unused \textsf{\small array}s/\textsf{\small string}s/\textsf{\small bytes} parameters in external functions.
Since \texttt{\footnotesize SigRec} recognizes such types of parameters by checking how they are used, it cannot recognize the unused parameters.
Missing such parameters may not be an important issue because they are not used. 
It is worth noting that \texttt{\footnotesize SigRec} can recognize all parameters in public functions and the basic-type parameters in the external functions, because such parameters will be read into the stack or the memory no matter whether or not they will be used.
\ignore{That is, by analyzing how parameters are loaded into the stack or the memory, \texttt{\footnotesize SigRec} can infer the types of unused parameters in the public mode and unused basic-type parameters in the external mode.}

\noindent\textbf{Case 4 $<602>$:} Since \texttt{\footnotesize SigRec} recovers the type of each parameter with the \textit{storage} modifier as \textsf{\small uint256}, its output may be incorrect if the storage variable referred by the parameter is not a \textsf{\small uint256} integer. 

\noindent\textbf{Case 5 $<1,123>$:} Our rules cannot handle a few scenarios. First, \texttt{\footnotesize SigRec} cannot recognize static \textsf{\small array}s in external functions, if the smart contract is compiled with optimization and the index of the array item being accessed is a constant number, because there are no bound checks which are necessary for inferring array structure (R2 and R3, \S \ref{rule_calldataload}). Second, \texttt{\footnotesize SigRec} cannot distinguish some types if it cannot obtain sufficient clues. 
For example, we cannot distinguish a \textsf{\small bytes} from a \textsf{\small string} if the smart contract does not access a single byte of the \textsf{\small bytes} value since R17 exploits the fact that a \textsf{\small bytes} allows accessing its individual byte but a \textsf{\small string} does not support such functionality. 
As another example, we cannot distinguish a \textsf{\small struct} whose items are all basic types from the same items which are not placed in a \textsf{\small struct}, due to the description about \textsf{\small struct} in \S 2.3.1 that the call data layout and the accessing pattern of a static \textsf{\small struct} are the same to that of the same items which are not placed in a \textsf{\small struct}.

\noindent\textbf{Answer}: \textit{The accuracy of \texttt{\footnotesize SigRec} is 98.7\% for Solidity and 97.8\% for Vyper.}

\begin{figure}
    \centering
    \includegraphics[width=0.4\textwidth]{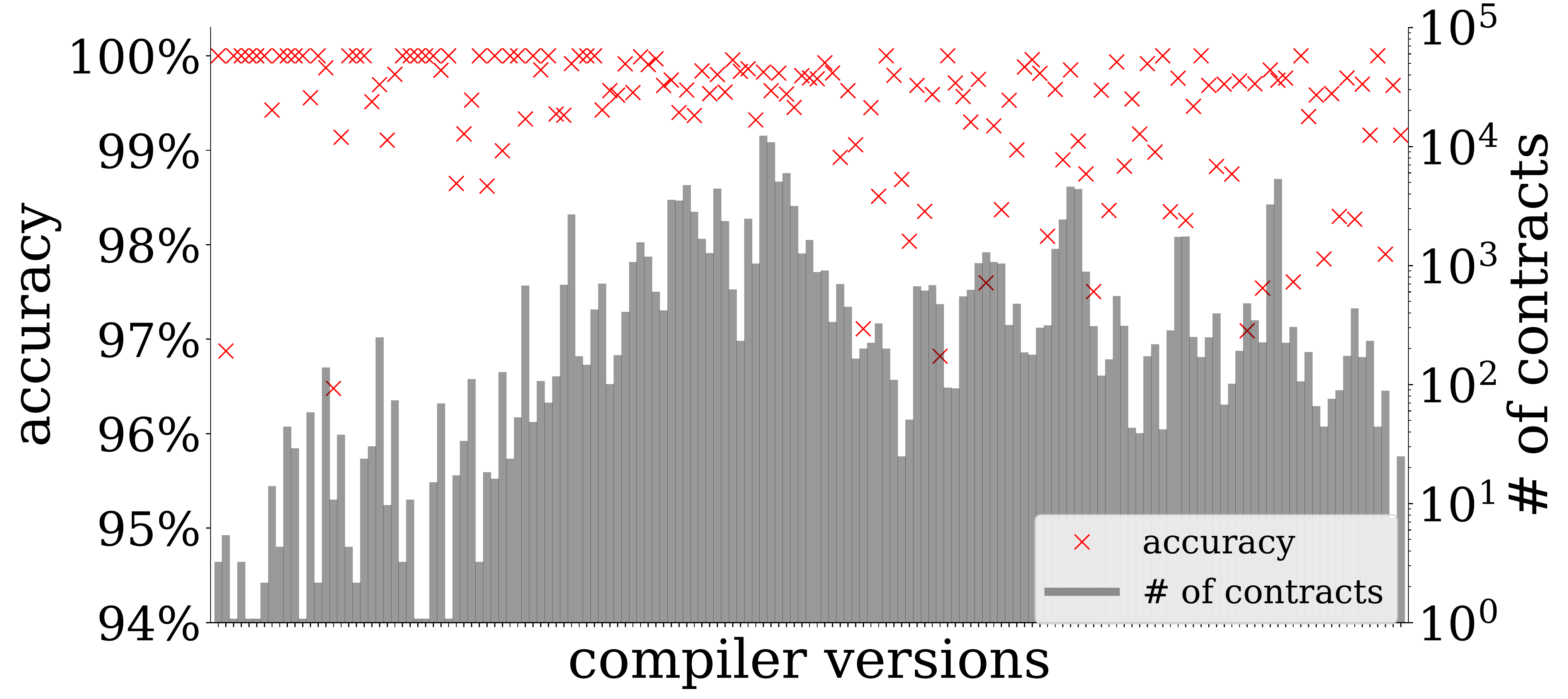}  
    \vspace{-2ex}
    \caption{Accuracies of \texttt{SigRec} for various Solidity compiler versions}   
    \label{fig_compiler}
\end{figure}
\hspace{3pt}
\begin{figure}
    \centering
    \includegraphics[width=0.4\textwidth]{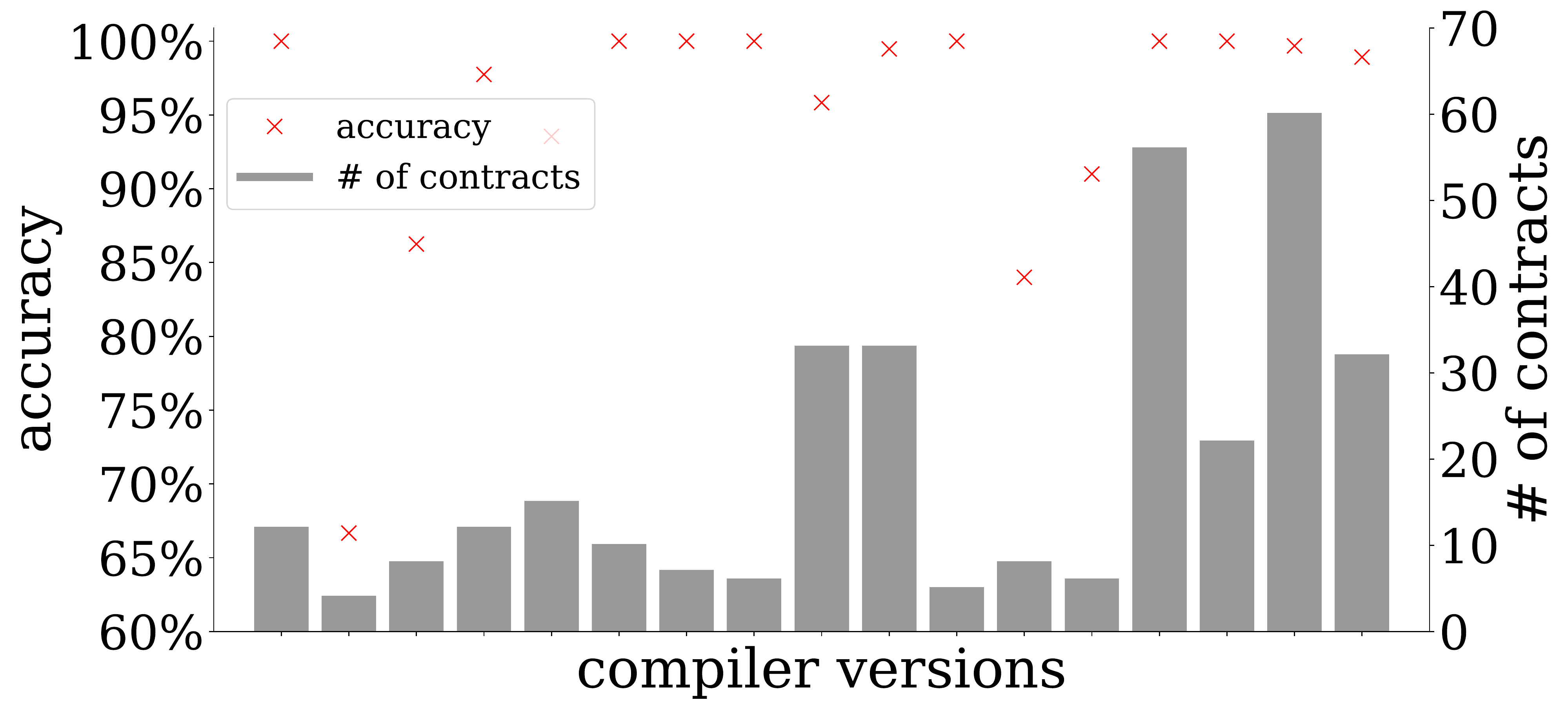}  
    \vspace{-2ex}
    \caption{Accuracies of \texttt{SigRec} for Vyper compiler versions}   
    \label{fig_vyper_compiler}
\end{figure}
\hspace{3pt}
\begin{figure}
    \centering
    \includegraphics[width=0.4\textwidth]{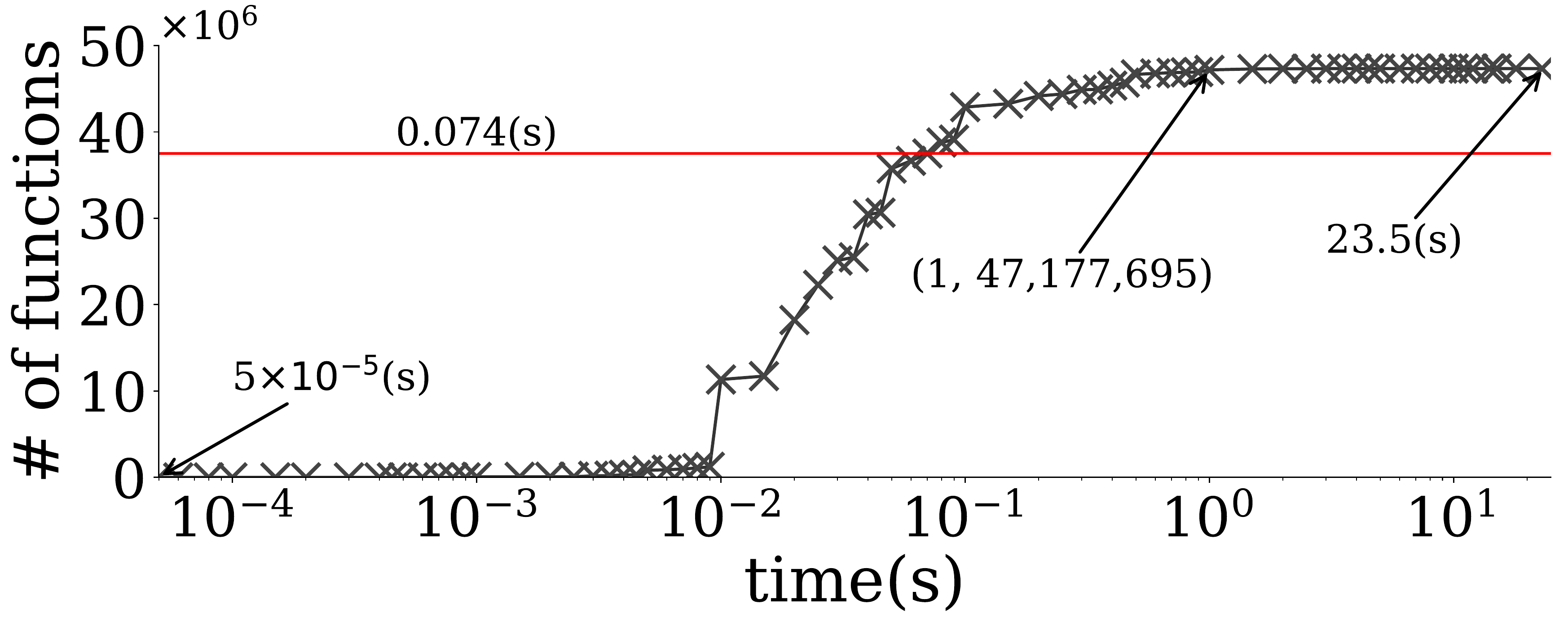}
    \vspace{-2ex}
    \caption{Time consumption to recover function signatures}   
    \label{fig_time}
\end{figure}
\hspace{3pt}
\begin{figure}
    \centering
    \includegraphics[width=0.45\textwidth]{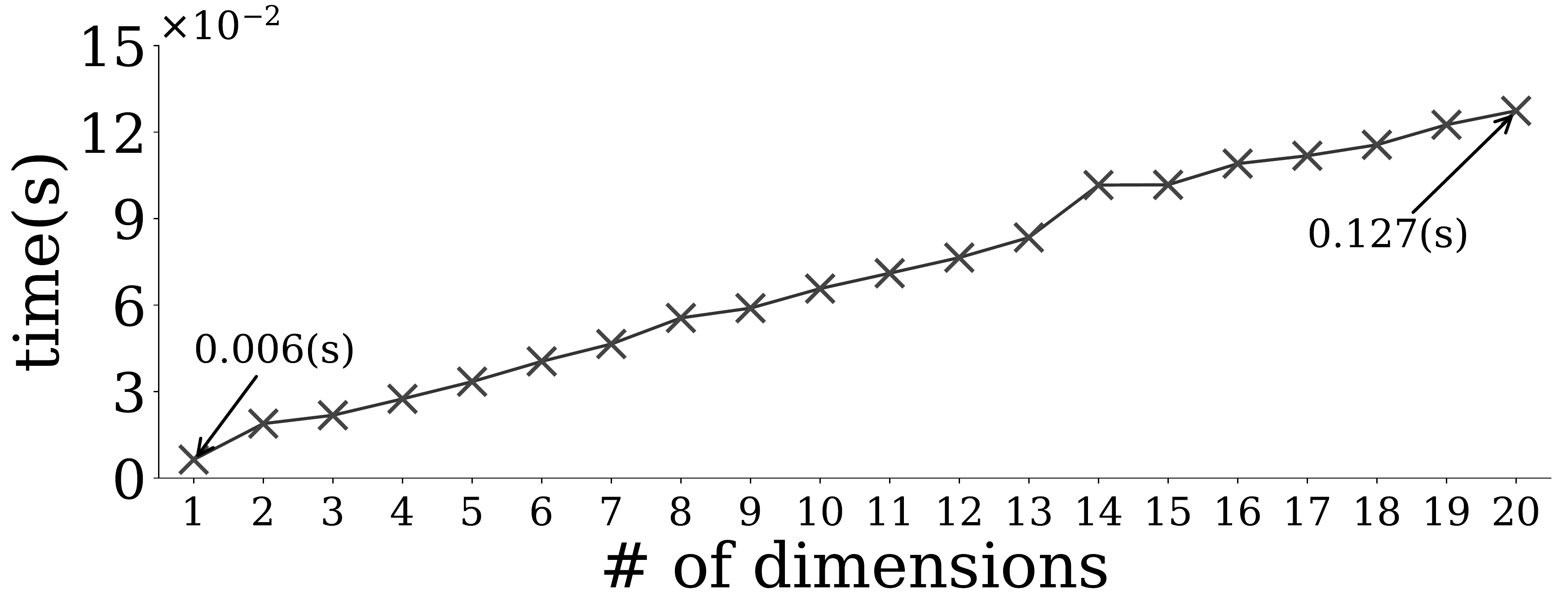}
    \vspace{-2ex}
    \caption{Time consumption to recover function signatures in different dimensions}   
    \label{fig_diff_dimension}
\end{figure}
\hspace{3pt}
\begin{figure*}
    \centering
	\vspace{-1ex}
    \includegraphics[width=0.85\textwidth]{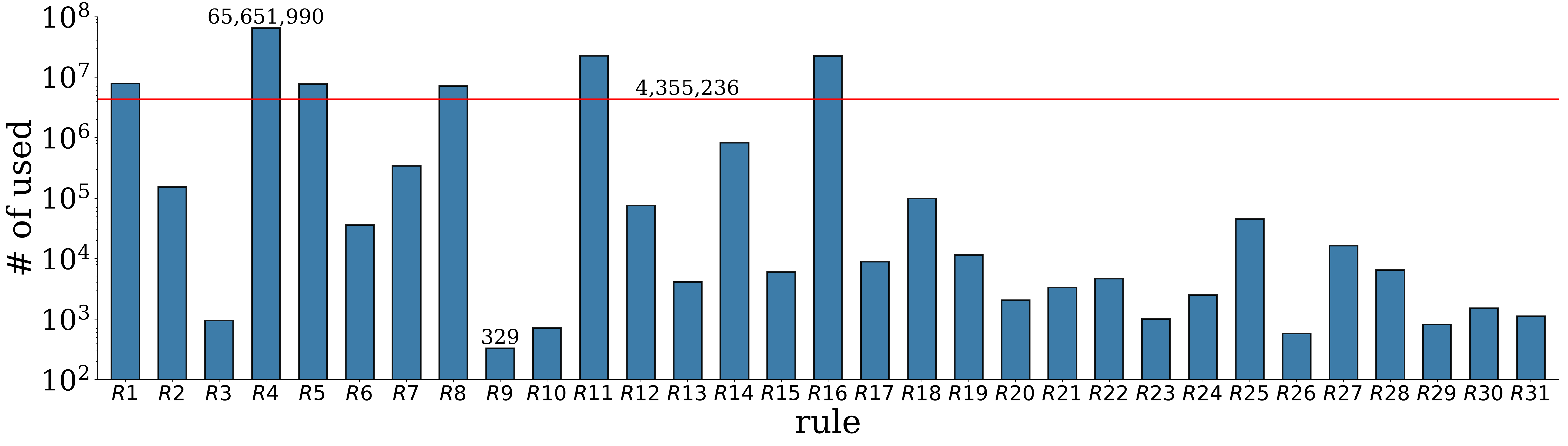}
	\vspace{-3ex}
    \caption{The number of times each rule is used} 
	\vspace{-2ex}  
    \label{fig_rule}
\end{figure*}

\subsection{RQ2: Will \texttt{SigRec} be affected by different compiler versions and optimizations?}
\label{sec_compiler}

We evaluate \texttt{\footnotesize SigRec} using the open-source smart contracts that are compiled by different versions of compilers w/o optimization. Since Etherscan lists the compiler version and whether or not a smart contract is optimized as well as the optimization level if any for each open-source smart contract~\cite{etherscan}, we eventually collect 119,126 and 278 unique open-source smart contracts in Solidity and Vyper, respectively. These smart contracts were compiled by 155 versions of Solidity compilers and 17 versions of Vyper compilers, e.g., Solidity from V 0.1.1 to V 0.8.0 and Vyper from V 0.1.0b4 to V 0.2.8. Note that we consider a compiler version with optimization and that without optimization as two different versions. 
Then, we compute the accuracy of \texttt{\footnotesize SigRec} for each compiler version. Fig. \ref{fig_compiler} shows the accuracies in ascending order of Solidity versions, including the number of smart contracts compiled by each version, ranging from 1 to 11,430. Fig. \ref{fig_vyper_compiler} shows the accuracies in ascending order of Vyper versions, which also includes the number of smart contracts compiled by each version, ranging from 1 to 57. We can see that the accuracy of \texttt{\footnotesize SigRec} is never lower than 96\% for all 155 Solidity compiler versions, and the accuracy of \texttt{\footnotesize SigRec} is more than 90\% in 12 out of 15 Vyper compiler versions. 
By investigating the three Vyper compiler versions that make the accuracy of our tool be lower than 90\%, we find that the relatively low accuracy is not caused by compiler features. Instead, the small number of Vyper contracts which are compiled by these three versions is the reason. Therefore, the accuracy does not show a downward trend with the evolving of compiler versions.

\noindent\textbf{Answer}: \textit{Compiler versions and optimizations bring minimum impact on the accuracy of \texttt{\footnotesize SigRec}. Its accuracy is never lower than 96\% for all 155 Solidity compiler versions and higher than 90\% in 80\% of Vyper compiler versions.}

\subsection{RQ3: How much time is required by \texttt{SigRec} to recover function signatures?}
\label{sec_efficiency}

We apply \texttt{\footnotesize SigRec} to all 47,329,149 public/external functions, and measure the time consumption for recovering each function. This experiment is conducted on a desktop equipped with an Intel Xeon E5-2609 CPU, 16GB main memory and 10TB hard disk. The results are shown in Fig. \ref{fig_time}, where each cross (\textit{x}, \textit{y}) indicates that \texttt{\footnotesize SigRec} uses no more than \textit{x} seconds to recover each of \textit{y} function signatures.
It shows that the time needed to recover a function signature ranges from $5\times 10^{-5}$ seconds to 23.5 seconds, and the average time is 0.074 seconds. For 99.7\% ($47,177,695/47,329,149$) of function signatures, \texttt{\footnotesize SigRec} needs no more than 1 second to recover each of them. 

We find three reasons for the function signatures which cost long analysis time. First, the analyzed function has many instructions. Second, the analyzed function signature contains any parameter types which will be confirmed after \texttt{\footnotesize SigRec} executes all the instructions of the function. Taking \textsf{\small uint256} as an example, please recall that we initially assume every basic type as a \textsf{\small uint256}, and then we change \textsf{\small uint256} to another type if the parameter is involved in a special instruction. In other words, \texttt{\footnotesize SigRec} recovers the type of parameter as \textsf{\small uint256} after it confirms that the parameter is not involved in any special instructions by running all instructions of the analyzed function. On the contrary, \texttt{\footnotesize SigRec} can identify some other parameter types quickly because it needs not to run all instructions of the analyzed function. For example, \texttt{\footnotesize SigRec} identifies an \textsf{\small int}$\langle (x+1)\times 8 \rangle$ when it encounters a \textsf{\footnotesize SIGEXTEND} instruction. 

Third, we find that recovering a higher dimensional \textsf{\small array} costs more time than a lower dimensional \textsf{\small array}, because the accessing code for a higher dimensional array contains more bound checks and a larger nested loop for reading array items.
To quantitatively study the effect of \textsf{\small array} dimension, we run \texttt{\footnotesize SigRec} to recover an \textsf{\small array} parameter whose dimension ranges from 1 to 20 and each \textsf{\small array} item is an \textsf{\small uint256}. Fig. \ref{fig_diff_dimension} shows that the time consumption increases linearly along with the increasing of the \textsf{\small array} dimension. Hence, the \textsf{\small array} dimension is not a major reason for long analysis time because we find that the \textsf{\small array} dimension is no larger than 3 in practice.

\noindent\textbf{Answer}: \textit{\texttt{\footnotesize SigRec} is very efficient. For 99.7\% of function signatures, it uses no more than 1 second to recover each of them}.
\subsection{RQ4: How frequently is each rule used?}
\label{sec_use}

After recovering 47,329,149 public/external functions in all smart contracts, we count how frequently each rule is used. As shown in Fig. \ref{fig_rule}, all rules have been used. On average, each rule is used for 4,355,236 times. R4 is the most frequently-used rule because the number of basic types in Solidity is more than that of other types. R9 is the least frequently-used rule, because we find that multidimensional static \textsf{\small array}s are infrequently used as parameters in public functions.

\noindent\textbf{Answer}: \textit{All rules have been used with different frequencies.}

\subsection{RQ5: Is \texttt{SigRec} superior to existing tools?}
\label{sec_compare}

We compare \texttt{\footnotesize SigRec} with five state-of-the-art decompilers, Gigahorse~\cite{gigahorse_web}, Eveem~\cite{eveem_web}, OSD~\cite{osd_web}, EBD~\cite{ebd_web}, and JEB~\cite{jeb_web} in terms of the accuracy of recovering function signatures.

\noindent\textbf{Datasets.}
We prepare three datasets for evaluation. Dataset 1 includes all unique closed-source smart contracts. 
Dataset 2 contains 1,000 synthesized functions. 
We construct the name of each function with 5 randomly-selected letters, and annotate it as a public function or an external function randomly. Each function takes in \textit{x} parameters ($1\le x\le 5$). For each parameter, we create its name with 5 randomly-selected letters and randomly select a \ignore{type that is supported by our approach as the} parameter type.
Each \textsf{\small array} parameter has at most three dimensions\ignore{, and the size of each dimension is at most 10 items}. Besides, each \textsf{\small array} parameter has at most five items. The body of each function contains statements to access each parameter, including \textsf{\small array} items and individual byte of a \textsf{\small bytes} and a \textsf{\small byte32}. We construct 100 smart contracts in Solidity, each of which includes 10 synthesized functions, and compile them into bytecode by Solidity 0.5.5 with the probability of 50\% to turn on optimization with the default optimization level.
Dataset 3 includes all unique open-source contracts obtained from Etherscan. 

\begin{table}[]
    \centering
	%\vspace{-4ex}
	\caption{Results of data set 1}
	\vspace{-2ex}
	\includegraphics[width=0.4\textwidth]{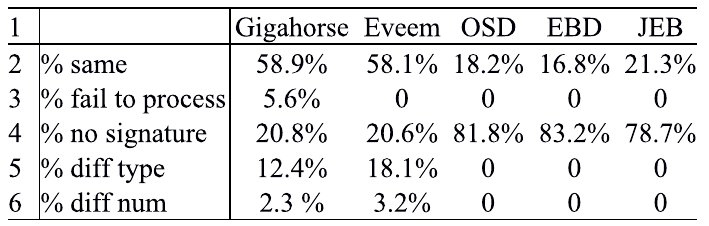}
	%\vspace{-2ex}
	\label{tab_compare_no_source}
\end{table}
\hspace{2pt}
\begin{table}[]
    \centering
	\vspace{-2ex}
	\caption{Results of data set 2}
	\vspace{-2ex}
	\includegraphics[width=0.4\textwidth]{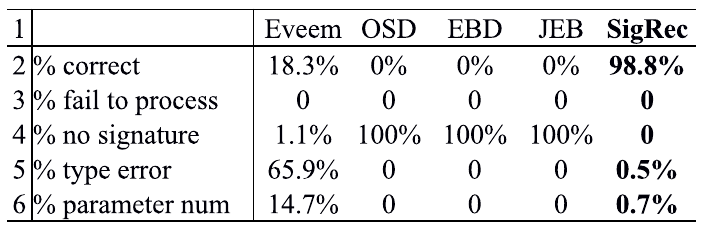}	
	%\vspace{-2ex}
	\label{tab_synthesize}
\end{table}
\hspace{2pt}
\begin{table}[]
    \centering
	\vspace{-2ex}
	\caption{Results of data set 3}
	\vspace{-2ex}
	\includegraphics[width=0.45\textwidth]{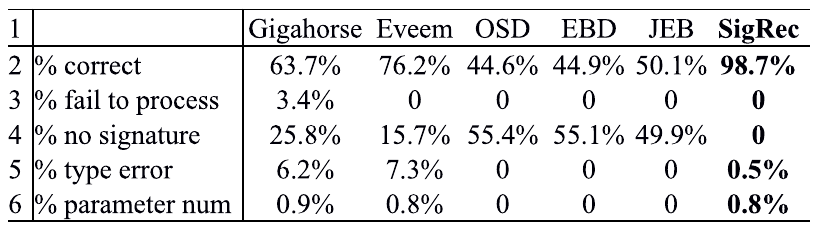}
	%\vspace{-2ex}
	\label{tab_compare_open_source}
\end{table}
\hspace{2pt}
\begin{table}[]
    \centering
	\vspace{-2ex}
	\caption{Results of \textsf{\small struct} and nested \textsf{\small array} in Solidity}
	\vspace{-2ex}
	\includegraphics[width=0.45\textwidth]{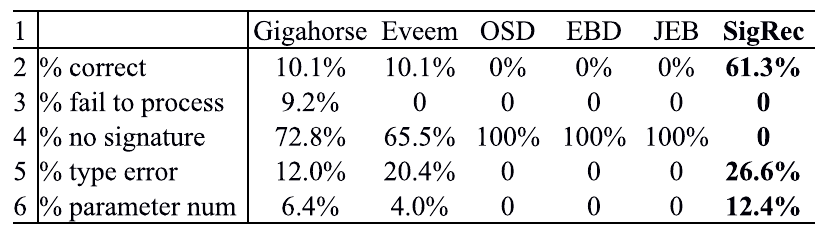}
	%\vspace{-2ex}
	\label{tab_compare_abiv2}
\end{table}
\hspace{2pt}
\begin{table}[]
    \centering
	\vspace{-2ex}
	\caption{Results of function signatures in Vyper}
	\vspace{-2ex}
	\includegraphics[width=0.45\textwidth]{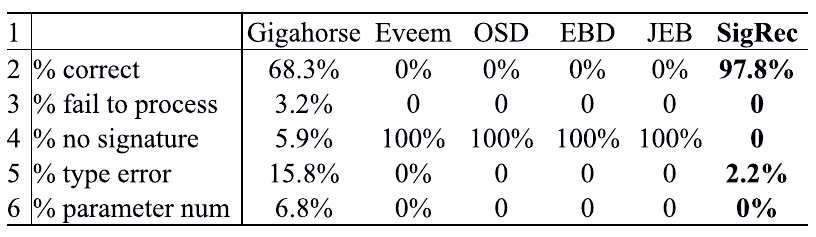}
	%\vspace{-2ex}
	\label{tab_compare_vyper}
\end{table}

\noindent\textbf{Results of dataset 1.}
Table \ref{tab_compare_no_source} lists the results of all closed-source smart contracts. Since there is no ground-truth for closed-source smart contracts, row 2 lists the ratio of function signatures that the five existing tools produce the same results as \texttt{\footnotesize SigRec}.
Since tools may abort abnormally when recovering some function signatures, Row 3 presents the ratio of such function signatures.
Row 4 presents the ratio of function signatures that tools just find their function ids but fail to recover their parameter lists. Row 5 presents the ratio of function signatures that the five tools find the same number of parameters as \texttt{\footnotesize SigRec}, but at least one parameter type recovered by them is different from what is recovered by \texttt{\footnotesize SigRec}. Row 6 presents the ratio of function signatures that the five tools find different parameter numbers from \texttt{\footnotesize SigRec}.

We have two observations from the results. 
First, the accuracies of OSD, EBD and JEB are low because if their databases do not record a function signature, they cannot recover it. Please note that the databases of OSD, EBD and JEB just cover 31.7\% of all function signatures in Ethereum public chain (Supplementary material A). Second, Gigahorse and Eveem outperform OSD, EBD and JEB because they try to infer parameter types if they cannot find function signatures from EFSD~\cite{gigahorse,eveem_web}. Unfortunately, they still fail to correctly recover a large proportion of function signatures.
Since Gigahorse and Eveem are not open-source,
we investigate their errors by randomly selecting 50 recovered function signatures with different numbers of parameters and 50 recovered function signatures with different parameter types from \texttt{\footnotesize SigRec} for each of them from data set 1.
After checking these 200 function signatures (100 ones are not recovered correctly by Gigahorse, and the other 100 ones are not recovered correctly by Eveem), we observe the following errors: 
(1) Gigahorse and Eveem report the wrong type. For example, Gigahorse regards the type of a \textsf{\small address}/\textsf{\small uint8}/\textsf{\small uint256[]}/\textsf{\small string}/\textsf{\small bytes} parameter as \textsf{\small uint256} for some function signatures. 
Moreover Gigahorse even reports the type of a \textsf{\small uint256} parameter as \textsf{\small uint2304}, which does not exist in smart contracts, in one function signature.
(2) Gigahorse mistakenly regards several consecutive parameters as one parameter of a \textit{nonexistent} type.
For example, for one function with four parameters whose types are \textsf{\small address[]}, \textsf{\small uint256[5]}, \textsf{\small uint256[5]} and \textsf{\small uint256[2]}, respectively, Gigahorse outputs only one parameter of \textsf{\small uint3328}, a nonexistent type. 
(3) Gighorse adds more parameters mistakenly.
(4) Gigahorse and Eveem miss some parameters\ignore{, e.g., \textsf{\small address}, \textsf{\small uint256[]}, \textsf{\small string}, \textsf{\small bytes}}. Supplementary material G lists all errors made by Gigahorse and Eveem from the 200 function signatures.

\noindent\textbf{Results of dataset 2.}
Table \ref{tab_synthesize} lists the results of recovering 1,000 synthesized function signatures by five tools. Row 2 presents the ratio of the function signatures that are correctly recovered. Row 3 and row 4 have the same meaning as the 3rd and 4th rows in Table \ref{tab_compare_no_source}.  Row 5 presents the ratio of function signatures that the tools can find the correct parameter numbers but fail to recover the types of at least one parameter. Row 6 presents the ratio of function signatures whose number of parameters cannot be correctly found by the tools. We do not evaluate Gigahorse with the synthesized function signatures because the web service of Gigahorse~\cite{gigahorse_web} takes in the addresses of deployed smart contracts rather than the bytecode of smart contracts while deploying 1,000 synthesized functions to Ethereum will cost much money. 

The results show that  \texttt{\footnotesize SigRec}'s accuracy is 98.8\%. Manual investigation reveals that the 8 incorrectly recovered function signatures belong to case 5 (\S \ref{sec_accuracy}). 
OSD, EBD and JEB recover 0 function signatures, because none of the synthesized function signatures are recorded in their databases. Eveem correctly recovers 183 synthesized function signatures thanks to its heuristic rules. However, Eveem cannot produce function signatures for 11 functions, outputs incorrect parameter types for 659 function signatures, and outputs incorrect parameter numbers for 147 function signatures. We manually investigate the errors made by Eveem and the results are similar with the investigation of the errors made by Eveem in dataset 1. We present the detailed results in Supplementary material G.  

\noindent\textbf{Results of dataset 3.}
Table \ref{tab_compare_open_source} lists the results. Row 2 to row 6 have the same meaning with the rows in Table \ref{tab_synthesize}. We have several
observations. First, \texttt{\footnotesize SigRec} outperforms the other tools by at least 22.5\% even the analyzed smart contracts are open-source. Second, Gigahorse is not stable because it aborts abnormally in processing 3.4\% of function signatures and it fails to recover some function signatures even they are recorded in EFSD. Third, the accuracies of OSD, EBD and JEB are lower than 51\%, even if these functions are implemented in open-source smart contracts.
That is, more than 49\% function signatures in open-source smart contracts are not recorded in existing function signature databases. Fourth, Eveem outperforms OSD, although they both query EFSD, because Eveem uses its simple rules to infer parameter types if it cannot find function signatures from EFSD. Finally, \texttt{\footnotesize SigRec} outperforms Eveem, because \texttt{\footnotesize SigRec} has a 
complete set of rules than Eveem and the powerful TASE
engine to infer parameter types. We also investigate the errors made by Gigahorse and Eveem by randomly selecting 200 incorrectly recovered function signatures. The results are similar to their errors in processing dataset 1. Supplementary material G contains the detailed results. 

\noindent\textbf{Recovery of \textsf{\small struct} and nested \textsf{\small array} in Solidity}. 
We compare \texttt{\footnotesize SigRec} and existing tools in terms of recovering \textsf{\small struct} and nested \textsf{\small array}, which are new parameter types introduced from V 0.4.19 and are supported by the experimental version of Solidity, so-called ABIEncoderV2 before V 0.8.0~\cite{solidity-changelog}. 
1,104 function signatures contain \textsf{\small struct} or nested \textsf{\small array} in dataset 3, and results are shown in Table \ref{tab_compare_abiv2}. We find that the accuracies of existing tools are no higher than 11\%, indicating than existing tools do not support these two new parameter types properly. Gigahorse and Eveem have the same accuracy in recovering \textsf{\small struct} and nested \textsf{\small array} because these 10.1\% of function signatures are recorded in EFSD. In other words, the rules built in these two tools cannot handle \textsf{\small struct} and nested \textsf{\small array}. The accuracy of \texttt{\footnotesize SigRec} is 61.3\%, which is much higher than existing tools. After investigating the function signatures recovered by \texttt{\footnotesize SigRec} incorrectly, we find that all of them belong to case 5 (\S \ref{sec_accuracy}). 
Although the accuracy of \texttt{\footnotesize SigRec} in recovering \textsf{\small struct} and nested \textsf{\small array} is not as high as the accuracy in recovering other types, we believe it is not a big limitation of \texttt{\footnotesize SigRec} because the function signatures taking in \textsf{\small struct} or nested \textsf{\small array} just account for 0.5\% of the total function signatures. We will derive more rules for recovering \textsf{\small struct} and nested \textsf{\small array} as our future work.

\noindent\textbf{Recovery of function signatures in Vyper contracts.} We compare \texttt{\footnotesize SigRec} and existing tools in recovering the function signatures in Vyper contracts, since Vyper is an alternative to Solidity and aims to provide better security than Solidity~\cite{kaleem2020vyper}.
1,076 function signatures appear in all 278 open-source unique Vyper contracts in dataset 3. Results are shown in Table \ref{tab_compare_vyper}. The accuracy of \texttt{\footnotesize SigRec} is 97.8\%, and we find that all the function signatures which are recovered by \texttt{\footnotesize SigRec} incorrectly belong to case 5 (\S \ref{sec_accuracy}). 
Eveem, OSD, EBD and JEB cannot handle Vyper contracts. The accuracy of Gigahorse is 68.3\%, and we then investigate the reason for its high accuracy. We find that Gigahorse recovers all Vyper types as \textsf{\small uint256} and fortunately, 68.3\% function signatures in Vyper contracts taking in \textsf{\small uint256} parameters only. 
Hence, existing tools are inadequate in recovering the function signatures in Vyper contracts.

\noindent\textbf{Importance of function signatures uncovered by existing databases}. 
289,123, accounting for 75.4\% of all unique function signatures are not included in EFSD, the most popular database for function signatures. To investigate whether these uncovered function signatures are important, we count the number of unique bytecode which contains at least one uncovered function signatures. The number of such unique bytecode is 148,268, which accounts for 40.2\% of the total unique bytecode, and thus we believe these uncovered function signatures are important. In other words, EFSD misses important function signatures which can be complemented by \texttt{\footnotesize SigRec}.

\noindent\textbf{Answer}: \textit{\texttt{\footnotesize SigRec} is much more accurate than the state-of-the-art tools in recovering function signatures without the need of function signature databases.}

\section{Applications of the Recovered Function Signatures}
\label{sec_app}
We use three applications to demonstrate the usefulness of \texttt{\footnotesize SigRec}, including detecting short address attacks, fuzzing smart contracts, and reverse engineering the bytecode of smart contracts. Due to page limit, we detail the first application and  briefly introduce the results of the last two applications here. Interested readers can refer to Supplementary material H, I for more details.

\subsection{Detecting Short Address Attacks}
We consider the actual arguments sent for function invocation as \textit{invalid} if they are not encoded according to the specification~\cite{solidity-sig}.
Although Web3 APIs can generate valid actual arguments, many invalid actual arguments can still be found in practice for various reasons, such as  malicious purposes~\cite{short}, confusion of parameters padding~\cite{confusion1, confusion2}, compiler bugs~\cite{encode_bug}, compatibility issues of Web3~\cite{web3_compatible1,web3_compatible2}, changes in the new versions of compiler~\cite{encode_change}, etc.
Invalid actual arguments may incur runtime exceptions, unexpected execution results, and even money stolen~\cite{short}.
We design ParChecker, the first tool for automatically detecting invalid actual arguments sent to Solidity smart contracts by using the recovered function signatures. We plan to extend ParChecker to support Vyper smart contracts as our future work. Please note that without function signatures, it is difficult to detect them because different parameter types have different padding schemes (\S \ref{sec_calldata}). For example, to detect short address attacks, we must focus on the functions with \textsf{\small address} type parameter(s).

ParChecker takes in the call data of a function invocation, and outputs (1) \textit{false} if the actual arguments are invalid, (2) \textit{true} otherwise. 
ParChecker first gets the function id from the call data, and looks for its function signature from the results of \texttt{\footnotesize SigRec}. Then, for each parameter in the function signature, ParChecker locates the actual argument in the call data and checks whether its padding is correct. Specifically, if the parameter is a basic type, ParChecker applies the rules in Table \ref{tab_check_rule} to check the correctness of padding. The rules are derived from the padding scheme of each parameter type described in \S \ref{sec_calldata}. 
If the parameter is a static \textsf{\small array}, ParChecker checks each item of the \textsf{\small array} according to the rules in Table \ref{tab_check_rule}. If the parameter is a dynamic \textsf{\small struct}, nested \textsf{\small array}, dynamic \textsf{\small array}, \textsf{\small bytes}, or \textsf{\small string}, ParChecker checks its structure and content. We use a \textsf{\small bytes} parameter as an example to illustrate the process. ParChecker first locates the \textit{offset} field and the \textit{num} field pointed by the \textit{offset} field in the call data. If either field cannot be found, the structure of the actual argument is invalid. Otherwise, ParChecker reads the \textit{num} field whose value is the length of the \textsf{\small bytes} before padding, termed by \textit{x}. After that, ParChecker reads $\lceil x/32\rceil \times 32$ bytes, termed by \textit{v}, after the \textit{num} field, and checks whether the lower-order $\lceil x/32\rceil \times 32-x$ bits of \textit{v} are zeros, where ``$\lceil \rceil$'' denotes rounding up to the next integer. If not, the content of the actual argument is invalid.

\begin{table}[ht]
	\centering
	\vspace{-3ex}
	\caption{Rules for checking basic types}
	\vspace{-2ex}
	\includegraphics[width=0.45\textwidth]{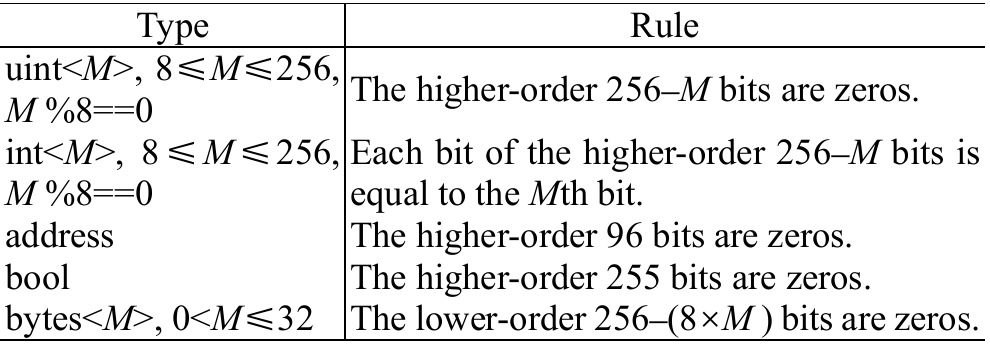}
	\vspace{-1ex}
	\label{tab_check_rule}
\end{table}

\begin{figure}[ht]
	\centering
	\vspace{-2ex}
	\includegraphics[width=0.48\textwidth]{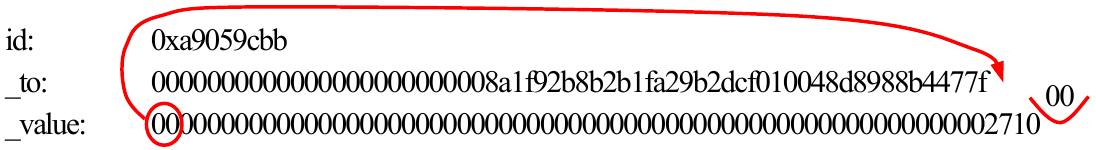}
	\vspace{-2ex}
	\caption{A short address attack} 
	\vspace{-1ex}
	\label{fig_short}
\end{figure}

We use ParChecker to analyze all transactions in 556,361 blocks starting from the block number 6,625,132. It finds 91,257,261 transactions with non-empty call data and detects 1,024,974 ($1\%=1,024,974/91,257,261$) transactions with invalid actual arguments. These problematic transactions contain 1,292 unique function ids, invoking 13,556 smart contracts. Among them, we look for the transactions launching short address attacks because they steal tokens~\cite{short}. 
We use such a transaction, which is detected by ParChecker and shown in Fig. \ref{fig_short}, to explain how a short address attack can steal tokens, and then present the detection result.

This transaction invokes the transfer() function of a token smart contract whose function id is 0xa9059cbb to transfer tokens. transfer() has two parameters: \textit{\_to} of the type \textsf{\small address} referring to the receiver, and \textit{\_value} of the type \textsf{\small uint256} denoting the amount of tokens. The attacker leaves off the trailing zeros of \textit{\_to} and then EVM pads \textit{\_to} to 32 bytes by the higher-order zeros of \textit{\_value}. After that, EVM adds zeros after the lowest byte of \textit{\_value} to the length of 32 bytes, and thus \textit{\_value} changes from 0x2710 to 0x271000. Attackers can launch such an attack to trick a victim exchange wallet to transfer much more tokens to the address (i.e., \textit{\_to}) controlled by them ~\cite{short}

We discover such attacks from the results of ParChecker by first locating the invocation of transfer(). Then, we check whether the length of actual parameters \textit{len} is shorter than 64 bytes, which is the length of a valid \textsf{\small address} plus a valid \textsf{\small uint256}. If so, for the last 32 bytes of the arguments, we check whether its highest $64-\textit{len}$ bytes are zeros. If that is the case, a short address attack is detected, because the highest $64-\textit{len}$ bytes will be used for complementing the short \textsf{\small address}. Eventually, we find 73 attacking transactions, which invoke 25 smart contracts. The detailed results are listed in Supplementary material J due to page limit. 

\noindent\textbf{Summary}: \textit{The function signatures recovered by \texttt{\footnotesize SigRec} can ease the detection of invalid actual arguments, in particular short address attacks.}

\subsection{Boosting the Performance of Existing Smart Contract Fuzzers} 
It is well-known that knowing parameter types is critical for fuzzing tools to generate high-quality inputs. Many existing fuzzing tools~\cite{contractfuzzer, sfuzz,ValentinICSE20,JingxuanCCS19} for discovering vulnerabilities in smart contracts assume the availability of function signatures. Unfortunately, such assumption does not always hold because existing tools recover function signatures with low accuracy (\S \ref{sec_compare}). Without function signatures, existing smart contract fuzzers may have to regard the list of parameters as a byte sequence and generate random byte sequences as input instead of applying specific mutation strategies for different parameter types. \texttt{\footnotesize SigRec} can recover function signatures for these fuzzing tools. 

To quantitatively evaluate how function signatures recovered by \texttt{\footnotesize SigRec} can benefit smart contract fuzzers, we develop a tool, named ContractFuzzer$^-$ which is the same with ContractFuzzer~\cite{contractfuzzer}, a state-of-the-art open-source smart contract fuzzer, except that ContractFuzzer$^-$ does not know function signatures and thus it produces random byte sequences as parameters. We compare ContractFuzzer which takes in the function signatures provided by \texttt{\footnotesize SigRec} and produces inputs by its default mutation strategies with ContractFuzzer$^-$ in analyzing 1,000 randomly selected smart contracts. Results show that with the function signatures recovered by \texttt{\footnotesize SigRec}, ContractFuzzer reveals 23\% more vulnerabilities and 25\% more vulnerable smart contracts than ContractFuzzer$^-$.

\noindent\textbf{Summary}: 
\textit{The function signatures recovered by \texttt{\footnotesize SigRec} help fuzzers adopt proper fuzzing strategies and find much more vulnerabilities}.

\subsection{Improving the Result of Reverse Engineering the Bytecode of Smart Contracts}
Erays~\cite{erays}, a reverse engineering tool for smart contracts, takes in the bytecode, and outputs register-based instructions which are more readable than EVM bytecode~\cite{erays}. Unfortunately, Erays recovers neither function signatures nor variable types, and its output contains lots of code produced by the compiler for accessing parameters, making it difficult to understand the program. 
We develop a tool named Erays$^+$ in 1,456 lines of Python code to improve the results of Erays by leveraging the function signatures recovered by \texttt{\footnotesize SigRec}. Specifically, Erays$^+$ adds the function signature for each public/external function, 
replacing meaningless variable names with meaningful parameter names if these variables are copied from parameters (e.g., Erays$^+$ replaces a variable \textit{x} with \emph{arg\rm{1}} indicating that \textit{x} the 1st parameter; Erays$^+$ also replaces a variable \textit{y} with \emph{num(arg\rm{1})} if \textit{y} is copied from the \textit{num} field of the 1st parameter), 
adding the type of the parameter to a variable if the parameter is assigned to the variable, and replacing the bulk of compiler-generated code for accessing parameters with simple assignment statements. Erays$^+$ now can leverage the function signatures taking in basics types, \textsf{\small static array}s, \textsf{\small dynamic array}s, \textsf{\small bytes}, and \textsf{\small string}s of Solidity, and we plan to enhance Erays$^+$ with the ability to handle \textsf{\small struct}s and \textsf{\small nested array}s of Solidity and Vyper types in our future work. Applying Erays$^+$ to the bytecode compiled from 53,166 unique open-source contracts, we find that Erays$^+$ improves the readability of the outputs of Erays in \textit{all} processed smart contracts. For each of them, the average numbers of added types, added parameter names, added \textit{num} names and removed code lines for accessing parameters are 5.5, 15, 3.4, and 15, respectively.

\noindent\textbf{Summary}: \textit{The function signatures provided by \texttt{\footnotesize SigRec} can be used for improving the readability of the results of reverse engineering tools for smart contracts}.

\section{Discussion}
\label{sec_discuss}
This section discusses the limitations of \texttt{\small SigRec} and possible solutions to be explored in future work. 
First, \texttt{\small SigRec} cannot recover the type of a parameter with the \textit{storage} modifier, because such parameter is a reference to a storage variable with any possible type and the address of the storage variable rather than its content is passed when passing such parameter~\cite{solidity}. The specialty makes the type of such parameter very challenging to be recognized, and we will extend our work to support such parameter in future work. 
Second, \texttt{\small SigRec} cannot recognize a function parameter if the clues from a function implementation are insufficient. For example, \texttt{\small SigRec} cannot distinguish a \textsf{\small bytes} from a \textsf{\small string}, if the smart contract does not access an individual byte of the \textsf{\small bytes} parameter. We will address this issue by exploiting the huge number of smart contracts, because one function signature may be found in many smart contracts with various function bodies that may provide sufficient clues. That is, one function signature may associate with many function bodies, and thus we may obtain sufficient clues from these function bodies.  
Third, the accuracy of \texttt{\small SigRec} to recover some parameter types (e.g., nested \textsf{\small array}) is not as high as the accuracy to recover other types. We will design more rules to infer these types with higher accuracy in future. 

Moreover, if a new parameter type or different accessing pattern is introduced, we will determine the rules for recognizing them through the steps described in \S \ref{rule_extract}. Please recall that there are five steps to derive rules and the first four have been automated so that it will not cost much time to derive new rules.
Similarly, malicious smart contracts may use obfuscation to hide their function signatures from being recognized by \texttt{\small SigRec}. A typical obfuscation technique is replacing the instruction sequence for accessing parameters which can be recognized by \texttt{\small SigRec} with a different instruction sequence with the same semantics which cannot be recognized by \texttt{\small SigRec}. We plan to propose general rules to resist obfuscation, where one rule can represent all possible instruction sequences with the same semantics.

\section{Related Work}
\label{sec_related}

The related studies about recovering function signatures in Smart Contracts have been presented in \S 1, so this section briefly introduces the related studies about reverse engineering of smart contracts and binaries.

\noindent\textbf{Reverse Engineering Smart Contracts.} Being a static analysis framework, Vandal decompiles EVM bytecode into IRs~\cite{vandal}. \ignore{Based on Vandal, Madmax detects gas-focused vulnerabilities~\cite{madmax}.} Porosity decompiles EVM bytecode into readable Solidity syntax contracts~\cite{porosity}. Erays transforms EVM bytecode into human readable expressions~\cite{erays}. EthIR translates EVM bytecode into a rule-based representation for further analysis~\cite{ethir}. Mythril decompiles EVM bytecode into IRs and performs symbolic execution on IRs~\cite{mythril}. 
GasReducer disassembles the bytecode into assembly code and detects 24 anti-patterns in the bytecode~\cite{chen2018towards}.
\cite{chen2019large} captures execution traces of smart contracts to evaluate the number of CFTs covered by traces that are not found by the selected tools.
GasChecker disassembles the bytecode into assembly code and performs symbolic execution on the assembly code for further analysis~\cite{chen2020gaschecker}.
However, none of them recover function signatures. As a necessary step to discover integer overflow bugs, Osiris applies heuristic rules that are similar to R11 and R13 to infer the types of integer variables~\cite{osiris}. However, Osiris can only recognize unsigned integers and signed integers. % only.

\noindent\textbf{Reverse Engineering Binaries.}
\label{sec_related_binary}
Reverse engineering techniques highly depend on the runtime architecture of the target programs. EVM bytecode is syntactically similar to Java bytecode. However, Java bytecode retains function signatures~\cite{decompile_java}, and thus does not need to recover function signatures. Next, we briefly introduce the differences between EVM and x86/x64, which demand a new technique to recover function signatures of smart contracts.\ignore{ make existing approaches for x86/x64 binaries ~\cite{caballero_ndss,lightweight,secondwrite,rewards,rewrite,zhang,esp,sekar_dsn,byteweight,rnn,fid,simple,tie,array,retypd,katz,xu,howard} useless in recovering function signatures of smart contracts.}

First, EVM has 130+ instructions including many blockchain-specific instructions, which have different semantics with x86/x64 instructions~\cite{yellow}. 
Second, EVM is a stack-based virtual machine without registers~\cite{yellow}. 
Third, the following concepts are unique to EVM and x86/x64 does not have such design. (1) The call data stores parameters, and only two EVM instructions \textsf{\footnotesize CALLDATALOAD} and \textsf{\footnotesize CALLDATACOPY} can read parameters from the call data~\cite{yellow}. 
(2) EVM maintains a special memory space named \emph{memory} to store some types of parameters, parameters to internal functions and execution results of smart contracts~\cite{yellow}. Moreover, every smart contract has a permanent storage space named \emph{storage} to record $<$key, value$>$ pairs~\cite{yellow}. 
(3) An intra-contract function invocation is compiled into a \textsf{\footnotesize JUMP} instruction, while an inter-contract function invocation is compiled into a \textsf{\footnotesize CALL}/\textsf{\footnotesize CALLCODE}/\textsf{\footnotesize DELEGATECALL}/\textsf{\footnotesize STATICCALL}\ignore{  invocation an invocation from a function to another function in the same contract is compiled into a \textsf{\footnotesize JUMP} instruction, while an invocation from a function of a smart contract to a function in another smart contract is often compiled into a \textsf{\footnotesize CALL} instruction}~\cite{yellow}. 

Related studies on x86/x64 binaries include recovering parameter types~\cite{caballero_ndss,lightweight}, recognizing parameters without recovering parameter types~\cite{secondwrite,rewrite,zhang,esp}, identifying function boundary~\cite{sekar_dsn,byteweight,rnn,fid}, and inferring variable types~\cite{rewards,simple,tie,array,retypd,katz,xu,howard}. Detailed description is in given Supplementary material K.

\section{Conclusion}
\label{sec_conclusion}
We propose and develop  \texttt{\footnotesize SigRec}, a novel approach to automatically recover function signatures in both Solidity smart contracts and Vyper smart contracts without the need of function signature databases. The extensive experimental results show that \texttt{\footnotesize SigRec} has very high accuracy across different compilers and various compiler versions. Experiments also show that \texttt{\footnotesize SigRec} is efficient and much more accurate than the state-of-the-art tools. 
Moreover, we demonstrate that the recovered function signatures are very useful in attack detection, fuzzing and reverse engineering of EVM bytecode. 

\section*{Acknowledgement}
This work was partly supported by National Key R\&D Program of China (2018YFB0804100), National Natural Science Foundation of China (61872057), HK RGC Project (No. 152193/19E), National Science Foundation under Grant No. 1951729, 1953813, and 1953893.
\bibliographystyle{IEEEtran}
\bibliography{ref}

\section*{Supplementary material}
\label{sec_appendix}
\subsection*{A. Function Signatures Recovered by Different Methods}
\begin{figure}[ht]
	\centering
	\vspace{-2ex}
	\includegraphics[width=0.45\textwidth]{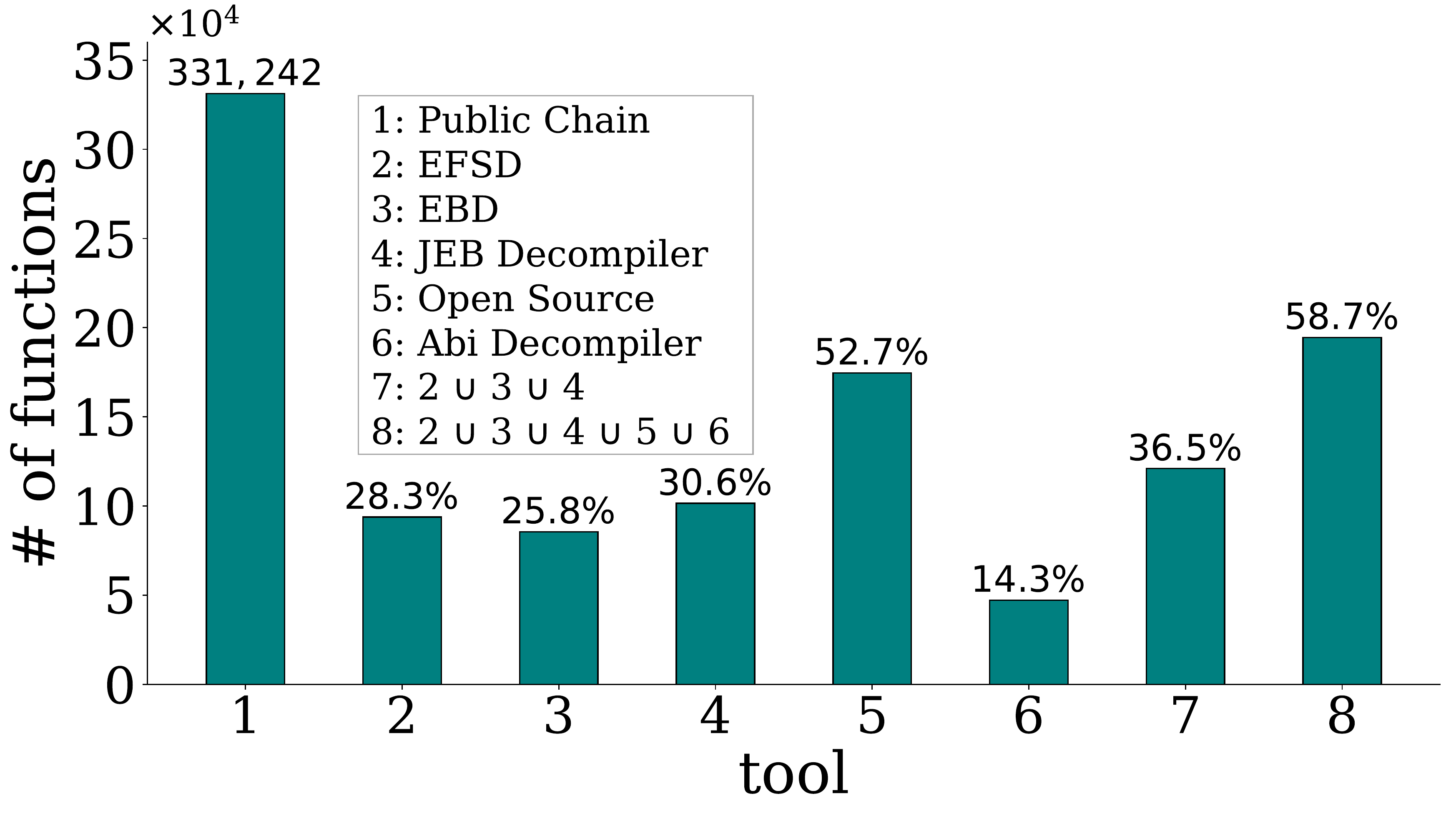}
	%\vspace{-2ex}
	\caption{Function signatures recovered by different methods.}
	%\vspace{-1ex}
	\label{fig_opentools}
\end{figure}

In Fig. \ref{fig_opentools}, the 1st bar gives the number (i.e., 383,522) of all unique function signatures in Ethereum public chain. The 2nd -- 6th bars present the numbers of function signatures that can be found from EFSD, EBD, JEB, the open-source smart contracts collected from Etherscan, and Abi Decompiler which is based on brute-force search for function signature recovery, respectively. The 7th bar presents the number of function signatures that can be found in databases, including EFSD, EBD and JEB. The 8th bar gives the number of all function signatures that can be found from EFSD, EBD, JEB, the open-source smart contracts in Etherscan, and Abi Decompiler. The ratio of the function signatures that can be found by each tool is presented above each bar.

EFSD records 24.6\% of the function signatures.
EBD's database contains 22.4\% of function signatures. The database of JEB covers 26.5\% of function signatures. Since function signatures can be easily extracted from open-source smart contracts, 55.0\% of function signatures can be obtained through this approach. Abi-decompiler synthesizes function signatures from a list of words and a list of types, and it can recover 12.3\% of the function signatures. Moreover, the union of them can recover 60.1\% of the function signatures, and hence the accuracy of \texttt{\footnotesize SigRec} is 38.6\% higher than the accuracy of the combination of all existing tools. %, which is 99.1\%.

\subsection*{B. Background}
\noindent\textbf{Account}. There are two kinds of accounts in Ethereum, namely external owned accounts (EOAs) and smart contracts~\cite{homestead,UnderstandingTOIT}. The major difference between them is that only the latter contains smart contract bytecode. A smart contract can be invoked by an EOA or a smart contract.

\noindent\textbf{Bytecode of smart contracts}. The bytecode of a smart contract has two parts: initialization bytecode and runtime bytecode~\cite{solidity,ChenINFOCOM18,dataether}. The former is only for contract deployment and thus cannot be invoked by others. Hence, we focus on the public/external functions in the latter. %Since it is easy to differentiate them, we can extract the runtime bytecode from the bytecode generated by a compiler.
The  \emph{gas} mechanism adopted by Ethereum defines the gas cost of each instruction\cite{yellow,chen2017under,chen2018towards,chen2020gaschecker,GasISPEC}. 

\noindent\textbf{Public/external function}. In a smart contract, only the functions declared as \emph{public} or \emph{external} can be invoked by others~\cite{homestead,chen2019tokenscope}.
The accessing patterns of some parameter types depend on whether the function is public or external 
% hao add
(\S 2 in manuscript~\cite{SigRec}). %\ref{sec_calldata}).
A smart contract can have a \emph{fallback} function, which is unnamed and will be called if the transaction does not specify the invoked function or the specified function does not exist in the invoked smart contract~\cite{solidity}.

\subsection*{C. Rules 5 -- 31}

\noindent\textbf{R5:} %
R5 is used to infer a one-dimensional dynamic \textsf{\small array}/\textsf{\small bytes}/\textsf{\small string} parameter in a public function. R5 will be used after R1 is fulfilled 
(Fig. 13 in manuscript~\cite{SigRec}).%\ref{fig_rule_tree}),
and therefore the \textsf{\footnotesize CALLDATACOPY} reads a dynamic \textsf{\small array}/\textsf{\small bytes}/\textsf{\small string}. Let $\textsf{\footnotesize CALLDATACOPY}(\textit{offset}_\textit{m}, x+36, \textit{len})$ denote that a \textsf{\footnotesize CALLDATACOPY} reads the 
data right after the \textit{num} field of a dynamic \textsf{\small array}/\textsf{\small bytes}/\textsf{\small string},
R5 adds one requirement: $\nexists\textsf{\footnotesize LT}\lhd \textsf{\footnotesize CALLDATACOPY}(\textit{offset}_\textit{m}, x+36, \textit{len})$, indicating that the \textsf{\footnotesize CALLDATACOPY} is not in a loop where \textsf{\footnotesize LT}s are loop guards. In this case, %and therefore
\textsf{\footnotesize CALLDATACOPY} reads a one-dimensional dynamic \textsf{\small array}/\textsf{\small bytes}/\textsf{\small string} in a public function.

\noindent\textbf{R6:} %
R6 is used to infer a one-dimensional static \textsf{\small array} in a public function. The $\textsf{\footnotesize CALLDATACOPY}(\textit{offset}_\textit{m}, \textit{offset}_\textit{c}, \textit{len})$ reads a static \textsf{\small array} in a public function, because $\textit{offset}_\textit{c}$ and $\textit{len}$ are constants, indicating that the location and the length of the \textsf{\small array} are known in compilation. If $\not\exists\textsf{\footnotesize LT}\lhd \textsf{\footnotesize CALLDATACOPY}(\textit{offset}_\textit{m}, \textit{offset}_\textit{c}$ holds, the \textsf{\small array} is one-dimensional, because the \textsf{\footnotesize CALLDATACOPY} is not inside a loop. Besides, the one-dimensional static \textsf{\small array} has $\textit{len}/32$ items because every array item is 32 bytes.

\noindent\textbf{R7:} 
R7 is used to infer a one-dimensional dynamic \textsf{\small array} in a public function. R7 will be used after R5 is fulfilled 
(Fig. 13 in manuscript~\cite{SigRec}), 
and thus the \textsf{\footnotesize CALLDATACOPY} reads a dynamic \textsf{\small array}/\textsf{\small bytes}/\textsf{\small string} in a public function. R7 has a requirement: $\textit{len}==32\times y$, where $y$ is the value of the \emph{num} field. This requirement indicates that the \textsf{\footnotesize CALLDATACOPY} reads an \textsf{\small array} since each array item is extended to 32 bytes.

\noindent\textbf{R8:} %
R8 is used to infer a \textsf{\small bytes}/\textsf{\small string} in a public function. It will  be used after R5 is fulfilled 
(Fig. 13 in manuscript~\cite{SigRec}), 
and thus the \textsf{\footnotesize CALLDATACOPY} reads a dynamic \textsf{\small array}/\textsf{\small bytes}/\textsf{\small string} in a public function. 
R8 has a requirement: $\textit{len}==32\times\lceil y/32\rceil$, where $y$ is the value of the \emph{num} field and ``$\lceil \rceil$'' denotes rounding up to the next integer. If it holds, the \textsf{\footnotesize CALLDATACOPY} reads a \textsf{\small bytes}/\textsf{\small string}. Such requirement is accordant with the layout of a \textsf{\small bytes}/\textsf{\small string}, which is extended so that its length is multiple of 32 bytes.

\noindent\textbf{R9:} 
R9 is used to infer an ($n+1$)-dimensional ($n>0$) static \textsf{\small array} in a public function. 
If its requirement $\textsf{\footnotesize LT}_n(i_n,\textit{num}_n)\lhd \textsf{\footnotesize LT}_{n-1}(i_{n-1},\textit{num}_{n-1})\lhd ... \lhd \textsf{\footnotesize LT}_1(i_1,\textit{num}_1)\lhd \textsf{\footnotesize CALLDATACOPY}(\textit{offset}_\textit{m}, \textit{offset}_\textit{c}, \textit{len})$, where $\textit{num}_n, ..., \textit{num}_1$ are constant numbers,  holds, the $\textsf{\footnotesize CALLDATACOPY}(\textit{offset}_\textit{m}, \textit{offset}_\textit{c}, \textit{len})$ reads from a multidimensional static \textsf{\small array}. 
The requirement indicates an $n$-layer nested loop to read an $(n+1)$-dimensional \textsf{\small array}, and hence $n$ \textsf{\footnotesize LT} instructions, which are $n$ loop guards, execute before the \textsf{\footnotesize CALLDATACOPY} (an example is shown in 
Listing 1 of manuscript~\cite{SigRec}. %\ref{code_array_copy}).
$\textit{num}_n, ..., \textit{num}_1$ are constant numbers because for a static \textsf{\small array}, the size of each dimension is known in compilation. Hence, the item numbers from the highest dimension to the lowest dimension are $\textit{num}_n, ..., \textit{num}_1, \textit{len}/32$, respectively.

\noindent\textbf{R10:} 
R10 is used to infer an ($n+1$)-dimensional ($n>0$) dynamic \textsf{\small array} in a public function. R10 will be applied after R1 is fulfilled 
% hao add
(Fig. 13 in manuscript~\cite{SigRec}), 
and thus the \textsf{\footnotesize CALLDATACOPY} reads a dynamic \textsf{\small array}/\textsf{\small bytes}/\textsf{\small string} in a public function. If its requirement $\textsf{\footnotesize LT}_n(i_n,y)\lhd \textsf{\footnotesize LT}_{n-1}(i_{n-1},\textit{num}_{n-1})\lhd ... \lhd \textsf{\footnotesize LT}_1(i_1,\textit{num}_1)\lhd \textsf{\footnotesize CALLDATACOPY}(\textit{offset}_\textit{m}, \textit{offset}_\textit{c}, \textit{len})$, where $y$ is the value of the \textit{num} field, holds, the \textsf{\footnotesize CALLDATACOPY} reads an ($n+1$)-dimensional dynamic \textsf{\small array} in a public function, because $n$ loop guards, belonging to a nested loop, execute before the \textsf{\footnotesize CALLDATACOPY} and the size of the highest dimension is given by the \textit{num} field. Hence, only the size of the highest dimension is dynamic and the numbers of items in the lower $n$ dimensions are $\textit{num}_{n-1}, ..., \textit{num}_1, \textit{len}/32$, which are constant values. 

\noindent\textbf{R11:} 
For an \textsf{\footnotesize AND}($op_1$, $op_2$), if one operand is a \textsf{\small uint256} parameter and the other one is a constant number whose higher-order side is $x$ zero-bytes ($0<x<32$), then the \textsf{\small uint256} parameter is refined to a \textsf{\small uint}$\langle 256-8\times x\rangle$. The rationale is that a \textsf{\small uint}$\langle M\rangle, 8\le M<256, M\%8==0$ will be extended on the higher-order side, and an \textsf{\footnotesize AND} is used for masking the value after extension before accessing its value.

\noindent\textbf{R12:} 
For an \textsf{\footnotesize AND}($op_1$, $op_2$), if one operand is a \textsf{\small uint256} parameter and the other one is a constant number whose lower-order side is $x$ zero-bytes ($0<x<32$), then the \textsf{\small uint256} parameter is adjusted to a \textsf{\small bytes}$\langle32-x\rangle$. The rational is that a \textsf{\small bytes}$\langle M\rangle$, $0<M<32$ will be extended on the lower-order side, and an \textsf{\footnotesize AND} is used for masking the extended value.

\noindent\textbf{R13:} 
For a \textsf{\footnotesize SIGEXTEND}($op$, $x$), if $op$ is a \textsf{\small uint256} parameter and $x$ is a constant number $0\le x<31$, then the type of $op$ is refined to \textsf{\small int}$\langle (x+1)\times 8 \rangle$. The rational is that \textsf{\footnotesize SIGNEXTEND} is used for sign extension of a signed integer and $x$ is the length (in bytes) of the signed integer minus 1~\cite{yellow}. %the constant parameter

\noindent\textbf{R14:} 
If two \textsf{\footnotesize ISZERO} instructions satisfy $x=\textsf{\footnotesize ISZERO}(op)$ and $y=\textsf{\footnotesize ISZERO}(x)$ and $op$ is a \textsf{\small uint256} parameter, the type of $op$ is refined to \textsf{\small bool}. The rational is that a \textsf{\small bool} value is extended to 32 bytes and two consecutive \textsf{\footnotesize ISZERO} instructions are used for masking.

\noindent\textbf{R15:} 
If a \textsf{\small uint256} parameter is an operand of an \textsf{\footnotesize SDIV}/\textsf{\footnotesize SMOD}/\textsf{\footnotesize SLT}/\textsf{\footnotesize SGT}, its type should be adjusted to \textsf{\small int256}, because such instructions take in signed integers~\cite{yellow}.

\noindent\textbf{R16:} 
If a \textsf{\small uint160} parameter is not involved in any mathematics instructions, its type is adjusted to \textsf{\small address}, because an \textsf{\small address} has 20 bytes and it should not be involved in mathematics computations.

\noindent\textbf{R17:} 
For a \textsf{\footnotesize BYTE}($op$)/\textsf{\footnotesize MSTORE8}($op$) which reads/writes 1 byte from/to a parameter $op$~\cite{yellow}, if the type of $op$ is a \textsf{\small string} or a \textsf{\small bytes}, we refine it into a \textsf{\small bytes} because a \textsf{\small bytes} supports byte reading and writing whereas a \textsf{\small string} does not support such operations~\cite{type}.

\noindent\textbf{R18:} 
For a \textsf{\footnotesize BYTE}($op$) which reads a byte from a parameter $op$, if the type of $op$ is \textsf{\small uint256}, we refine it into \textsf{\small bytes32}. The reason is that a \textsf{\small bytes32} supports byte reading by executing a \textsf{\footnotesize BYTE}~\cite{yellow}, while an \textsf{\footnotesize AND} is used for extracting a byte from a \textsf{\small uint256}.

\noindent\textbf{R19:} R19 is used to distinguish the two types, \textsf{\small struct} and nested \textsf{\small array} from other Solidity types. %, and R19 should be applied after R20 isn't satisfied. 
R19 depends on one requirement $v$1, which is defined as $\textit{offset}_1$ = \textsf{\footnotesize CALLDATALOAD}$_1$($\textit{loc}_1$) $\lhd$ $\textit{offset}_2$ = \textsf{\footnotesize CALLDATALOAD}$_2$($\textit{loc}_2$) $\lhd...\lhd$ $\textit{offset}_n$ = \textsf{\footnotesize CALLDATALOAD}$_n$($\textit{loc}_n$), where $n>2$, and 
($\textit{loc}_1$ is a constant number) $\wedge \, (exp(\textit{loc}_2)\diamond (\textit{offset}_1)) \, \wedge ... \wedge \, (exp(\textit{loc}_n)\diamond (\textit{offset}_{n-1}))$. 
$v$1 means that the read location of $\textit{offset}_1$ is a constant value, and the read location of $\textit{offset}_2$ is computed by $\textit{offset}_1$, and so on. When $x>2$, the value of $\textit{offset}_x$ is computed by nested \textsf{\footnotesize CALLDATALOAD}s for $\textit{offset}_1$. For example, the value of $\textit{offset}_3$ is \textsf{\footnotesize CALLDATALOAD}(\textsf{\footnotesize CALLDATALOAD}($\textit{offset}_1$+)+). 
That means, for an item in the parameter, which value is \textsf{\footnotesize CALLDATALOAD}($\textit{offset}_1$+), it is the \textit{offset} field for a value \textsf{\footnotesize CALLDATALOAD}(\textsf{\footnotesize CALLDATALOAD}($\textit{offset}_1$+)+).
So the explanation for $v$1 is that there is a read location which is computed from the \textit{offset} field of the item inside the parameter. 
So if $v$1 is satisfied, the parameter contains an item with dynamic size, and please recall that only the nested \textsf{\small array} and \textsf{\small struct} can satisfy $v$1. 

\noindent\textbf{R20:} R20 is used to distinguish Vyper bytecode from Solidity bytecode.
R20 depends on one requirement $v$1, which is defined as $x$ = \textsf{\footnotesize CALLDATALOAD}(0) $\wedge$ \textsf{\footnotesize MSTORE}($\textit{offset}_m, x$). If $v$1 is satisfied, this bytecode is a Vyper bytecode.  
We observe that at the beginning of the Vyper bytecode, Vyper will apply \textsf{\footnotesize CALLDATALOAD} and \textsf{\footnotesize MSTORE} to store the first 4 bytes of call data indicating the function id to the memory.
Differently, Solidity will store the 4 bytes function id to the stack from the call data by only a \textsf{\footnotesize CALLDATALOAD}.

\noindent\textbf{R21:} R21 is used to infer the \textsf{\small struct} type in Solidity, which is applied after R19 is satisfied. 
R21 depends on one requirement $v$1, which is defined as $\textit{offset}_1$=\textsf{\footnotesize CALLDATALOAD}(\textit{loc})$\lhd \nexists$ LT(\textit{i}, \textsf{\footnotesize CALLDATALOAD}($\textit{offset}_1+0x4$)) $\lhd$ \textsf{\footnotesize CALLDATALOAD}($\textit{offset}_1 +0x4+0x32\times i$), where $\textit{loc}$ is a constant value.
$v$1 means that the read location of $\textit{offset}_1$ is a constant value, there is a \textsf{\footnotesize CALLDATALOAD} that its read location is computed by \textit{i} and $\textit{offset}_1$, while the $\textit{offset}_1$ is the \textit{offset} field of the parameter, and there is no comparison with \textit{i} and \textsf{\footnotesize CALLDATALOAD}($\textit{offset}_i+0x4$).
The explanation for $v$1 is that there is no \textit{num} field for this parameter, and \textit{num} field is used for bound check for accessing the item in parameter.
So if $v$1 is satisfied, the parameter is a \textsf{\small struct} type parameter, please recall that there is no \textit{num} field for the \textsf{\small struct} type.

\noindent\textbf{R22:} R22 is used to infer the nested \textsf{\small array} type in Solidity, which is applied after R19 is satisfied. 
R22 depends on only one requirement $v$1.
$v$1 is defined as \textsf{\footnotesize LT}($i_n, \textit{num}_n)$$\lhd$ \textsf{\footnotesize LT}($i_{n-1}$, $\textit{num}_{n-1}$)$\lhd$ ... $\lhd$ \textsf{\footnotesize LT}($i_1$, $\textit{num}_1$)$\lhd$ \textsf{\footnotesize CALLDATALOAD}$_{n+1}$(\textit{loc}), while (exp(\textit{loc})$\diamond$ (\textit{offset}$_n$)) and 
$\textit{num}_x$ can be constant or variable number, and at least one of $\textit{num}_1$ ... $\textit{num}_{n-1}$ should be variable.
$v$1 means that there are \textit{n} bound checks before read from a position computed from $\textit{offset}_n$.
So if $v$1 is satisfied, this parameter is a \textit{n}-dimension nested \textsf{\small array}, please recall that for each dimension, there is a bound check. If $\textit{num}_x$ is a constant number, the corresponding dimension has $\textit{num}_x$ items. On the contrary, if $\textit{num}_x$ is a variable, the corresponding dimension is dynamic.

\noindent\textbf{R23:} R23 is used to infer the fixed-size \textsf{\small bytes} and \textsf{\small string} in Vyper. R23 will be applied after R20 is satisfied. 
R23 depends on one requirement $v$1, which is defined as $x$ = \textsf{\footnotesize CALLDATALOAD}($\textit{loc}$) $\wedge$ \textsf{\footnotesize CALLDATACOPY}($\textit{offset}_m, \textit{offset}_c, \textit{len}$), while \textit{len} is a constant number and $\textit{offset}_c$ is computed by \textit{x}. If $v$1 is satisfied, this parameter is a fixed-size \textsf{\small bytes} or \textsf{\small string}.
The explanation for $v$1 is that Vyper uses a \textsf{\footnotesize CALLDATACOPY} to copy the item number field and the whole value from call data to memory for fixed-size \textsf{\small bytes} and \textsf{\small string} type, and the size of fixed-size \textsf{\small bytes} and \textsf{\small string} is known before compilation as \textit{maxLen}. So the copy length \textit{len} is a constant value and computed by $32 + \textit{maxLen}$ and the copy position in call data is computed by \textit{offset} of this value as $x$.

\noindent\textbf{R24:} R24 is used to infer the fixed-size \textsf{\small list} type in Vyper. The requirements of R24 are the same as those of R3, but R24 should be applied after R20 is satisfied.

\noindent\textbf{R25:} R25 is used to infer the \textsf{\small uint256} type in Vyper, %and R25 is similar to R4 in Solidity. R25 can 
which is applied after R20 is satisfied.
$x$ is regarded as a \textsf{\small uint256} in Vyper, if R23 and R24 are not fulfilled. That means that without sufficient hints we just
know that the length of $x$ is 32 bytes and thus we currently consider a 32-bytes parameter as a \textsf{\small uint256}. We will refine it to a specific
type after applying other rules to get more hints. 

\noindent\textbf{R26:} R26 is used to distinguish fixed-size \textsf{\small bytes} from fixed-size \textsf{\small string}. If a parameter type is fixed-size \textsf{\small bytes} or fixed-size \textsf{\small string} and the parameter is an operand of a \textsf{\footnotesize BYTE} instruction or a \textsf{\footnotesize MSTORE8} instruction, we adjust the parameter type to fixed-size \textsf{\small bytes} because a fixed-size \textsf{\small bytes} supports to access its individual byte whereas a fixed-size \textsf{\small string} does not support such operation.
R26 is applied after R23 is satisfied.

\noindent\textbf{R27:} R27 is used to infer the \textsf{\small address} type in Vyper, which will be applied after R25 is satisfied. For a \textsf{\footnotesize LT}($op_1, op_2$), if $op_1$ is a \textsf{\small uint256} parameter and the value of $op_2$ is $2^{160}$, then the uint256 parameter is refined to an \textsf{\small address} type parameter, due to the existence of the bound check for the \textsf{\small address} type. 

\noindent\textbf{R28:} R28 is used to infer the \textsf{\small int128} type in Vyper. R28 should be applied after R25 is satisfied. R28 depends on two requirements, $v$1 and $v$2. The parameter type is int128 when both of the two requirements are satisfied. $v$1 requires a \textsf{\footnotesize SLT}($op_1, op_2$) instruction where $op_1$ is a \textsf{\small uint256} parameter and $op_2$ is $2^{127}-1$. $v$2 requires a \textsf{\footnotesize SGT}($op_1, op_3$) instruction where $op_1$ is a \textsf{\small uint256} parameter and $op_3$ is $-2^{127}$. The two requirements indicate the bound checks before accessing an \textsf{\small int128} parameter. 

\noindent\textbf{R29:} R29 is used to infer the decimal type in Vyper. R29 should be applied after R25 is satisfied. R29 depends on two requirements, $v$1 and $v$2. The parameter type is decimal when both of the two requirements are satisfied. $v$1 requires a \textsf{\footnotesize SLT}($op_1, op_2$) instruction where $op_1$ is a \textsf{\small uint256} parameter and $op_2$ is the 10 decimal value of $2^{127}-1$. $v$2 requires a \textsf{\footnotesize SGT}($op_1, op_3$) instruction where $op_1$ is a \textsf{\small uint256} parameter and $op_3$ is the 10 decimal value of $-2^{127}$. The two requirements indicate the bound checks before accessing a decimal parameter. It is worth noting that the precision of 10 decimal value of $2^{127}-1$ and $-2^{127}$ is different with it in \textsf{\small int128} type.

\noindent\textbf{R30:} R30 is used to infer the \textsf{\small bool} type in Vyper, which will be applied after R25 is satisfied. For a \textsf{\footnotesize LT}($op_1, op_2$), if $op_1$ is a \textsf{\small uint256} parameter and the value of $op_2$ is 2, then the \textsf{\small uint256} type is refined to a \textsf{\small bool} type, due to the existence of the bound check before accessing the \textsf{\small bool} parameter.

\noindent\textbf{R31:} R31 is used to infer the \textsf{\small bytes32} type in Vyper. The requirements of R31 is the same as those of R18, but R31 should be applied after R25 is satisfied.

\subsection*{D. Examples of Using Rules}
\noindent \textbf{1. An Example of Using R2:}
\lstset{
  xleftmargin=10pt
}
\vspace{1ex}
%\vspace{-1ex}
\begin{lstlisting}[caption={The instructions to read an item from a dynamic \textsf{array}}, label={code_example_r2}]
offset = CALLDATALOAD(0x4)
num = CALLDATALOAD(offset + 0x4) //num = 2
LT //i2 < 2, or else abort
LT //i1 < 3, or else abort
loc =  offset + 4 + 32 + 32 * 3
x = CALLDATALOAD(loc) //x[1][0]
\end{lstlisting}
%\vspace{-2ex}

Given an external function with a two-dimensional dynamic \textsf{\small array} parameter \textit{x}[3][], a transaction for invoking it contains the actual argument \textit{x}[3][2].% to invoke this function.
Listing \ref{code_example_r2} shows the instructions to read \textit{x}[1][0]. For the ease of presentation, we omit irrelevant instructions and merge some instructions into one statement.
Line 1 and Line 2 read the values of the \textit{offset} field and the \textit{num} field, respectively. Since the size of the highest dimension is two, the value of the \textit{num} field is two. Line 3 and Line 4 are two bound checks for two dimensions. The location of \textit{x}[1][0] is computed in Line 5, and we explain how the location is computed in more detail. Since \textsf{\small array} items are located right after the \textit{num} field, the smart contract first locates the \textit{num} field, which is pointed by the \textit{offset} field (i.e., $\emph{offset}+4$, where 4 is the length of the function id). 
\textit{x}[0][0] is at $\emph{offset}+4+32$, because the length of the \textit{num} field is 32 bytes. Since \textit{x}[1][0] is the first item of the second \textit{x}[3], \textit{x}[1][0] is located $32\times 3$ bytes after \textit{x}[0][0].
From the instructions, we can see that the requirements of R2 are held, and therefore \texttt{\footnotesize SigRec} infers that the parameter is a dynamic \textsf{\small array}. 
More specifically, the symbolic expression of \textit{loc} is computed by adding the value of the \textit{offset} field with a number which contains the multiplication of 32 (Line 5). 
Besides, the two \textsf{\footnotesize LT}s are passed before reading the \emph{num} field (Line 6). From the two \textsf{\footnotesize LT}s, we know that the \textsf{\small array} has two dimensions. From Line 4, we know that the size of its lowest dimension is three.

\vspace{1ex}
\noindent \textbf{2. An Example of Using R3:}
\vspace{1ex}
%We use an example to explain this rule. 

Given an external function with a two-dimensional static \textsf{\small array} parameter \textit{x}[3][2], Listing  \ref{code_example_r3} lists the instructions to read \textit{x}[1][0]. 
Lines 1 and 2 are bound checks. Line 3 computes the location to be read. More precisely, \textit{x}[0][0] follows the function id and hence it is located at 0x4. 
Since \textit{x}[1][0] is the first item of the second \textit{x}[3], \textit{x}[1][0] is located $32\times 3$ bytes after \textit{x}[0][0]. 
Line 4 reads \textit{x}[1][0]. From the instructions, we learn that the requirements of R3 are held, and therefore the parameter is a static \textsf{\small array}. 
More precisely, the read location is not affected by the \textit{offset} field (Line 3). Besides, two \textsf{\footnotesize LT}s are passed before reading the \textsf{\small array} item (Line 4). Moreover, the two \textsf{\footnotesize LT}s suggest that the \textsf{\small array} has two dimensions, and the sizes of its highest dimension and lowest dimension are two and three, respectively.
%\vspace{-1ex}
\begin{lstlisting}[caption={The instructions to read an item from a static array}, label={code_example_r3}]
LT //i2 < 2, or else abort
LT //i1 < 3, or else abort
loc =  4 + 32 * 3
x = CALLDATALOAD(loc) //x[1][0]
\end{lstlisting}

\subsection*{E. Block \& Function Recognition}
To recognize blocks and public/external functions, \texttt{\footnotesize SigRec} %first disassembles EVM bytecode and then 
partitions the disassembled code into blocks, and then extracts function ids and identifies the first block of every public/external function by parsing the dispatch routine which directs contract execution to the invoked function. 

\noindent \textbf{Block Recognition.} 
A block starts with the first instruction of the bytecode or the instruction right after the last instruction of a block. Besides, every \textsf{\footnotesize JUMPDEST} instruction begins a block since every jump instruction will go to a \textsf{\footnotesize JUMPDEST} instruction~\cite{yellow,chen2019large}. The last instruction of a block should be a \textsf{\footnotesize JUMP}/\textsf{\footnotesize JUMPI}/\textsf{\footnotesize STOP}/\textsf{\footnotesize RETURN}/\textsf{\footnotesize REVERT}/\textsf{\footnotesize INVALID}. \textsf{\footnotesize JUMP} and \textsf{\footnotesize JUMPI} are unconditional jump and conditional jump, respectively~\cite{yellow}. \textsf{\footnotesize STOP}, \textsf{\footnotesize RETURN}, \textsf{\footnotesize REVERT} and \textsf{\footnotesize INVALID} halt the execution of the smart contract~\cite{yellow}.

\noindent \textbf{Function Recognition.} \texttt{\footnotesize SigRec} extracts function ids and identifies the first block of every public/external function by exploiting the fact that the invoked smart contract extracts the function id from the call data and jumps to the first block of the invoked function.
Solidity and Vyper generate different bytecode for such function dispatch, so \texttt{\footnotesize SigRec} handles the two compilers differently.

\noindent \textit{Solidity.} \texttt{\footnotesize SigRec}
locates a function id and the first block if the first 4 bytes of the call data are compared with a 4-bytes constant number and a \textsf{\footnotesize JUMPI} instruction follows the comparison.
Note that the 4-bytes constant number is the function id and the jump target of \textsf{\footnotesize JUMPI} is the first block. 
Listing \ref{code_func_recognition} presents an example. Line 1 reads the first 32-bytes from the call data. Line 3 extracts the 4-bytes function id. Line 5 pushes a 4-bytes constant number 0x16d93ade on the stack. Line 6 compares them. Line 7 pushes a constant number 0x55 on the stack. Line 8 jumps to 0x55 if the first 4-bytes of the call data is equal to 0x16d93ade. Hence, the function id is 0x16d93ade and the first block of the corresponding function locates at 0x55 from the beginning of the bytecode.

\begin{lstlisting}[caption={A function with id 0x16d93ade}, label={code_func_recognition}]
CALLDATALOAD //read the 1st 32 bytes from the call data.
......
DIV //extract 4-byte function id
.....
PUSH4 0x16d93ade //push the function id on the stack.
EQ //whether they are equal?
PUSH1 0x55 //push the location of 1st block on the stack
JUMPI //jump to 1st block if function 0x16d93ade is called
\end{lstlisting}

\noindent \textit{Vyper.} \texttt{\footnotesize SigRec}
locates a function id and the first block if the first 4 bytes of the call data are compared with a 4-byte constant number and a \textsf{\footnotesize JUMPI} instruction follows the comparison.
Note that the 4-bytes constant number is the function id and the first block follows the \textsf{\footnotesize JUMPI} instruction, which is different from Solidity.
Listing \ref{vyper_code_func_recognition} presents an example. Line 1 reads the first 32-bytes from the call data to stack. Line 3 store the first 4 bytes of the call data to memory. Line 5 pushes a 4-bytes constant number 0xd178231c on the stack. Line 7 read the first 4 bytes of the call data from memory to stack. Line 8 compares them. Line 11 jumps to 0x043c if the first 4-bytes of the call data is unequal to 0xd178231c. Hence, the function id is 0xd178231c and the first block of the corresponding function locates follows the \textsf{\footnotesize JUMPI} instruction in Line 11.

\begin{lstlisting}[caption={A function with id 0xd178231c}, label={vyper_code_func_recognition}]
CALLDATALOAD//read the 1st 32 bytes from the call data
......
MSTORE //store the 1st 4 bytes of the call data to memory
......
PUSH4 0xd178231c //push the function id on the stack
PUSH1 0x00
MLOAD //read the 1st 4 bytes of the call data from memory
EQ //whether they are equal?
ISZERO
PUSH2 0x043c
JUMPI //jump to the next comparison if unequal
\end{lstlisting}.

\subsection*{F. Reasons for using TASE instead of CSE and other methods}
We propose TASE instead of using other methods (e.g.,  abstract interpretation~\cite{divine}, value set analysis~\cite{secondwrite,rewrite,byteweight}, data dependence analysis~\cite{zhang}, taint analysis~\cite{rewards}) to recover function signatures due to three reasons.  First, symbolic execution is more suitable than other methods for our task, because it can identify how a variable is computed from the parameters through symbolic expressions and such information is required to apply our rules.

Second, conventional symbolic execution (CSE) is much slower than TASE and suffers from path explosion, and thus is not proper for recovering the ever-growing number of function signatures. More precisely, there are four differences between TASE and CSE. (1) TASE takes into account the rules when symbolically executing type-related instructions (e.g.,  \textsf{\footnotesize CALLDATALOAD}) to infer parameter types. These rules are proposed by this work and specific to EVM bytecode, and thus no CSE tools can achieve the same purpose.
(2) TASE does not need to conduct the time-consuming checking of the branch feasibility. Instead, it explores both branches because it aims to recover parameter types and thus does not need to care about the program logic of smart contracts. By contrast, CSE usually spends lots of time in checking branch feasibility.
(3) For each public/external function, TASE executes all its blocks only once starting from the first block. That is, TASE does not unfold loops, because executing each block once is enough for inferring parameter types. Therefore, there is no path explosion in TASE. In contrast, CSE usually attempts to explore program paths, thus suffering from severe path explosion.
(4) TASE can cover all blocks except dead code since it explores all branches and does not get stuck into loops. In contrast, CSE may leave some reachable blocks unexplored due to the difficulty of checking branch feasibility and path explosion.

Third, we employ TASE instead of taint analysis because of two reasons. (1) some rules need to check whether a parameter is involved in a specific instruction rather than whether the variables are affected by any parameters. For instance, R11 requires that an operand is a \textsf{\small uint256} parameter, and checks whether the parameter is involved in an \textsf{\footnotesize AND} instruction. Symbolic execution knows how an operand is computed from parameters (e.g., directly copy or computed by arithmetic instructions) whereas taint analysis just knows whether an operand is affected by parameters (i.e., tainted or not). (2) some rules need symbolic expressions. For example, R2 checks whether the symbolic expression of the location contains the summation of the \textit{offset} value and the multiplication of 32. However, taint analysis does not collect symbolic expressions.

\subsection*{G. Errors Made by Gigahorse and Eveem}
\noindent\textbf{1. Dataset 1:}
\vspace{1ex}

Table \ref{tab_type_error_dataset1} presents the number of function signatures from dataset 1 (i.e., closed-source smart contracts) whose parameter types are incorrectly recovered by Gigahorse and Eveem. For each such function signature, this table also presents the correct type and the recovered type. If one recovered function signature has several incorrect parameter types, we count it repeatedly. 
\begin{table}[ht!]
	\centering
	%\vspace{-3ex}
	\caption{Errors of incorrect parameter types from dataset 1 recovered by Gigahorse and Eveem}
	%\vspace{-2ex}
	\includegraphics[width=0.4 \textwidth]{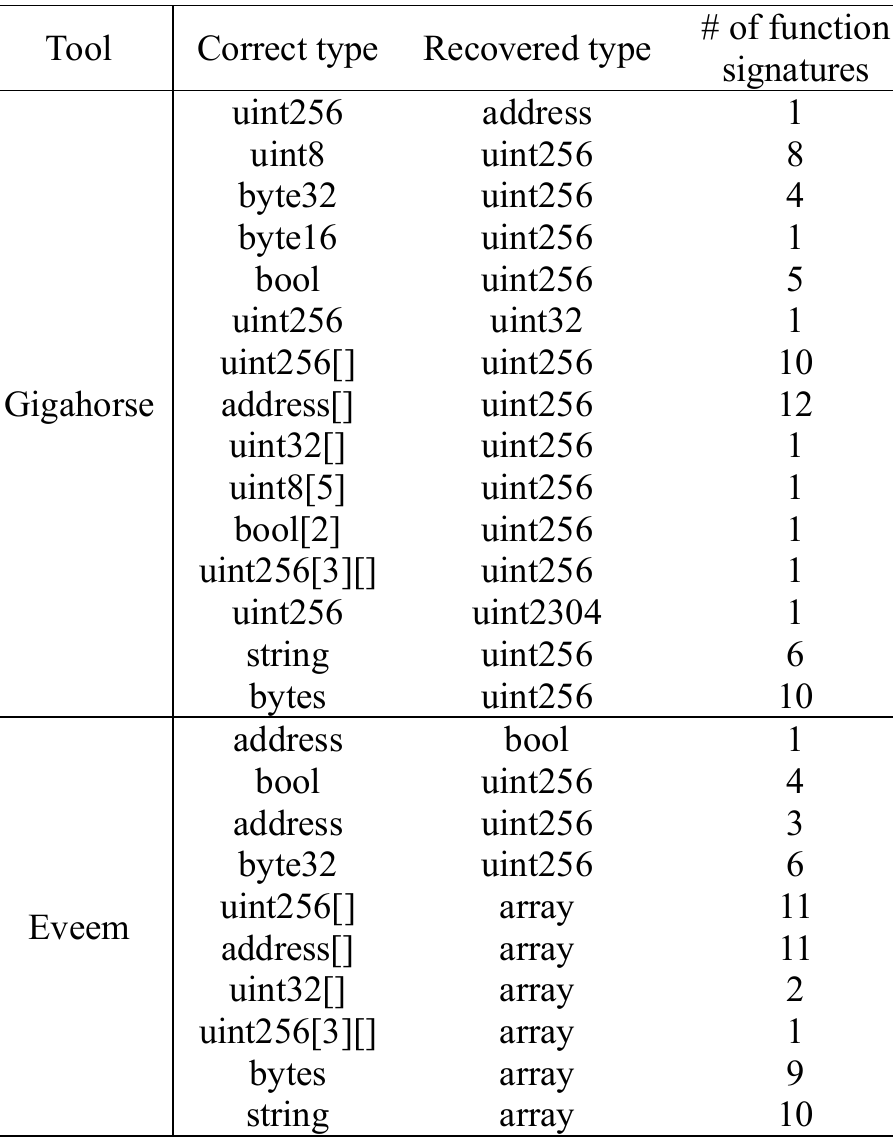}
	%\vspace{-2ex}
	\label{tab_type_error_dataset1}
\end{table}

\begin{table*}[ht!]
	\centering
	%\vspace{-3ex}
	\caption{Errors of incorrect parameter numbers from dataset 1 found by Gigahorse and Eveem}
	%\vspace{-2ex}
	\includegraphics[width=0.65\textwidth]{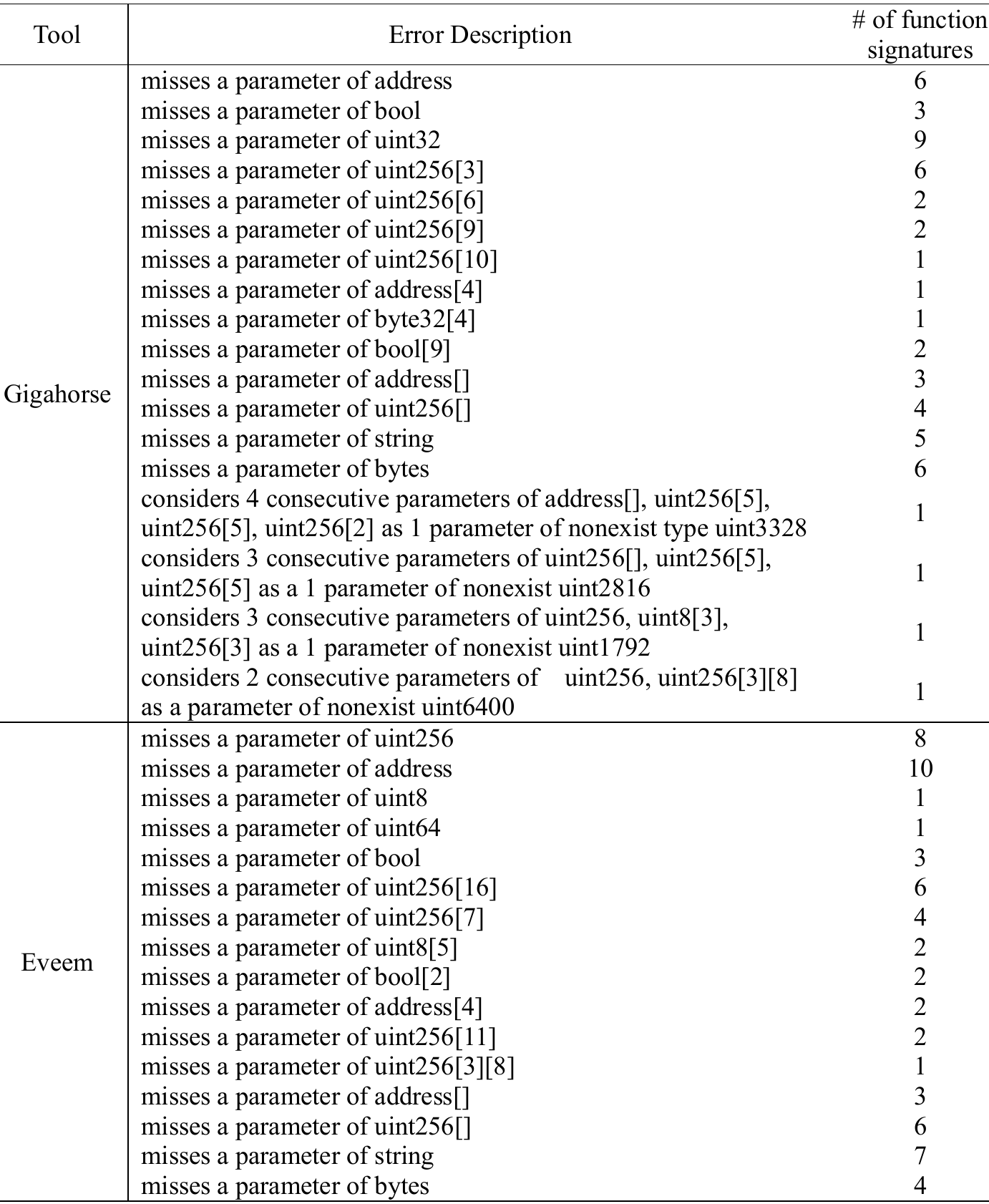}
	%\vspace{-2ex}
	\label{tab_number_error_dataset1}
\end{table*}

Table \ref{tab_number_error_dataset1} lists the number of function signatures from dataset 1 whose number of parameters are incorrectly determined by Gigahorse and Eveem. The correct parameter numbers and types are determined by manually investigating the EVM instructions for accessing parameters. This table also describes why the number of parameters is incorrect. If a recovered function signature misses or adds several parameters, we count it repeatedly.

\vspace{1ex}
\noindent\textbf{2. Dataset 2:}
\vspace{1ex}

Table \ref{tab_error_dataset2} lists the errors made by Eveem in recovering synthesized function signatures. This table also presents the number of function signatures incorrectly recovered by Eveem for each kind of error. We count the number repeatedly, if one incorrectly recovered function signature contains several incorrect recovered parameter types, misses or adds several parameters.

\vspace{1ex}
\noindent\textbf{3. Dataset 3:}
\vspace{1ex}

Table \ref{tab_type_error_dataset3} presents the number of function signatures from dataset 3 (i.e., open-source smart contracts) whose parameter types are incorrectly recovered by Gigahorse and Eveem. For each such function signature, this table also lists the correct type and the recovered type. If one recovered function signature has several incorrect parameter types, we count it repeatedly. 

\begin{table}[ht!]
	\centering
	%\vspace{-3ex}
	\caption{Errors made by Eveem when processing dataset 2}
	%\vspace{-2ex}
	\includegraphics[width=0.49\textwidth]{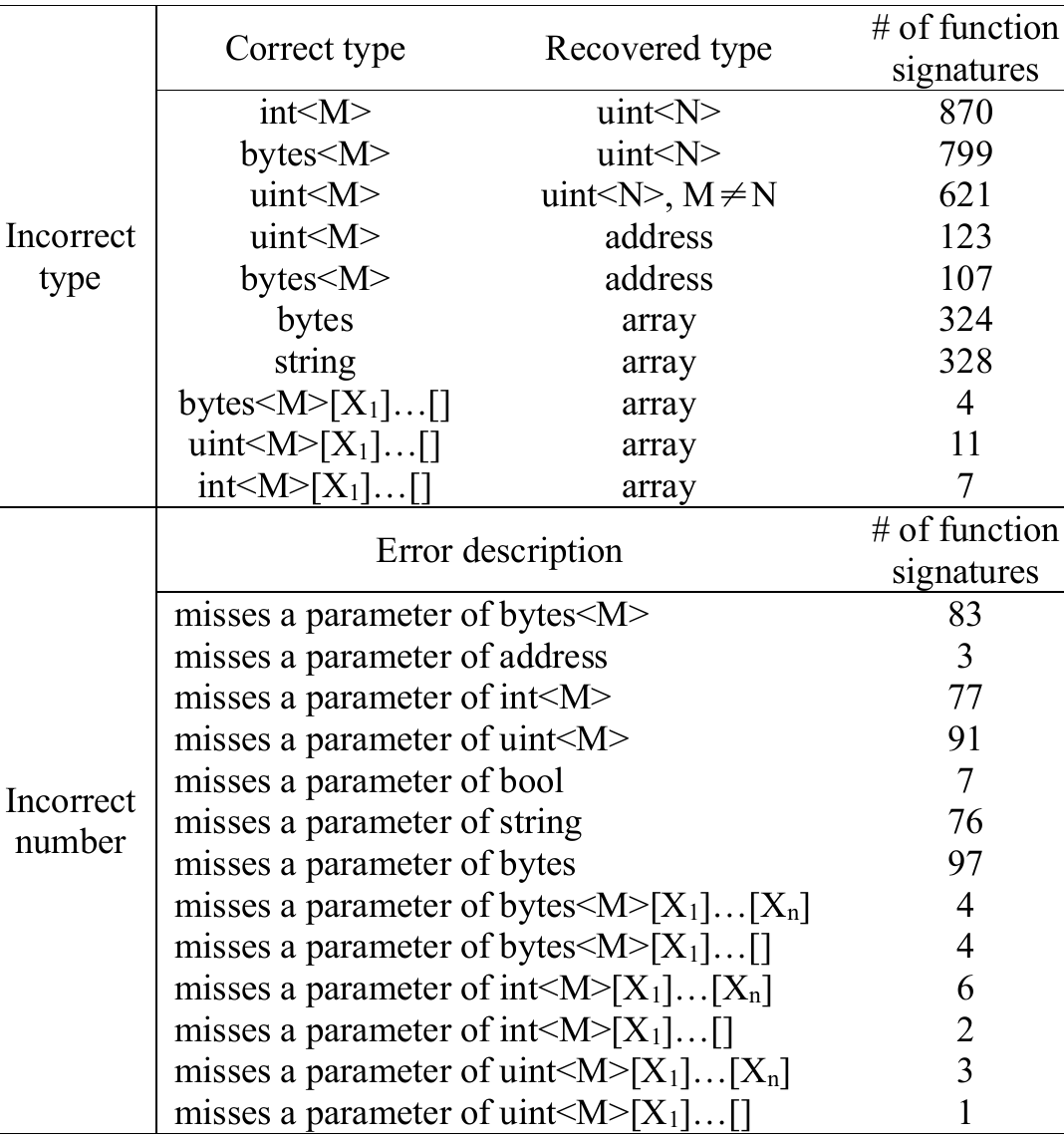}
	%\vspace{-2ex}
	\label{tab_error_dataset2}
\end{table}

\begin{table}[ht!]
	\centering
	%\vspace{-3ex}
	\caption{Errors of incorrect parameter types from dataset 3 recovered by Gigahorse and Eveem}
	%\vspace{-2ex}
	\includegraphics[width=0.41\textwidth]{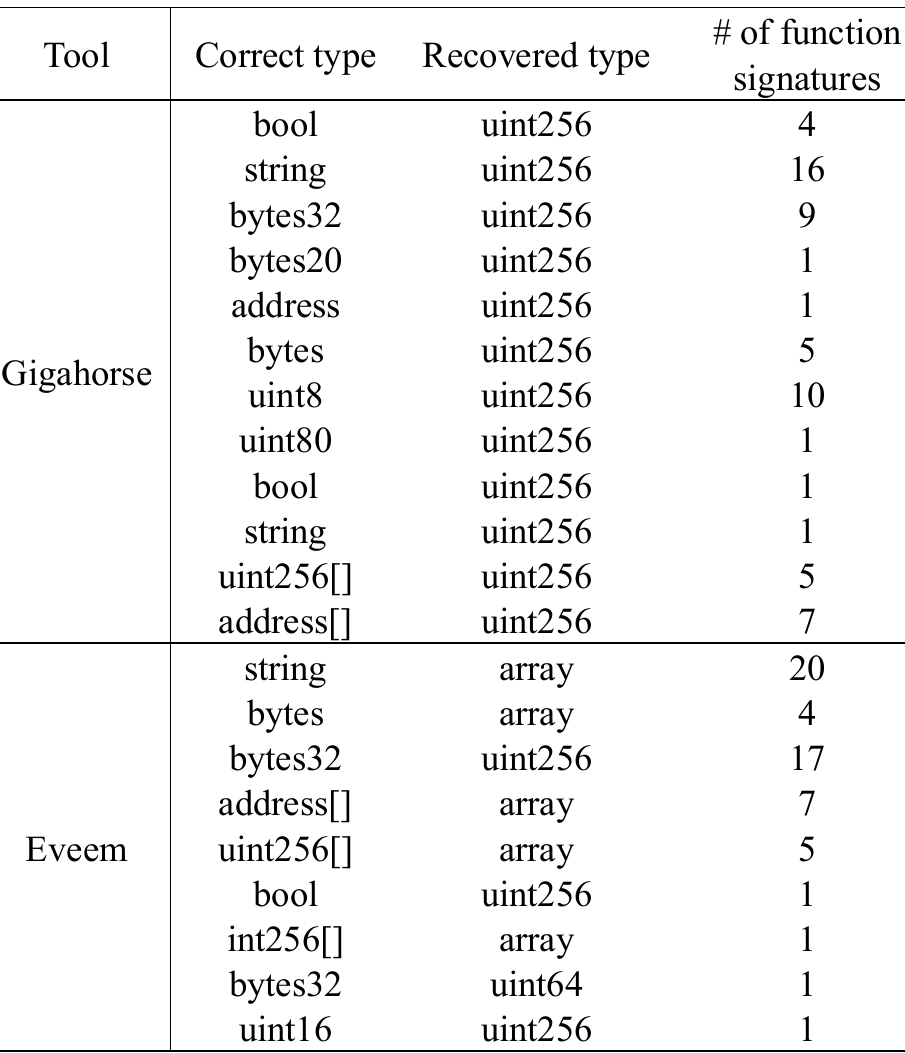}
	%\vspace{-2ex}
	\label{tab_type_error_dataset3}
\end{table}

\begin{table*}[ht!]
	\centering
	%\vspace{-3ex}
	\caption{Errors of incorrect parameter numbers from dataset 3 found by Gigahorse and Eveem}
	%\vspace{-2ex}
	\includegraphics[width=0.75\textwidth]{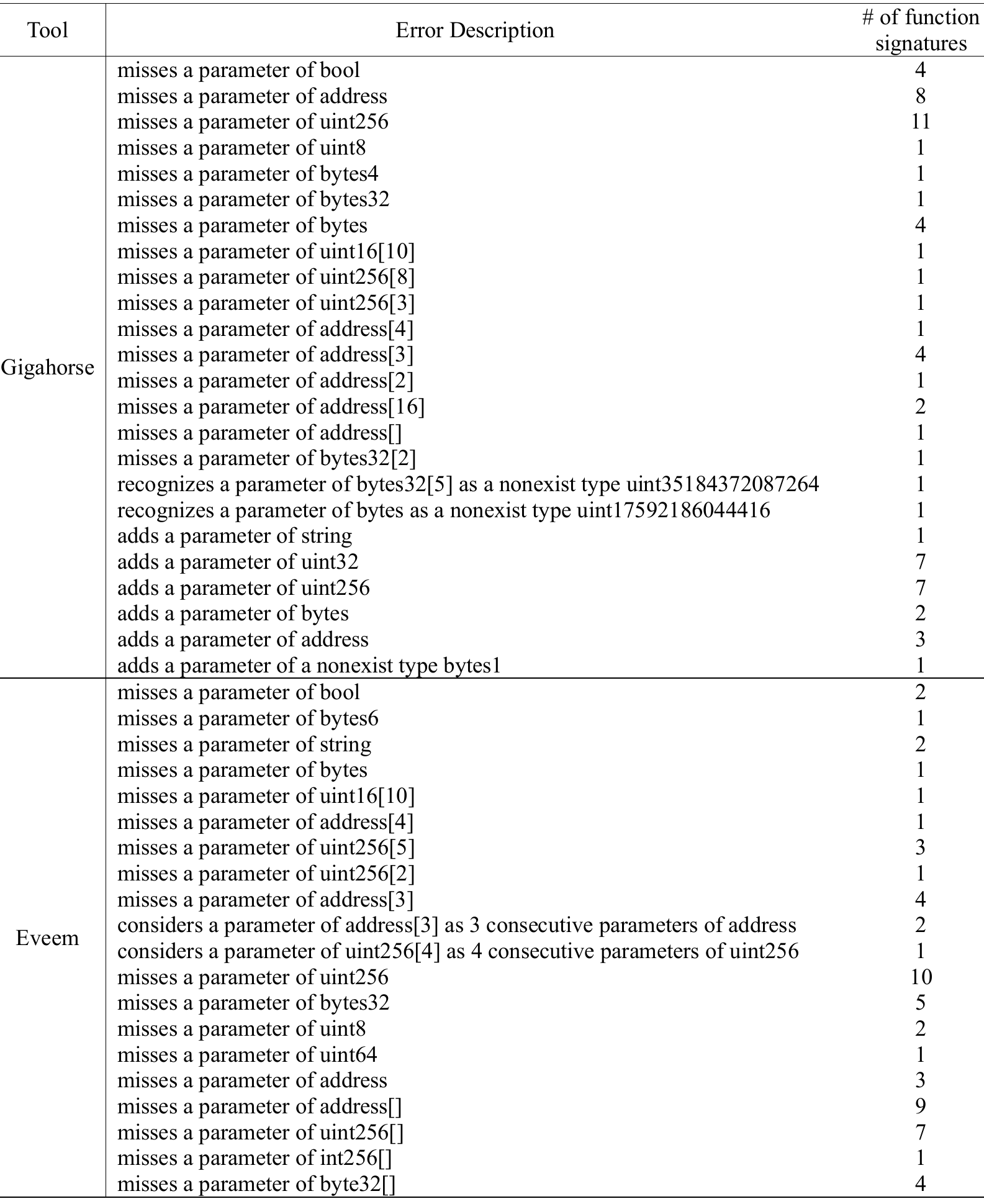}
	%\vspace{-2ex}
	\label{tab_number_error_dataset3}
\end{table*}

Table \ref{tab_number_error_dataset3} lists the number of function signatures from dataset 3 whose number of parameters are incorrectly determined by Gigahorse and Eveem. This table also describes why the number of parameters is incorrect. If a recovered function signature misses or adds several parameters, we count it repeatedly.

\subsection*{H. Boosting the Performance of Existing Smart Contracts Fuzzers}
All existing fuzzing tools\cite{contractfuzzer, sfuzz,ValentinICSE20,JingxuanCCS19} to discover vulnerabilities in smart contracts assume the availability of function signatures. Unfortunately, they are not available for most of the deployed smart contracts. Without function signatures, existing smart contract fuzzers may have to regard the list of parameters as a byte sequence and generate random byte sequences as input instead of applying specific mutation strategies for different parameter types. 

To evaluate how function signatures recovered by \texttt{\footnotesize SigRec} can benefit smart contract fuzzers, we develop a tool, named ContractFuzzer$^-$ which is the same as ContractFuzzer~\cite{contractfuzzer}, a state-of-the-art open-source smart contract fuzzer, except that ContractFuzzer$^-$ does not know function signatures. 
With function signatures, ContractFuzzer knows the type of each parameter and then adopts specific mutation strategy. In contrast, without knowing function signatures, ContractFuzzer$^-$ regards the list of parameters as a byte sequence and generates random byte sequences like a traditional black-box fuzzer~\cite{random}.
To make a fair comparison, we ensure that the length of arguments generated by ContractFuzzer is equal to the length of arguments generated by ContractFuzzer$^-$. More precisely, ContractFuzzer$^-$ first applies the same mutation strategy as ContractFuzzer to produce a list of arguments so that the length of arguments is obtained. Then, it replaces the list with a randomly-generated byte sequence of the same length.

\begin{figure}[ht]
	\centering
	\vspace{-1ex}
	\includegraphics[width=0.4\textwidth]{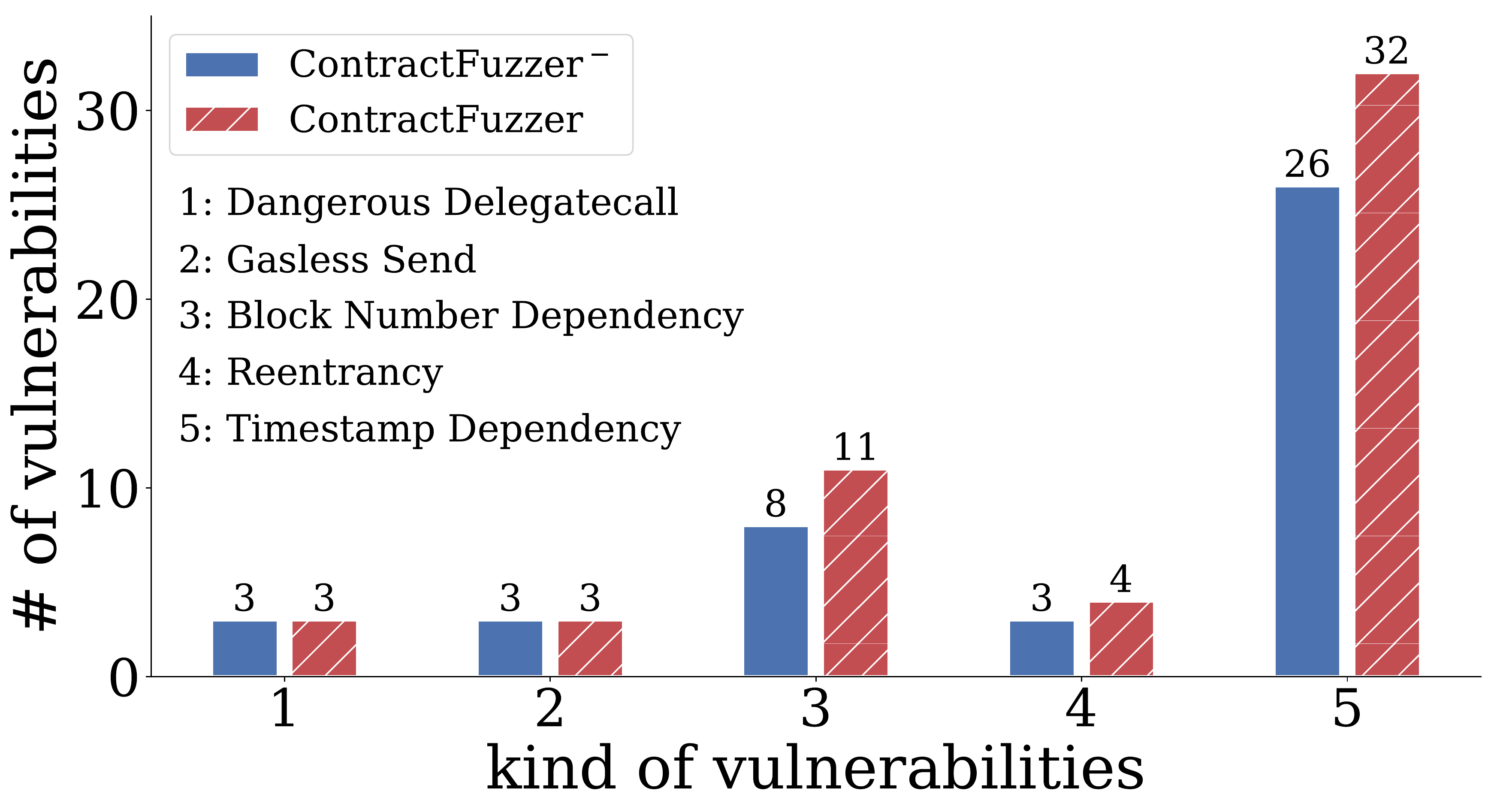}
	\vspace{-1ex}
	\caption{Vulnerabilities found by ContractFuzzer and ContractFuzzer$^-$}
	\vspace{-2ex}
	\label{fig_fuzz}
\end{figure}

We randomly select 1,000 smart contracts for this evaluation instead of conducting a larger scale experiment because ContractFuzzer has to maintain an Ethereum private chain and therefore it runs quite slow. We feed ContractFuzzer with the  function signatures recovered by \texttt{\footnotesize SigRec} for all the public/external functions of the 1,000 smart contracts.
ContractFuzzer discovers 53 vulnerabilities belonging to 5 kinds of vulnerabilities in 40 smart contracts.
By contrast, ContractFuzzer$^-$ only finds 43 vulnerabilities in 32 smart contracts.
Manual investigation shows that all vulnerable smart contracts and vulnerabilities found by ContractFuzzer$^-$ are also discovered by ContractFuzzer. Therefore, with the help of function signatures, ContractFuzzer discovers 25\% ($(40-32)/32$) more vulnerable smart contracts and 23\% ($(53-43)/43$) more vulnerabilities, as shown in Fig. \ref{fig_fuzz}. By further investigating why ContractFuzzer$^-$ misses the 10 vulnerabilities, we find that it is due to the lack of function signatures and detail the observations as follows. 

\noindent\textbf{Invalid parameters.} Some parameter types have structural layouts. For example, a \textsf{\small bytes} has an \emph{offset} field which must point to the \emph{num} field. Unfortunately, since ContractFuzzer$^-$ is not aware of such structure, it just randomly generates a byte sequence as arguments, thus making it almost impossible to generate a valid structural input. Note that a smart contract halts immediately when reading an invalid input. Two missed vulnerabilities are due to this reason.

Listing \ref{code_invalid_parameter} presents one such missed vulnerability. It is a reentrancy bug allowing an attacker to steal money from the smart contract. More precisely, \textit{\_to.call.value}() will call the fallback function of \textit{\_to} since no function is specified. If the fallback function of \textit{\_to} calls the function \textit{checkForward}() (Line 1) again, the money of this vulnerable smart contract will be sent to \textit{\_to} repeatedly by executing Line 6. \textit{checkForward}() is a public function because a function is public by default before Solidity 0.5.0~\cite{default_public} though it can be specified as being public, external, internal, or private. ContractFuzzer$^-$ cannot find the vulnerability since the contract will halt the execution at Line 1 due to reading the invalid input, \textit{\_data}.
%\vspace{-1ex}
\begin{lstlisting}[caption={A missed vulnerability due to an invalid parameter}, label={code_invalid_parameter}]
function checkForward(bytes _data) constant returns(bool, bool) {
  return _forward(allowedForwards[sha3(_data[0], _data[1], _data[2], _data[3])], _data);
}
function _forward(address _to, bytes _data) internal  returns(bool, bool) {
......
  if (!_to.call.value(msg.value)(_data)) {//vulnerability
    _returnFee(msg.sender, fee);
    return (false, _safeFalse());
  }
  return (true, _applyRefund(startGas + additionalGas));
}
\end{lstlisting}

\noindent\textbf{Improper unsigned integer generation.} ContractFuzzer adopts a special mutation strategy for unsigned integers to generate small unsigned integers. In particular, it divides the input space of an unsigned integer into regions, and randomly selects integers from each region. For instance, for a \textsf{\small uint256}, ContractFuzzer divides the input spaces into $[0, 2^8-1]$, $[0, 2^{16}-1]$, ..., $[0, 2^{256}-1]$.
However, it is difficult for ContractFuzzer$^-$ to generate small unsigned integers since it treats an unsigned integer as a 32-bytes byte sequence and it sets each bit to 0 or 1 randomly. For example, if ContractFuzzer$^-$ needs to produce an unsigned integer smaller than $2^8$, it must set each bit in the 31 higher-order bytes to zeros, and the probability is $(1/2)^{31\times 8}$. Consequently, ContractFuzzer$^-$ cannot find the vulnerability that can only be triggered by small unsigned integers. Seven missed vulnerabilities are due to this reason.%

Listing \ref{code_always-big} presents one such vulnerability. \textit{startBlock} is set to the current block number (Line 6), and therefore the comparison at Line 11 depends on the block number. ContractFuzzer detects a vulnerability because the following requirements are satisfied~\cite{contractfuzzer}. First, the miner can control the comparison result at Line 11 because it can control the block number~\cite{homestead}. Second, Line 12, which is control dependent on Line 11, modifies the global variable, \textit{endBlock}. Line 12 is control dependent on Line 11 because the execution of Line 11 determines whether Line 12 can be executed. For example, if the condition at Line 11 does not hold, the smart contract will halt immediately, and thus Line 12 will not be executed. Unfortunately, ContractFuzzer$^-$ fails to discover the vulnerability because it is difficult for it to generate a small unsigned integer \textit{\_block} that can pass the check at Line 10. Note that \textsf{\small uint} is shorthand of \textsf{\small uint256}~\cite{type}.

%\vspace{-1ex}
\begin{lstlisting}[caption={A missed vulnerability due to improper \textsf{\small uint} generation}, label={code_always-big}]
uint public startBlock;
uint public endBlock;
......
function start(uint _block) external onlyOwner() {
  require(_block < 54000);
  startBlock = block.number;
  endBlock = safeAdd(startBlock, _block);
}
function adjustDuration(uint _block) external onlyOwner() {
  require(_block <= 72000);
  require(_block > safeSub(block.number, startBlock));
  endBlock = safeAdd(startBlock, _block);//vulnerability
}
\end{lstlisting}

\noindent\textbf{Improper \textsf{\small bool} value generation.} ContractFuzzer randomly sets a \textsf{\small bool} value to true or false. Please recall that a \textsf{\small bool} argument is extended to 32 bytes, and hence ContractFuzzer$^-$ considers a \textsf{\small bool} argument as a 32-bytes byte sequence.
Consequently, the probability for ContractFuzzer$^-$ to set a \textsf{\small bool} value to false is just $(1/2)^{32\times 8}$ (i.e., all bits of the byte sequence are set to 0). Therefore, ContractFuzzer$^-$ cannot find the vulnerability which can only be triggered by a bool parameter set to false. One undiscovered vulnerability is due to this reason as shown in Listing \ref{code_always-true}.
%\vspace{-1ex}

\textit{now} means the current timestamp, and therefore the comparison at Line 7 depends on the current timestamp. ContractFuzzer detects the vulnerability because the following requirements are satisfied. First, the miner can control the comparison result at Line 7 because it can control the timestamp~\cite{homestead}. Second, the miner can affect the global variable, \textit{rateSale} at Line 8 which is control dependent on the comparison at Line 7. However, ContractFuzzer$^-$ always sets \textit{isGlobalPause} to true (Line 4) because it always sets the \textsf{\small bool} parameter \textit{\_state} (Line 3) to true. Therefore, the last two expressions $\rm{\textit{now}}>\rm{\textit{startSale}}$ and $\rm{\textit{now}}<\rm{\textit{finishSale}}$ will not be evaluated (Line 7) since \textit{isGlobalPause} is already true. That is why ContractFuzzer$^-$ misses the bug. 

\begin{lstlisting}[caption={A missed vulnerability due to improper \textsf{\small bool} generation}, label={code_always-true}]
bool public isGlobalPause=false;
uint public rateSale = 400*10**18;
function globalPause(bool _state) public onlyOwner {
  isGlobalPause = _state;
}
function changeRateSale(uint _tokenAmount) public onlyOwner {
  require(isGlobalPause || (now > startSale && now < finishSale));
  rateSale = _tokenAmount;
}
\end{lstlisting}

\noindent\textbf{Summary}: 
\textit{\texttt{\footnotesize SigRec} empowers existing smart contract fuzzers to analyze all deployed smart contracts by providing function signatures. 
The function signatures recovered by \texttt{\footnotesize SigRec} help fuzzers adopt proper fuzzing strategies and find much more vulnerabilities}.

\subsection*{I. Improving the Result of Reverse Engineering the Bytecode of Smart Contracts}
We demonstrate the usefulness of \texttt{\footnotesize SigRec} to reverse engineering the bytecode of smart contracts by enhancing Erays \cite{erays}. 
We believe that other reverse engineering tools will also benefit from \texttt{\footnotesize SigRec}.
Erays takes in the bytecode, and outputs a register-based instructions which are more readable than EVM bytecode~\cite{erays}. Unfortunately, Erays  recover neither function signatures nor variable types, and its results contain lots of code produced by the compiler for accessing parameters, making it difficult to understand the program.

We develop a tool named Erays$^+$, which first uses Erays to reverse engineering the bytecode and then converts the outputs of Erays into a more readable form. Erays$^+$ adds 1,456 lines of Python to Erays. The enhancement consists of five parts.
First, Erays$^+$ adds the function signature for each public/external function. 

Second, Erays$^+$ replaces meaningless variable names with meaningful parameter names if these variables are copied from parameters (e.g., Erays$^+$ replaces a variable \textit{x} with \emph{arg\rm{1}} indicating that \textit{x} the 1st parameter; Erays$^+$ also replaces a variable \textit{y} with \emph{num(arg\rm{1})} if \textit{y} is copied from the \textit{num} field of the 1st parameter)

Third, for a assignment statement, if a parameter is assigned to a variable, Erays$^+$ adds the type of the parameter to the variable.
Fourth, if a variable refers to the \emph{num} field of a dynamic \textsf{\small array}, or a \textsf{\small bytes}, or a \textsf{\small string}, Erays$^+$ replaces its meaningless variable name with a meaningful name, \emph{num(argN)} indicating the \textit{num} field of the \textit{N}st parameter. 
Finally, Erays$^+$ replaces the bulk of code generated by the compiler for accessing parameters with simple assignment statements. To recognize such code, we first synthesize a set of contracts containing synthesized functions which take in various parameter types in the same way described in 
\S 5.6 of manuscript~\cite{SigRec} %\ref{sec_compare}
and then compile these synthesized contracts into EVM bytecode using all major versions of Solidity. After reverse engineering of these EVM bytecode by Erays, we learn the code patterns. Hence, Erays$^+$ 
can recognize the code for accessing parameters by scanning the results of Erays according to the code patterns. Please note that Erays$^+$ is used for demonstrating the usefulness of \texttt{\footnotesize SigRec} to reverse engineering the bytecode of smart contracts, rather than a full-fledged decompiler.

\begin{lstlisting}[caption={The source code of an example contract}, label={code_reverse_example}]
contract EraysABI{
  string s;
  function f1(uint[2][3][4][] data, string name) public {
    uint item = data[3][2][0][1];
    s = name;
  }
}
\end{lstlisting}

Listing \ref{code_reverse_example} shows a contract which defines a \textsf{\small string} \textit{s} (Line 2) and a public function \textit{f}1 (Line 3). \textit{f}1 takes in a multidimensional dynamic \textsf{\small array} \textit{data}, and a \textsf{\small string} \textit{name}. \textit{f}1 gets access to one array item (Line 4) and the \textsf{\small string} \textit{name} (Line 5). 
We compile this contract using Solidity v 0.4.22 and use Erays to reverse engineer the bytecode. The result, as shown in Listing \ref{code_reverse_erays}, is difficult to interpret because it consists of 112 lines of instructions and complicated program constructs (e.g., nested \textit{while}, \textit{if}...\textit{else}, \textit{break}) that cannot be found in the source code. Line 1 guarantees that no Ether can be sent to \textit{f}1. Lines 2 -- 58 read the 1st parameter (i.e., the multidimensional dynamic \textsf{\small array}, \textit{data}) into the memory by a nested loop with 3 layers (Lines 13 -- 58). Lines 59 -- 65 read the 2nd parameter (i.e., the \textsf{\small string}, \textit{name}) into the memory by a \textsf{\footnotesize CALLDATACOPY}. Line 70 reads the \textsf{\small array} item from the memory to the stack, which corresponds to Line 4 in the source code. Before accessing the \textsf{\small array} item, several bound checks (Lines 66 -- 69) are performed to prevent array overrun. In particular, Line 66 reads the size of the highest dimension which can only be obtained at runtime, and checks whether the index (i.e., 3) is smaller than the size. Lines 71 to 111 copies the \textsf{\small string} (i.e., \textit{name}) from the memory to the string \textit{s}. The last line halts the execution. We can see that most code is produced by the compiler for accessing parameters.

\begin{lstlisting}[caption={Reverse engineering result outputted by Erays}, label={code_reverse_erays}, breaklines,columns=flexible]
assert(0 == msg.value)
$s5 = 0x4 + c[0x4]
$t = c[$s5]
$s7 = $t
$s8 = $m
$m = $m + (0x20 + (0x20 * $t))
$t = 0x20 + $s5
$s5 = $s8
$s6 = $t
m[$s8] = $s7
$s9 = 0x0
$s10 = 0x20 + $s8
while (0x1) {
  if ($s9 >= $s7)
    break
  $s13 = $m
  $m = 0x80 + $m
  $s12 = (0x300 * $s9) + $s6
  $s14 = 0x0
  $s15 = $s13
  while (0x1) {
    if ($s14 >= 0x4)
      break
    $s18 = $m
    $m = 0x60 + $m
    $s17 = (0xc0 * $s14) + $s12
    $s19 = 0x0
    $s20 = $s18
    while (0x1) {
      if ($s19 >= 0x3)
        break
      $s23 = $m
      $m = 0x40 + $m
      calldatacopy($s23, (0x40 * $s19) + $s17, 0x40)
      m[$s20] = $s23
      $t = $s19
      $s19 = 0x20 + $s20
      $s20 = 0x1 + $t
      $t = $s19
      $s19 = $s20
      $s20 = $t
    }
    m[$s15] = $s18
    $t = $s14
    $s14 = 0x20 + $s15
    $s15 = 0x1 + $t
    $t = $s14
    $s14 = $s15
    $s15 = $t
  }
  m[$s10] = $s13
  $t = $s9
  $s9 = 0x20 + $s10
  $s10 = 0x1 + $t
  $t = $s9
  $s9 = $s10
  $s10 = $t
}
$s6 = 0x4 + c[0x24]
$t = c[$s6]
$s9 = $m
$m = $m + (0x20 + (0x20 * ((0x1f + $t) / 0x20)))
m[$s9] = $t
calldatacopy(0x20 + $s9, 0x20 + $s6, $t)
$s3 = $s9
assert(0x3 < m[$s5])
assert(0x1)
assert(0x1)
assert(0x1)
$s4 = m[0x20 + m[m[0x40 + m[0x60 + (0x20 + $s5)]]]]
$s8 = m[$s9]
$s10 = s[0x0]
m[0x0] = 0x0
$s9 = sha3(0x0, 0x20)
$t = 0x20 + $s3
$s7 = $s9 + ((0x1f + ((((0x100 * (0 == (0x1 & $s10))) - 0x1) & $s10) / 0x2)) / 0x20)
$s10 = $t
if (0x1f >= $s8){
  s[0x0] = ($s8 + $s8)|(0xffffffffffffffffffffffffffffffffffffffffffffffffffffffffffffff00 & m[$t])
} else {
  s[0x0] = 0x1 + ($s8 + $s8)
  if ($s8){
    $t = $s8
    $s8 = $s10
    $s10 = $s10 + $t
    while (0x1) {
      if ($s10 <= $s8)
        break
      s[$s9] = m[$s8]
      $t = $s8
      $s8 = $s10
      $s10 = 0x20 + $t
      $t = $s8
      $s8 = $s10
      $s10 = $t
      $t = $s9
      $s9 = $s10
      $s10 = 0x1 + $t
      $t = $s9
      $s9 = $s10
      $s10 = $t
    }
  }
}
$s10 = $s9
while (0x1) {
  if ($s7 <= $s10)
    break
  s[$s10] = 0x0
  $s10 = 0x1 + $s10
}
stop()
\end{lstlisting}

\begin{lstlisting}[caption={Reverse engineering result outputted by Erays$^+$}, label={code_reverse_erays_plus}]
0xd780c374(uint256[2][3][4][] arg1,string arg2) {
  assert(0 == msg.value)
  assert(0x3 < num(arg1))
  assert(0x1)
  assert(0x1)
  assert(0x1)
  uint256 $s4 = arg1[3][2][0][1]
  string s = arg2
  stop()
}
\end{lstlisting}

Listing \ref{code_reverse_erays_plus} shows the result of Erays$^+$ from the same bytecode. Erays$^+$ adds the function signature provided by \texttt{\footnotesize SigRec} (Line 1). 
Since the recovered function signature does not include function name and parameter names, 
Erays$^+$ uses the function id and uses \textit{argN} to represent the \textit{N}th parameter. 
Lines 7 and 8 in Listing \ref{code_reverse_erays_plus} correspond to Lines 4 and 5 in the source code. Besides, Erays$^+$ retains the code for checking Ether transfer (Line 2), array overrun (Lines 3 -- 6) and terminating execution (Line 10) which are produced by the compiler. Please note that Erays simplifies the bound checks for the lower dimensions to true (Lines 4 -- 6) because the sizes of the lower dimensions are known and thus the outcomes of these bound checks can be determined without execution. 
We then explain how Erays$^+$ simplifies the result of Erays. First, Erays$^+$ replaces a meaningless variable name \textit{m}[\$\textit{s}5] with \textit{num}(\textit{arg}1) denoting the size of the highest dimension of \textit{arg}1 which is a multidimensional \textsf{\small array}. 
Second, Erays$^+$ replaces 58 lines of code in Listing \ref{code_reverse_erays} for reading one item of \textit{arg}1 with a single line of code in Listing \ref{code_reverse_erays_plus} (Line 7). In the meanwhile, Erays$^+$ adds a type to \$\textit{s}4 because Erays$^+$ knows that \$\textit{s}4 is an array item of the first parameter.
Then, Erays$^+$ replaces 48 lines of code in Listing \ref{code_reverse_erays} for reading \textit{arg}2 with a single line of code in Listing \ref{code_reverse_erays_plus} (Line 8). Erays$^+$ also adds a type to \textit{s} because it is directly copied from the second parameter. Therefore, the results of Erays$^+$ are much easier to interpret than the original outputs of Erays.

\begin{figure}[ht]
	\centering
	\vspace{-1ex}
	\includegraphics[width=0.45\textwidth]{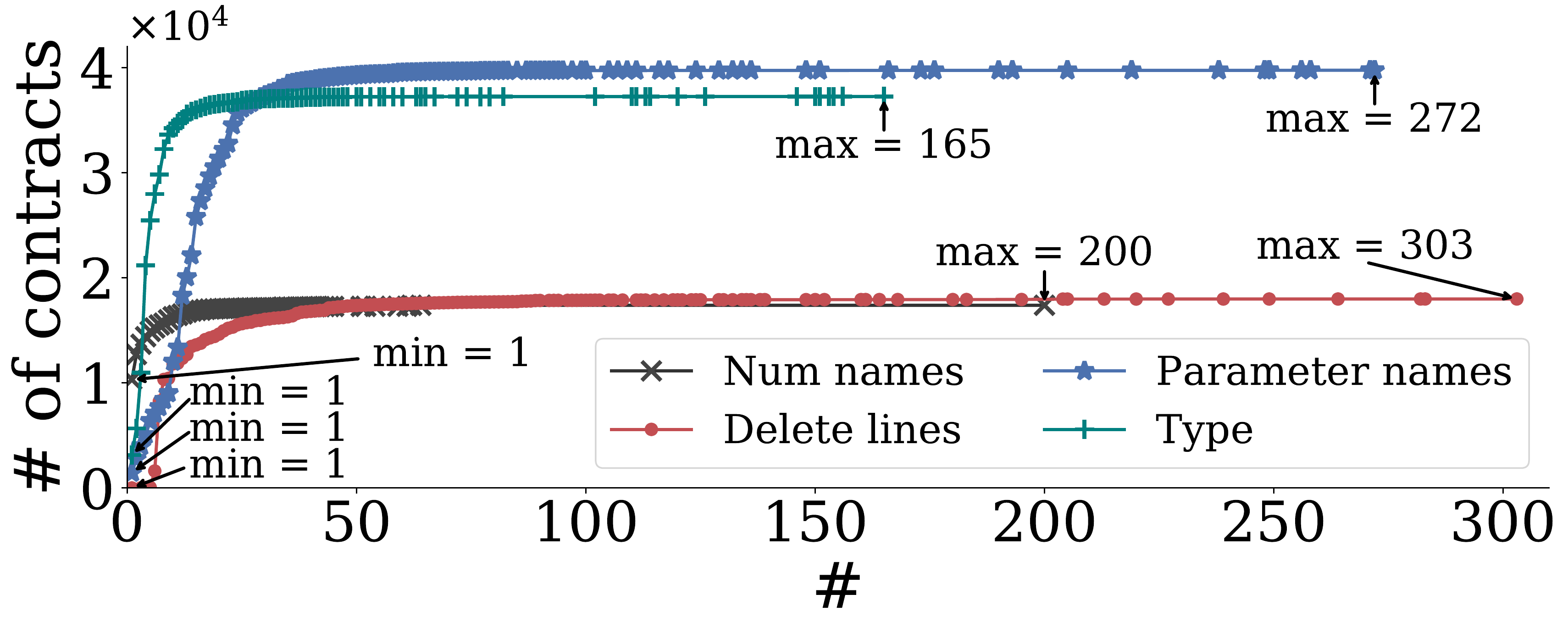}
	\vspace{-2ex}
	\caption{The numbers of smart contracts whose reverse-engineering results are improved by Erays$^+$}
	\vspace{-2ex}
	\label{fig_cdf_erays}
\end{figure}

%results
%We then use all open-source contracts that have been deployed in Ethereum to evaluate the effectiveness of Erays$^+$. 
We use Erays$^+$ to process 53,166 unique open-source contracts and obtain the results from 39,935 out of them (Erays crashes in processing the other 13,231 contracts). We find that Erays$^+$ improves the readability of the outputs of all 39,935 contracts. The numbers of smart contracts whose reverse engineering results can be improved by Erays$^+$ in terms of adding types, adding parameter names, adding \textit{num} names, and removing compiler-generated code are 37,249, 39,740, 17,381 and 17,983, respectively.
Fig. \ref{fig_cdf_erays} depicts the cumulative distribution function (CDF) plots for each case. Each green point ($x$, $y$) in this figure denotes that there are $y$ smart contracts whose reverse engineering results are improved by Erays$^+$ in terms of adding types, for each, the number of added types is no more than $x$. The other colors of points can be interpreted in the same way. We find that the numbers of added types, added parameter names, added \textit{num} names, removed code lines range from 1 to 165, 1 to 272, 1 to 200, and 1 to 303, respectively.

\noindent\textbf{Summary}: \textit{The information provided by \texttt{\footnotesize SigRec} including function signatures can obviously improve the readability of the results of reverse engineering tools for smart contracts}

\subsection*{J. Short Address Attacks Detected by ParCheck}
Table \ref{tab_short_detail} presents the addresses of the victim smart contracts and the number of attacking transactions.

\begin{table}[ht!]
	\centering
	%\vspace{-2ex}
	\caption{Short address attacks detected by ParChecker}
	%\vspace{-2ex}
	\includegraphics[width=0.45\textwidth]{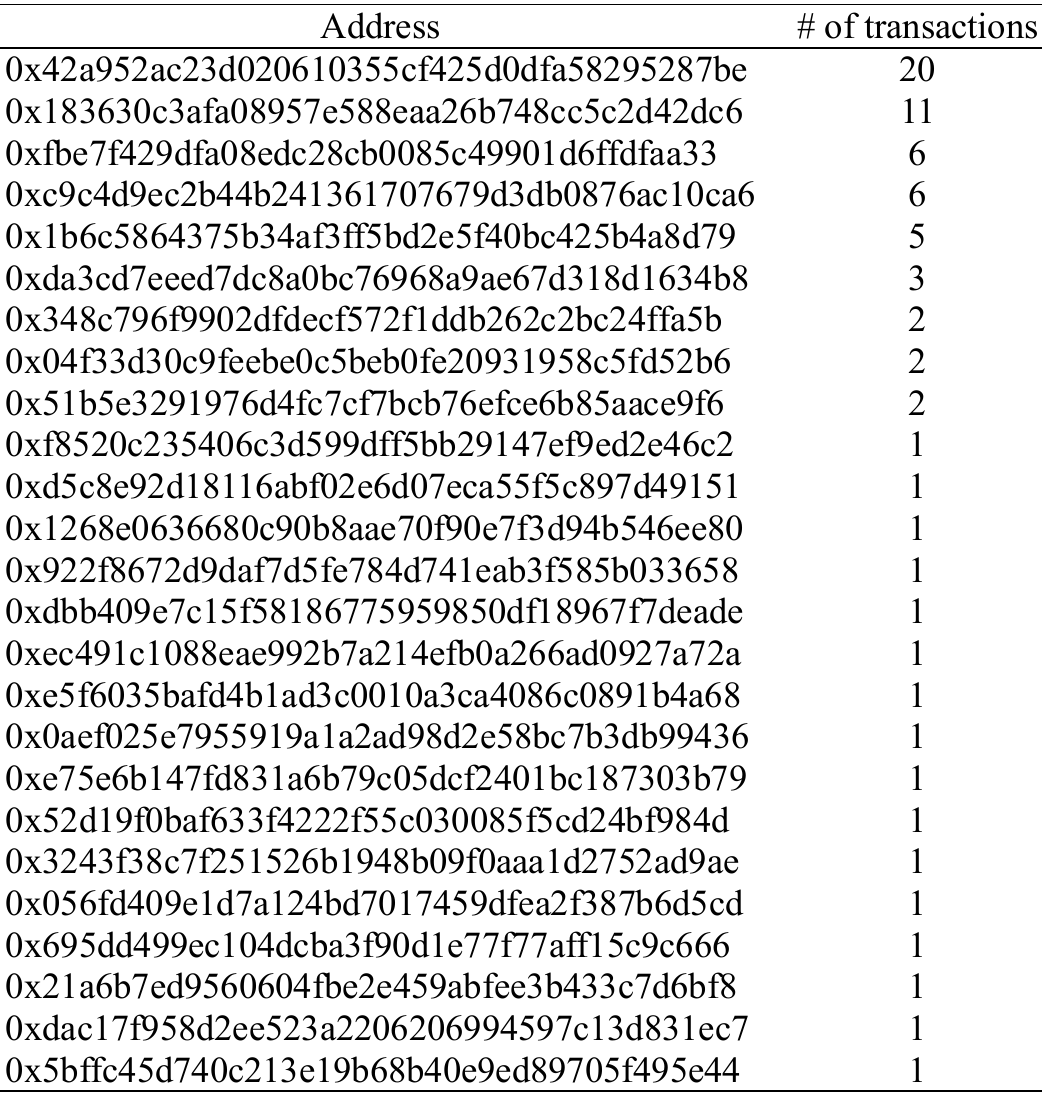}
	%\vspace{-2ex}
	\label{tab_short_detail}
\end{table}

\subsection*{K. Related Work on Reverse Engineering Binaries}
Caballero et al. propose to recover function interfaces by dynamic analysis~\cite{caballero_ndss}. Their approach applies several architecture-specific rules, e.g., it ignores any access to ESP to exclude unnecessary input locations~\cite{caballero_ndss}. Different from our work, their recovered function interfaces provide the information of the parameter location (i.e., on the stack, in a register or in the memory), parameter length, whether the parameter is a pointer or not, etc., rather than the type list declared in the source code. Go\"{e}r et al. propose a set of architecture-specific heuristic rules (e.g., floats are passed through \%xmm0 to \%xmm7) to infer parameter number and parameter types~\cite{lightweight}. However, the recovered type is either an integer, a float, or an address~\cite{lightweight} rather than the accurate type. 

SecondWrite proposes a modified value set analysis to recognize parameters~\cite{secondwrite,rewrite}. For a call instruction \textit{i}, if a location \textit{l} is defined at somewhere before \textit{i} and is used after \textit{i}, and then Zhang et al.'s work considers the location \textit{l} as a parameter of \textit{i}~\cite{zhang}. Balakrishnan and Reps consider the stack items with positive offsets to the location pointed by ESP as parameters~\cite{esp}. These studies~\cite{rewrite,zhang, esp} just recognize parameters, but do not recover parameter types.
Qiao and Sekar apply static analysis to discover function boundaries~\cite{sekar_dsn}. BYTEWEIGHT learns signatures for function starts using a weighted prefix tree, recognizes function starts by matching binary fragments with the signatures, and applies value set analysis to recognize function bodies and discover function boundaries~\cite{byteweight}. Shin et al. train a recurrent neural network to take bytes of the binary as input, and predict, for each location, whether a function boundary is
present at that location~\cite{rnn}. FID applies symbolic execution to generate semantic information, learns the function recognition model, and recognizes function boundaries by machine learning~\cite{fid}. However, these studies~\cite{sekar_dsn,byteweight,rnn,fid} just discover function boundaries, but do not recover function signatures.

REWARDS applies architecture-specific rules which take advantage of system calls, standard library calls and type-revealing instructions (e.g., MOVS moves a string) to infer variable types~\cite{rewards}. Guilfanov infers variable types from standard library function prototypes~\cite{simple}. Besides standard library functions, TIE extracts rules from instructions and applies these rules to infer variable types~\cite{tie}. Troshina et al. extract architecture-specific rules (e.g., the address computed by $m+j\times4$ is used for accessing the array item \textit{m}[\textit{j}]) from the accessing patterns of structs and arrays, and apply the rules to discover structs and arrays from binaries~\cite{array}. Retypd is a powerful type system to recover variable types, and it is more accurate than TIE, REWARDS, and SecondWrite~\cite{retypd}. Considering the code which accesses a variable as a feature to identify the variable type, Katz et al. uses statistical language models to predict variable types~\cite{katz} and Xu et al. use machine learning to recover variable types~\cite{xu}. Howard recovers data structures (e.g., array) by applying rules (e.g. if \textit{A} is the address of a structure, *($A+4$) presumably accesses a field in this structure) to execution traces~\cite{howard}. However, these rules for other architectures cannot be used to recover the function signatures of smart contracts due to the significant differences between those architectures and EVM.

\end{document}